\documentclass{aastex62}
\usepackage{natbib}
\usepackage{color}
\usepackage{lineno}
\usepackage{chngpage}
\usepackage{multirow}
\renewcommand\thetable{\arabic{table}}
\received{--}
\revised{--}
\accepted{--}
\submitjournal{ApJ}

\shorttitle{The intrinsic properties for Fermi TeV blazars}
\shortauthors{Zhou et al.}

\begin{document}
\title{The Intrinsic Properties of Multiwavelength Energy Spectra for Fermi Teraelectronvolt Blazars}

\correspondingauthor{Y.G. Zheng}\email{ynzyg@ynu.edu.cn}

\author{R.X. Zhou}
\affiliation{Department of Physics, Yunnan Normal University, Kunming, Yunnan, 650092, People's Republic of China}

\author{Y.G. Zheng}
\affiliation{Department of Physics, Yunnan Normal University, Kunming, Yunnan, 650092, People's Republic of China}

\author{K.R. Zhu}
\affiliation{Department of Physics, Yunnan Normal University, Kunming, Yunnan, 650092, People's Republic of China}

\author[0000-0002-9071-5469]{S.J. Kang}
\affiliation{School of Physics and Electrical Engineering, Liupanshui Normal University, Liupanshui, Guizhou, 553004, People's Republic of China}



\begin{abstract}
In this paper, we have selected a sample of 64 teraelectronvolt blazars, with redshift, from those classified in the fourth Fermi Large Area Telescope source catalog\footnote{\url{https://fermi.gsfc.nasa.gov/ssc/data/access/lat/8yr_catalog/}}. We have obtained the values of the relevant physical parameters by performing a log-parabolic fitting of the average-state multiwavelength spectral energy distributions. We estimate the range of the radiation zone parameters, such as the Doppler factor (${D}$), the magnetic field strength ($B$), the radiative zone radius ($R$) and the peak Lorentz factor (${\gamma _{\rm p}}$) of nonthermal electrons. Here, we show that (1) there is a strong linear positive correlation between the intrinsic synchrotron peak frequency and the intrinsic inverse Compton scattering (ICs) peak frequency among different types of blazars; (2) if radio bands are excluded, the spectral index of each band is negatively correlated with the intrinsic peak frequency; (3) there is a strong linear negative correlation between the curvature at the peak and the intrinsic peak frequency of the synchrotron bump, and a weak positive correlation between the curvature at the peak and the intrinsic peak frequency of the ICs bump; (4) there is a strong linear positive correlation between the intrinsic ICs peak luminosity and intrinsic $\gamma$-ray luminosity and between the intrinsic ICs peak frequency and peak Lorentz factor; (5) there is a strong negative linear correlation between $\rm log{\;B}$ and $\rm log{\;\gamma_{p}}$; and (6) there is no correlation between $\rm log{\;R}$ and $\rm log{\;\gamma_{p}}$.
\end{abstract}




\section{Introduction}\label{sect:intro}

Blazars are one of the most extreme subclasses of active galactic nuclei (AGNs), characterized by high luminosity, high polarization, rapid flux variability, radio core dominance, apparent superluminal motions and strong $\gamma$-ray emission (\citealt{1995PASP..107..803U,2014Ap&SS.352..819Y,2016ApJS..226...20F}). Blazars are usually classified as BL Lacertae (BL Lac) objects or flat-spectrum radio quasars (FSRQs). They are distinctive in BL Lac objects characterized by weak or no emission lines (equivalent Width of a line $\rm \leq 5{\AA}$) whereas FSRQs show broad, strong emission lines. Their continuous nonthermal radiation is thought to arise from a relativistic jet oriented in the direction of the observer (\citealt{1986ApJ...310..317G}). The beaming effect is significant to the observational properties of blazars (\citealt{2013PASJ...65...25F,2017RAA....17...66L}) and embodies the relativistic property of a jet. The Doppler factors of most objects have been estimated in literature (\citealt{1999ApJ...521..493L,2009A&A...494..527H,2013RAA....13..259F,2014RAA....14.1135F,2017RAA....17...66L}).

The multiwavelength spectral energy distribution (SED) of BL Lac objects generally displays a double broad bump structure in the $\log\nu-\log\nu f_{\nu}$ frame (\citealt{2018RAA....18...56K,2018RAA....18..120L,2019AcASn..60....7Z,2020MNRAS.491.2771C}). The low-energy bump is located in the range of radio to X-ray frequencies, which is generally believed to be generated by the synchrotron radiation of extreme relativistic electrons in the jet. However, the high-energy bump is located in the frequency range of megaelectronvolts to gigaelectronvolts, and its generation mechanism remains an outstanding issue. Within the lepton origin model, the high-energy bump is generated by the inverse Compton scattering (ICs) process of the extreme relativistic electron. If the soft photons scattered on the inverse Compton come from the synchrotron radiation soft photons inside the jet, the process is known as the synchrotron self-Compton (SSC) process (\citealt{1992ApJ...397L...5M,1997A&A...320...19M,1998MNRAS.299..433F,2018RAA....18...56K,2019ApJ...873....7Z}). At present, the SSC model is able to explain the multiwavelength SED of blazars very well and SSC is generally considered to be the dominant factor in $\gamma$-ray radiation (\citealt{1998MNRAS.299..433F,2019ApJ...873....7Z}). Based on a nonthermal relativistic electron energy distribution (EED), we can reproduce the multiwavelength SED of blazars by the SSC model (\citealt{2018PASP..130h3001Z,2019ApJ...873....7Z,2020MNRAS.499.1188Z}).
It is a significant research topic that links the parameters corresponding to the observation to the parameters corresponding to the radiation region of the jet.

Different SEDs are produced by different particle distributions in the jet radiation region of blazars. The photon spectrum of the blazars depends on the electron spectrum, which is directly related to the physical conditions and processes in the radiation region of the jet. Therefore, we can determine the EED form from to the observed photon spectrum. The study of SED properties is essential to theoretical models and we can obtain important information related to the physical properties of the radiation region through the parameters. The shape and position of the two bumps of the SED in the $\log\nu-\log\nu f_{\nu}$ coordinate system and the mechanism that controls them have been a subject of discussion for some years. \cite{2010ApJ...716...30A} proposed that the synchrotron peak frequency $\nu _{\rm peak}^{\rm syn}$ could be estimated by effective spectral indices $\alpha_{\rm ox}$ and $\alpha_{\rm ro}$, and gave the empirical relations. \cite{2016ApJS..226...20F} presented a least-squares fitting method to estimate the peak frequency $\nu _{\rm peak}^{\rm syn}$ by fitting the broadband SED, and the parameters of the spectral curvature and the peak flux were also obtained. \cite{2021RAA....21....8Z} used an extrapolation model to estimate the peak location of the synchrotron SED component. According to the observation results of the peak frequency of the synchrotron SED component, BL Lac objects could be classified as either low- or high-frequency-peaked (\citealt{1995MNRAS.277.1477P}); more recently they have been classified as low-synchrotron-peaked (LBLs; e.g., $\nu _{\rm peak}^{\rm syn} < {10^{14}}$Hz), intermediate-synchrotron-peaked (IBLs; e.g., ${10^{14}} < \nu _{\rm peak}^{\rm syn} < {10^{15}}$Hz), or high-synchrotron-peaked (HBLs, e.g., $\nu _{\rm peak}^{\rm syn} > {10^{15}}$Hz) sources (\citealt{2010ApJ...716...30A}).

Curvature is a critical spectral parameter that can provide important information enabling us to study particle acceleration mechanisms. \cite{2011ApJ...739...66T} presented the statistical acceleration case, using the multiplicative model of the central limit theorem to generate the EED in the form of a logarithmic parabola. It consists of two mechanisms: energy-dependent acceleration probability and fluctuation of fractional acceleration gain. The energy distribution of the electrons produced by the former is
\begin{equation}
\log \left( {\frac{{{N_e}(\gamma )}}{{cons.}}} \right) =  - r{\log ^2}\left( {\frac{\gamma }{{{\gamma _0}}}} \right) - s\log \left( {\frac{\gamma }{{{\gamma _0}}}} \right)
\label{Eq:1}
\end{equation}
where $\gamma _{\rm 0}$ is the initial electron Lorentz factor, $r = q/(2\log \varepsilon )$ is the EED curvature term, $s =  - 2r/q\log (g/{\gamma _0}) - (q - 2)/2$, and $\varepsilon  = cons.$ is constant. ${p_a} = g/{\gamma ^q}$ denotes the energy-dependent acceleration probabilities.
The energy distribution of the electrons produced by the latter is
\begin{equation}
\log \left( {\frac{{{N_e}(\gamma )}}{G}} \right) =  - \frac{{{c_e}}}{{2\sigma _\gamma ^2}}{\log ^2}\left( {\frac{\gamma }{{{\gamma _0}}}} \right) - (\mu /\sigma _\gamma ^2 - 1)\log \left( {\frac{\gamma }{{{\gamma _0}}}} \right)
\label{Eq:2}
\end{equation}
where $G = {10^{K - \log {\gamma _0} - {\mu ^2}/(2{c_e}\sigma _\gamma ^2)}}$ is a constant term, ${c_e} = 1/{\log _{10}}e$ is the logarithm conversion constant, $\sigma _\gamma ^2 \approx {n_s}{({\sigma _\varepsilon }/\bar \varepsilon )^2}$, $\mu  = {n_s}[{c_e}\log \bar \varepsilon  - {({\sigma _\varepsilon }/\bar \varepsilon )^2}/2]$, and the coefficient of the second term can be defined through the curvature term of the EED: $r = {c_e}/(2\sigma _\gamma ^2)$.

The data of the fourth Fermi Large Area Telescope (LAT) Source Catalog (4FGL), which is by far the largest sample of $\gamma$-ray sources, has been released. There are a total of 5064 $\gamma$-ray sources in 4FGL catalog\footnote{\url{https://fermi.gsfc.nasa.gov/ssc/data/access/lat/8yr_catalog/gll_psc_v21.fit}} (\citealt{2020ApJS..247...33A}). Based on 4FGL, the fourth catalog of AGN  data\footnote{\url{https://doi.org/10.3847/1538-4357/ab791e}} (4LAC) has been released (\citealt{2020ApJ...892..105A}). The 4LAC catalog contains 3207 $\gamma$-ray sources: 2863 high Galactic latitude ($\left| b \right| > 10^\circ $) and 344 low Galactic latitude ($\left| b \right| \le 10^\circ $) $\gamma$-ray sources, most of which (3137) appertain to blazars (\citealt{2020ApJ...892..105A}). The blazars in 4LAC catalog include 694 FSRQs, 1131 BL Lac objects, and 1312 blazar candidates of uncertain type (BCUs). In addition to the 4LAC, we can collect a large amount of multiband data from Space Science Data Center (SSDC\footnote{\url{https://tools.ssdc.asi.it/SED/}}) and the NASA/IPAC Extragalactic Database (NED\footnote{\url{http://ned.ipac.caltech.edu/}}) covering the range of radio to $\gamma$-ray bands. These samples provide new opportunities and challenges for the study of the multiwavelength radiation characteristics and spectral shapes of Fermi blazars.

At present, for the very high energy (VHE; $\rm E > 0.1$TeV) $\gamma$-ray (TeV) band, 228 extragalactic objects can be obtained from the online TeVCat catalog\footnote{\url{http://tevcat.uchicago.edu/}}, which has been compiled from observations by ground-based telescopes (\citealt{2008ICRC....3.1341W}). The radiation of 186 of these sources in the gigaelectronvolt energy band has been detected by Fermi-LAT; 73 of these are TeV blazars (\citealt{2020ApJS..247...33A}). These blazars include 62 TeV BL Lac objects, 7 TeV FSRQs, and 4 BCUs. TeV sources are bright and although rare, they play a crucial role in the study of high-energy (or VHE) radiation mechanisms and spectral shapes, extragalactic background light, and intergalactic magnetic fields (\citealt{2010ApJ...712..238F,2012Sci...338.1190A,2017A&A...608A..68F}).

In recent years, with the increase of the number of TeV sources, some researchers have studied their energy spectrum structure and discussed the differences in the multiwavelength spectrum shapes of different source types (\citealt{2012ApJ...752..157Z,2014ApJ...788..179C,2016RAA....16..103L,2018RAA....18..120L,2018RAA....18...56K,
2018Ap&SS.363..179Z,2019AcASn..60....7Z}). In this work, we have collected the average-state multifrequency data based on all 73 TeV blazars in 4LAC and fit the SED with the log-parabolic model to obtain the relevant physical parameters. Our objective is to summarize the differences in the multiwavelength radiation characteristics of HBLs, IBLs, LBLs, and FSRQs through statistical analysis of sample parameters and to estimate the parameters of the jet radiation zone. The paper is structured as follows: In Section \ref{sect:model}, we introduce the fitting model and parameters. In Section \ref{sect:sample}, we give the sample and data processing. In Section \ref{sect:results}, the results are given. In Section \ref{sect:discusion}, the discussion and conclusion are presented. Throughout the paper, we assume the Hubble constant to be $H_{0}=75$ km s$^{-1}$ Mpc$^{-1}$, the matter energy density to be $\Omega_{\rm M}=0.27$, the radiation energy density to be $\Omega_{\rm r}=0$, and the dimensionless cosmological constant to be $\Omega_{\Lambda}=0.73$.

\section{The models and parameters} \label{sect:model}
\subsection{The log-parabolic model} \label{sect: LP model}
It is worth noting that we can employ a reasonably smooth interpolation (or extrapolation) model to estimate the peak frequency of the two bumps of the SED component (\citealt{2004A&A...413..489M,2005ApJ...630..130B,2009A&A...501..879T,2010ApJ...724.1509U,2021RAA....21....8Z}). Based on this method, we can also obtain the curvature of the energy spectrum, the spectral index of a given frequency, and the peak luminosity.

In the $\nu- f_{\nu}$ frame, we introduce a log-parabolic model with (\citealt{2016ApJ...819..156B})
\begin{equation}
  f(\nu)=k\biggl(\frac{\nu}{\nu_{*}}\biggr)^{-\alpha_{*}-\beta\log(\nu/\nu_{*})}\;
\label{Eq:3}
\end{equation}
where $\alpha_{*}$ and $\beta$ are free parameters, $\nu_{*}$ is the reference frequency, $\alpha_{*}$ equals the local power-law photon index, and $\beta$ is the curvature of the energy spectral bump, which represents the deviation of the spectral slope away from $\nu_{*}$ (\citealt{2021RAA....21....8Z}). This expression can be written as
\begin{equation}
  \log \nu f_{\nu}={\hat A}(\log\nu)^{2} + {\hat B} \log\nu + {\hat C}\;
\label{Eq:4}
\end{equation}
in the $\log\nu-\log\nu f_{\nu}$ frame. In this expression, ${\hat A}=-\beta$, ${\hat B}=1-\alpha_{*}+2\beta\log\nu_{*}$, and ${\hat C}=\log k+\alpha_{*}\log\nu_{*}-\beta(\log\nu_{*})^{2}$.

Generally, nonthermal radiation of blazars can be described by a typical power law:
\begin{equation}
  f_{\nu}\propto\nu^{-\alpha}\;
\label{Eq:5}
\end{equation}
where $ f_{\nu}$ is the flux at frequency $\nu $ and $\alpha $ is the spectral index (\citealt{2016ApJS..226...20F}). By transformation of Equation (\ref{Eq:5}) into $\log\nu-\log\nu f_{\nu}$ frame, we can obtain the relation
\begin{equation}
  \log \nu f_{\nu}=k+(1-\alpha)\log\nu\;
\label{Eq:6}
\end{equation}
where $\textit{k}$ is a constant. Comparing Equation (\ref{Eq:4}) and Equation (\ref{Eq:6}), we find that at $\nu=\nu_{0}$, their slope values are equal, that is,
\begin{equation}
\left[\frac{d(\log \nu
f_{\nu})}{d(\log\nu)}\right]_{\nu=\nu_{0}}=2{\hat A}(\log\nu_{0})+ {\hat B} = 1-\alpha\;
\label{Eq:7}
\end{equation}
Thus, the spectral index at frequency $\nu_{0}$ is
\begin{equation}
\alpha=1-[2{\hat A}(\log\nu_{0})+ {\hat B}]\;
\label{Eq:8}
\end{equation}
and the peak frequency of the synchrotron bump and of the ICs bump is
\begin{equation}
\log\nu_{\rm peak}=-\frac{{\hat B}}{2{\hat A}}\;
\label{Eq:9}
\end{equation}
Based on the definition of the curvature term of the EED in Equation (\ref{Eq:1}) and Equation (\ref{Eq:2}), we can estimate the energy spectral curvature of the synchrotron bump and of the ICs bump at the peak as
\begin{equation}
\beta  =  - \hat A
\label{Eq:10}
\end{equation}
The peak flux can be calculated as
\begin{equation}
{f_{\rm peak}} = k{\left( {\frac{{{\nu _{\rm peak}}}}{{{\nu _ * }}}} \right)^{ - {\alpha _  *  } - \beta \log ({\nu _{\rm peak}}/{\nu _ * })}}\;
\label{Eq:11}
\end{equation}
and the peak luminosity can be calculated as
\begin{equation}
{L_{\rm peak}} = 4\pi d_L^2{\nu _{\rm peak}}{f_{\rm peak}}\;
\label{Eq:12}
\end{equation}
where ${d_L}$ is the luminosity distance (\citealt{2015IJMPA..3045020F,2016ApJS..226...20F,2017ApJ...835L..38F}).

Equation (\ref{Eq:4}) is just a phenomenological fit of the data, which does not take into account the internal physical mechanisms due to injection and the cooling effects. Here, we focus only on the statistical and physical significance of the fitting parameters.

\subsection{The intrinsic physical parameters} \label{sect:parameters}
To obtain the intrinsic physical parameters, we need to transform them from the observer's coordinate system into the co-moving coordinate system, to introduce Doppler corrections and redshift corrections (\citealt{1999qagn.book.....K}). The intrinsic physical parameters are described by the equations (\citealt{2001PASJ...53..469X,2013PASJ...65...25F,2015IJMPA..3045020F}):
\begin{equation}
\nu^\prime  = \frac{{(1 + z)}}{D }{\nu}\\
\label{Eq:13}
\end{equation}
\begin{equation}
{f^\prime }({\nu ^\prime }) = {\left( {1 + z} \right)^{{\alpha _\nu } - 1}}{\left( {\frac{1}{{{D_{{\rm{ }}}}}}} \right)^{(3 + {\alpha _\nu })}}f(\nu )\\
\label{Eq:14}
\end{equation}
\begin{equation}
{L^\prime }({\nu ^\prime }) = {\left( {1 + z} \right)^{{\alpha _\nu }}}{\left( {\frac{1}{{{D_{{\mathop{ }} }}}}} \right)^{(4 + {\alpha _\nu })}}L(\nu )
\label{Eq:15}
\end{equation}
where the physical quantities with a prime are calculated in the co-moving frame and the observed quantities are without a prime. The spectral index $\alpha_\nu$ is estimated according to the method in Section \ref{sect: LP model}.

In order to further analyze the physical mechanism and obtain essential radiation parameters, we relate the electron spectrum parameters to the photon spectrum parameters obtained, as in Section \ref{sect: LP model} (\citealt{2006A&A...448..861M,2009A&A...504..821P,2011ApJ...736..128P,2014ApJ...788..179C,2016ApJS..226...20F,
2018ApJS..235...39C}). In the Thomson scattering region, we have ($\rm See \ Appendix \ref{sect:appendix}\  for\ details$)
\begin{equation}
\left\{ \begin{array}{l}
{D_{{\rm{Th}}}} = {\left( {247.2179} \right)^{\frac{1}{{4\xi }}}}{(1 + z)^{\frac{{1 - \xi }}{\xi }}}{\left( {\frac{{{{10}^{45}}{\rm{erg}}\;{{\rm{s}}^{{\rm{ - 1}}}}}}{{L_{{\rm{peak}},{\rm{Th}}}^{\rm{ic}}}}} \right)^{\frac{1}{{4\xi }}}}{\left( {\frac{{L_{{\rm{peak}}}^{\rm{syn}}}}{{{{10}^{45}}{\rm{erg}}\;{{\rm{s}}^{{\rm{ - 1}}}}}}} \right)^{\frac{1}{{2\xi }}}}\\
\qquad \ \ \ \ \ \ \ \ \ \ \times{\left( {\frac{{\nu _{{\rm{peak}},{\rm{Th}}}^{\rm ic}}}{{{{10}^{23}}{\rm{Hz}}}}} \right)^{\frac{1}{{2\xi }}}}{\left( {\frac{{{{10}^{15}}{\rm{Hz}}}}{{\nu _{{\rm{peak}}}^{\rm{syn}}}}} \right)^{\frac{1}{\xi }}}{\left( {\frac{{1\;{\rm{day}}}}{{\Delta t}}} \right)^{\frac{1}{{2\xi }}}}{\left( {{\beta _{{\rm{syn}}}}} \right)^{\frac{1}{{8\xi }}}}\\
B = 13.4109{\left( {\frac{{\nu _{{\rm{peak}}}^{\prime {\rm{syn}}}}}{{{{10}^{15}}{\rm{Hz}}}}} \right)^2}\left( {\frac{{{{10}^{23}}{\rm{Hz}}}}{{\nu _{{\rm{peak}},{\rm{Th}}}^{\prime {\rm{ic}}}}}} \right)\\
R = 2.7235 \times {10^{16}}\left( {\frac{{\nu _{{\rm{peak}},{\rm{Th}}}^{\prime {\rm{ic}}}}}{{{{10}^{23}}{\rm{Hz}}}}} \right){\left( {\frac{{{{10}^{15}}{\rm{Hz}}}}{{\nu _{{\rm{peak}}}^{\prime {\rm{syn}}}}}} \right)^2}\left( {\frac{{L_{{\rm{peak}}}^{\prime {\rm{syn}}}}}{{{{10}^{45}}{\rm{erg}}\;{{\rm{s}}^{{\rm{ - 1}}}}}}} \right){\left( {\frac{{{{10}^{45}}{\rm{erg}}\;{{\rm{s}}^{{\rm{ - 1}}}}}}{{L_{{\rm{peak}},{\rm{Th}}}^{\prime {\rm{ic}}}}}} \right)^{\frac{1}{2}}}\\
{\gamma _{\rm{p}}} = 8660.3{\left( {\frac{{\nu _{{\rm{peak}},{\rm{Th}}}^{\prime {\rm{ic}}}}}{{{{10}^{23}}{\rm{Hz}}}}\cdot\frac{{{{10}^{15}}{\rm{Hz}}}}{{\nu _{{\rm{peak}}}^{\prime {\rm{syn}}}}}} \right)^{\frac{1}{2}}}{10^{ - \frac{1}{{10{\beta _{{\rm{syn}}}}}}}}
\end{array} \right.
\label{Eq:16}
\end{equation}

In the Klein-Nishina (KN) scattering region, we have
\begin{equation}
\left\{ \begin{array}{l}
{D_{{\rm{KN}}}} = {(0.0295)^{\frac{1}{\zeta }}}{\left( {1 + z} \right)^{\frac{{\zeta  - 2}}{\zeta }}}{\left( {\frac{{L_{{\rm{peak}}}^{\rm syn}}}{{{{10}^{45}}{\rm{erg}}\;{{\rm{s}}^{{\rm{ - 1}}}}}}} \right)^{\frac{1}{\zeta }}}{\left( {\frac{{{{10}^{45}}{\rm{erg}}\;{{\rm{s}}^{{\rm{ - 1}}}}}}{{L_{{\rm{peak}},{\rm{KN}}}^{\rm{ic}}}}} \right)^{\frac{1}{{2\zeta }}}}\\
\qquad \ \ \ \ \ \ \times{\left( {\frac{{{{10}^{15}}{\rm{Hz}}}}{{\nu _{{\rm{peak}}}^{\rm{syn}}}}} \right)^{\frac{3}{{2\zeta }}}}{\left( {\frac{{\nu _{{\rm{peak}},{\rm{KN}}}^{\rm ic}}}{{{{10}^{23}}{\rm{Hz}}}}} \right)^{\frac{1}{\zeta }}}{\left( {\frac{{1{\rm{day}}}}{{\Delta t}}} \right)^{\frac{1}{\zeta }}}{10^{\frac{2}{{5{\beta _{{\rm{syn}}}}\zeta }}}}{\left( {{\beta _{{\rm{syn}}}}} \right)^{ - \frac{1}{{4\zeta }}}}\\
{B} = 1.5356 \times {10^3}\left( {\frac{{\nu _{{\rm{peak}}}^{\prime {\rm{syn}}}}}{{{{10}^{15}}{\rm{Hz}}}}} \right){\left( {\frac{{{{10}^{23}}{\rm{Hz}}}}{{\nu _{{\rm{peak}},{\rm{KN}}}^{\prime {\rm{ic}}}}}} \right)^2}{10^{ - \frac{2}{{5{\beta _{{\rm{syn}}}}}}}}\\
{R} = 7.6456 \times {10^{13}}{\left( {{\beta _{{\rm{syn}}}}} \right)^{ - \frac{1}{4}}}{10^{\frac{2}{{5{\beta _{{\rm{syn}}}}}}}}\left( {\frac{{\nu _{{\rm{peak}},{\rm{KN}}}^{\prime \rm ic}}}{{{{10}^{23}}{\rm{Hz}}}}} \right){\left( {\frac{{{{10}^{15}}{\rm{Hz}}}}{{\nu _{{\rm{peak}}}^{\prime {\rm{syn}}}}}} \right)^{\frac{3}{2}}}\left( {\frac{{L_{{\rm{peak}}}^{\prime {\rm{syn}}}}}{{{{10}^{45}}{\rm{erg}}\;{{\rm{s}}^{{\rm{ - 1}}}}}}} \right){\left( {\frac{{{{10}^{45}}{\rm{erg}}\;{{\rm{s}}^{{\rm{ - 1}}}}}}{{L_{{\rm{peak}},{\rm{KN}}}^{\prime {\rm{ic}}}}}} \right)^{\frac{1}{2}}}\\
{\gamma _{{\rm{p}}}} = 809.332\left( {\frac{{\nu _{{\rm{peak}},{\rm{KN}}}^{\prime {\rm{ic}}}}}{{{{10}^{23}}{\rm{Hz}}}}} \right){10^{\frac{1}{{10{\beta _{{\rm{syn}}}}}}}}
\end{array} \right.
\label{Eq:17}
\end{equation}
where $D$ is the Doppler factor; $\gamma_{\rm {\rm p}}$ is the peak Lorentz factor of nonthermal electrons; $R$ is the radius of the emission region in $\rm centimeters$, $B$ is the magnetic field of the jet in $\rm gauss$; $z$ is the redshift; $\nu _{\rm peak}^{\rm syn}$ and $\nu _{\rm peak}^{\rm ic}$ are the peak frequencies of synchrotron radiation and the inverse Compton spectrum, respectively, in $\rm hertz$; $L _{\rm peak}^{\rm syn}$ and $L _{\rm peak}^{\rm ic}$ are, respectively, the peak luminosity of synchrotron radiation and the inverse Compton energy spectrum in ergs per second; ${\beta _{\rm syn}}$ is the energy spectral curvature of the synchrotron bump; $\xi  = 1 + \frac{{2{\alpha _{\nu _{\rm peak}^{\rm syn}}} - {\alpha _{\nu _{\rm peak}^{\rm ic}}}}}{4}$, $\zeta  = \frac{{5 + 2{\alpha _{\nu _{\rm peak}^{\rm syn}}} - {\alpha _{\nu _{\rm peak}^{\rm ic}}}}}{2}$; $\alpha _{\nu_{\rm peak}} =  - \frac{{\partial \log {f_\nu }}}{{\partial \log \nu }} = 1$ denotes the spectral indices at the peak frequency; and $\Delta t$ is the variability timescale. For simplicity, we use the average value of variability timescale, $\frac{{\Delta t}}{{\left( {1 + z} \right)}} \approx 1{\rm{ }}\;\rm{day}$, in the source frame (\citealt{1997ARA&A..35..445U,2010ApJ...716...30A,2013MNRAS.430.1324N,2014ApJS..215....5K,2015ApJ...807...51Z,
2018ApJS..235...39C}).

Through the log-parabolic model fitting, 12 fitting parameters can be obtained as $\beta_{\rm syn}$, $\beta_{\rm ic}$, $\log \nu _{\rm peak}^{\rm syn}$, $\log \nu _{\rm peak}^{\rm ic}$, $\log L_{\rm peak}^{\rm syn}$, $\log L_{\rm peak}^{\rm ic}$, and the spectral index $\alpha_{\nu}$ of each band: the radio (1.4 GHz), infrared (2.2 $\mu$m), optical (J band, 1250 nm), ultraviolet (250 nm), soft X-ray (1 keV) and $\gamma$-ray (20 GeV) bands (\citealt{1998MNRAS.299..433F,2014SCPMA..57.2007N,2017Ap&SS.362...22Y}). We can also calculate the physical parameters indirectly, including $D$, $R$, $B$ and $\gamma_{\rm p}$. In addition, we can calculate the $\gamma$-ray luminosity at $E = 20{\rm{ }}\;\rm{GeV}$ by the following formula (\citealt{2012ApJ...761..125F,2015IJMPA..3045020F,2014SCPMA..57.2007N,2017Ap&SS.362...22Y}):
\begin{equation}
L(E) = 4\pi d_L^2\nu f(E)\\
\label{Eq:18}
\end{equation}
where $f(E)$ is the $\gamma$-ray flux density, which can be expressed as (\citealt{2014SCPMA..57.2007N,2017Ap&SS.362...22Y})
\begin{equation}
f(E) = 6.626 \times {10^{ - 4}}\frac{{1 - \Gamma }}{{E_U^{1 - \Gamma } - E_L^{1 - \Gamma }}} \cdot N \cdot {E^{1 - \Gamma }}\quad({\rm{Jy}})\\
\label{Eq:19}
\end{equation}
where $N = {N_0}\frac{{E_U^{1 - \Gamma } - E_L^{1 - \Gamma }}}{{1 - \Gamma }}$ is the integral photon flux in units of photon count per square centimeter per second, which can be obtained from 4FGL. $\Gamma$ is the $\gamma$-ray photon spectrum index, and ${E_L} = 1{\rm{ }}\;\rm{GeV}$ and ${E_U} = 100{\rm{ }}\;\rm{GeV}$ are the lower and the upper integrating limits.

\section{Sample selection and data processing} \label{sect:sample}
The 4LAC samples list 3137 $\gamma$-ray blazars (\citealt{2020ApJ...892..105A}) with 694 FSRQs, 1131 BL Lac objects, and 1312 BCUs. We selected all 73 Fermi TeV blazars from the 4LAC samples, including 65 blazars with redshift values. In this sample, there are 57 TeV BL Lac objects with redshifts, 7 TeV FSRQs with redshifts and 1 TeV BCU with redshifts (\citealt{2020ApJ...892..105A}).

In order to reasonably estimate the physical parameters and obtain an accurate energy spectrum distribution diagram, we collected a large number of multifrequency data from the SSDC\footnote{{The SSDC SED Builder contains radio to TeV band data from several missions and experiments together with catalogs and archival data, such as the NED, the DB database (including CLASSCAT, Fermi1FGL, 2FGL, MAGIC, VERITAS, and IPC), TOMASS data, and USNOA2 data.}} database and 4FGL, respectively, including the flux density and the flux densities error of the radio, infrared, optical, ultraviolet, soft X-ray, and $\gamma$-ray bands (\citealt{2001A&A...366...62A,2010A&A...520A..83H,2013MNRAS.434.1889H,2015ApJ...802...65A,2015ApJ...812...60B,
2015NIMPA.770...42S,2018RAA....18..120L,2019AcASn..60....7Z,2020ApJS..247...33A}). In the calculation, we grouped and filtered the SED observation data of each source as follows: (1) For multiple data points of the same frequency, we took the average value of the corresponding flux. For the flux without error, we took 1\% of the observed radio and optical flux, 2\% of the observed ultraviolet and X-ray flux, and 20\% of the observed $\gamma$-ray flux as the errors of the data points whose errors were available (\citealt{2012ApJ...752..157Z,2014A&A...567A.135A,2017ApJS..228....1Z,2018RAA....18..120L}). (2) Since NED collected observations using very different methods and literature from those used by other databases, we removed the data whose $\nu >{10^{16}}$ Hz from NED according to \cite{2018RAA....18..120L}. (3) Data with the upper limit of energy, duplicates, and flux errors greater than the flux value were excluded. (4) We considered that in the ultraviolet (or optical) band, some SEDs of AGNs may have a bump formed by tori, accretion disks, or host galaxies (the bump of thermal radiation). For this part of the sample, $13.5 < \log \nu  < 15 $ observations were ignored for SED processing (\citealt{1978Natur.272..706S,1987ApJ...321..305C,2012MNRAS.420.2899G,2019AcASn..60....7Z}).

In our sample, the log-parabolic model was used to fit the average-state multifrequency SED of 65 TeV blazars. Since the log-parabolic shape held only close to the peak of the SED, we repeatedly adjusted the parameters according to Equation (\ref{Eq:4}), so that the reduced chi-square value of each source was close to 0 when fitting the data. The sample data and fitting results are shown in Table~\ref{tab:1}. Figures 1 and 2 show the SED fitting diagram. In Table 1, Column (1) gives both the 4FGL source name and the TeVCat name. Column (2) gives the class designation for the source and the redshift (z), where class represents the type of sample: H, I, L, F and B represent sources belonging to HBLs, IBLs, LBLs, FSRQs, and BCUs, respectively. Column (3) gives the energy spectral curvature of the synchrotron bump and ICs bump at the peak: $\beta_{\rm syn}$ and $\beta_{\rm ic}$. Column (4) gives the synchrotron peak frequency and ICs peak frequency in units of $\rm hertz$: $\log \nu _{\rm peak}^{\rm syn}$ and $\log \nu _{\rm peak}^{\rm ic}$. Column (5) gives the spectral indices in the radio ($1.4\rm GHz$) and infrared ($2.2\mu {\rm m}$) bands: $\alpha_{\rm r}$ and $\alpha_{\rm ir}$. Column (6) gives the spectral indices in the optical ($1250\rm nm$) and ultraviolet ($250\rm nm$) bands: $\alpha_{\rm o}$ and $\alpha_{\rm uv}$. Column (7) gives the spectral indices in the soft X-ray ($1\ \rm keV$) and $\gamma$-ray ($20\ \rm GeV$) bands: $\alpha_{\rm x}$ and ${\alpha _{\rm \gamma }}$. Column (8) gives the synchrotron peak luminosity and ICs peak luminosity in units of ergs per second: $\log L_{\rm peak}^{\rm syn}$ and $\log L_{\rm peak}^{\rm ic}$. Column (9) gives the radio Doppler factor $D_{\rm r}$ and $\gamma$-ray luminosity $\log L_{\rm \gamma}$ ($20{\rm{ }}\;\rm{GeV}$) in units of ergs per second; the radio Doppler factor ($D_{\rm r}$) was collected from the following literature: \cite{2013PASJ...65...25F}, \cite{2013RAA....13..259F}, \cite{2009PASJ...61..639F}, \cite{2014RAA....14.1135F}, \cite{2017ApJ...835L..38F}, \cite{2017RAA....17...66L}, \cite{2009A&A...494..527H}, \cite{1993ApJ...407...65G}, \cite{1999A&A...341...74H}, \cite{2007A&A...466...63W}, \cite{2010A&A...512A..24S}, \cite{1999ApJ...521..493L}, and \cite{2009RAA.....9..168W}. Column (10) gives the reduced ${\chi ^2}$ values for the energy spectrum of the synchrotron bump and the energy spectrum of the ICs bump of our fits: $\chi _{\rm syn}^2$ and $\chi _{\rm ic}^2$, which can be calculated by ${\chi ^2} = \frac{1}{{N - dof}}\sum\limits_{{\rm{\rm i = 1}}}^{\rm N} {{{({{\hat y}_{\rm{i}}} - {y_{\rm{i}}})}^2}}$, where $N$ is the number of observed data points; $dof$ is the degrees of freedom, i.e., the number of free parameters used for the model; The ${{\hat y}_{\rm{i}}}$ denotes the expected values from the model; and $y_{\rm i}$ denotes the observed data (\citealt{2014sdmm.book.....I,2004A&A...413..489M}). In addition, $\sigma$ represents the uncertainty of the parameter in Table~\ref{tab:1}, which can be calculated by the error transfer formula according to the equation (\citealt{2012msma.book.....F}), i.e., $\sigma (y) = \sqrt {{{\sum\limits_{\rm i = 1}^m {\left( {\frac{{\partial y}}{{\partial {x_{\rm i}}}}} \right)} }^2}{{\left( {\sigma ({x_{\rm i}})} \right)}^2} + 2\sum\limits_{\rm i,\rm j = 1,\rm i \ne \rm j}^m {\frac{{\partial y}}{{\partial {x_{\rm i}}}}\frac{{\partial y}}{{\partial {x_{\rm j}}}}{\rho _{\rm ij}}\sigma ({x_{\rm i}})\sigma ({x_{\rm j}})} }$, where ${\rho _{{\rm{ij}}}}$ is the correlation coefficient.

Based on the calculation results of the synchrotron radiation peak frequency (Equation (\ref{Eq:10})), we further classified the TeV blazars. The sources were classified as LBLs (i.e., $\nu _{\rm peak}^{\rm syn} < {10^{14}}$Hz), IBLs (i.e., ${10^{14}} < \nu _{\rm peak}^{\rm syn} < {10^{15}}$Hz), HBLs (i.e., $\nu _{\rm peak}^{\rm syn} > {10^{15}}$Hz), or FSRQs according to \cite{2010ApJ...716...30A}. There was only one source, PGC 2402248, classified as a BCU in our sample, which had little impact on the statistical results, so we removed it in our correlation test and regression analysis.

According to the fitting results in Table~\ref{tab:1}, we can use Equations (\ref{Eq:16}) and (\ref{Eq:17}) to estimate the physical parameters of the jet radiation zone of the TeV blazars, and the results are listed in Table~\ref{tab:2}.

\begin{table*}[!htp]
\setlength{\abovecaptionskip}{0 cm}
\setlength{\belowcaptionskip}{0.1cm}
\begin{center}
\caption{Sample Data and Fitting Results}
\label{tab:1}
\centering
\small
\setlength{\LTleft}{-5.5cm} \setlength{\LTright}{0 cm} 
\begin{adjustwidth}{-1cm}{-1cm}
\resizebox{\textwidth}{!}{
\begin{tabular}{ccccccccccc} \hline
\renewcommand\arraystretch{2}
\centering
{4FGL Name}& class & $\beta_{\rm syn}$/($\sigma \times {10^{ - 4}}$)& $\log \nu _{\rm peak}^{\rm syn}$ /$\sigma $ & $\alpha_{\rm r}$/$\sigma$\ & $\alpha_{\rm o}$/$\sigma$& $\alpha_{\rm x}$/$\sigma$ & $\log L_{\rm peak}^{\rm syn}$/$\sigma$ & $D_{\rm r}$ &$\chi _{\rm syn}^2$ \\
Other Name & z & $\beta_{\rm ic}$/($\sigma \times {10^{ - 4}}$)  & $\log \nu _{\rm peak}^{\rm ic}$ /$\sigma $ & $\alpha_{\rm ir}$/$\sigma$ & $\alpha_{\rm uv}$/$\sigma$ & ${\alpha _{\rm \gamma }}$/$\sigma$ & $\log L_{\rm peak}^{\rm ic}$/$\sigma$ & $\log L_{\rm \gamma}$/$\sigma$ &$\chi _{\rm ic}^2$ \\
\normalsize(1) & \normalsize(2) & \normalsize(3) &\normalsize(4) & \normalsize(5) &  \normalsize(6) &  \normalsize(7) &  \normalsize(8) & \normalsize(9) & \normalsize(10) \\
\hline 
J0013.9-1854	&	$\rm	H	$	&	0.07 	/	2.63 	&	16.91 	/	0.12 	&	-0.03 	/	0.01 	&	0.67 	/	0.01 	&	1.08 	/	0.02 	&	43.90 	/	0.17 	&		&	0.11 	\\
$\rm	SHBL\;J001355.9-185406 $	&	0.095 	&	0.25 	/	2.38 	&	24.22 	/	0.04 	&	0.63 	/	0.01 	&	0.76 	/	0.02 	&	1.23 	/	0.02 	&	43.26 	/	0.06 	&	42.94 	/	0.85 	&	0.33 	\\
J0033.5-1921	&	$\rm	H	$	&	0.10 	/	0.92 	&	15.81 	/	0.03 	&	-0.33 	/	0.01 	&	0.72 	/	0.01 	&	1.33 	/	0.01 	&	45.81 	/	0.04 	&	6.43 	&	0.01 	\\
$\rm	KUV\;00311-1938    $	&	0.610 	&	0.24 	/	0.60 	&	24.58 	/	0.01 	&	0.67 	/	0.01 	&	0.85 	/	0.01 	&	1.05 	/	0.01 	&	45.85 	/	0.02 	&	45.83 	/	1.44 	&	0.02 	\\
J0112.1+2245	&	$\rm	I	$	&	0.12 	/	2.24 	&	14.41 	/	0.05 	&	-0.27 	/	0.01 	&	1.00 	/	0.01 	&	1.74 	/	0.01 	&	45.21 	/	0.07 	&	9.10 	&	0.08 	\\
$\rm	S2\;0109+22  $	&	0.265 	&	0.10 	/	0.45 	&	23.56 	/	0.02 	&	0.93 	/	0.01 	&	1.16 	/	0.01 	&	1.22 	/	0.01 	&	45.26 	/	0.03 	&	45.22 	/	1.69 	&	0.01 	\\
J0152.6+0147	&	$\rm	H	$	&	0.07 	/	1.23 	&	16.25 	/	0.05 	&	-0.05 	/	0.01 	&	0.72 	/	0.01 	&	1.18 	/	0.01 	&	43.76 	/	0.07 	&	3.00 	&	0.02 	\\
$\rm	RGB\;J0152+017   $	&	0.080 	&	0.22 	/	0.87 	&	24.75 	/	0.02 	&	0.69 	/	0.01 	&	0.83 	/	0.01 	&	0.97 	/	0.01 	&	43.64 	/	0.03 	&	43.30 	/	1.15 	&	0.09 	\\
J0214.3+5145	&	$\rm	H	$	&	0.06 	/	1.33 	&	16.50 	/	0.07 	&	0.16 	/	0.01 	&	0.76 	/	0.01 	&	1.11 	/	0.01 	&	43.39 	/	0.10 	&		&	0.04 	\\
$\rm	TXS\;0210+515   	$	&	0.049 	&	0.66 	/	5.06 	&	24.27 	/	0.03 	&	0.73 	/	0.01 	&	0.84 	/	0.01 	&	1.55 	/	0.04 	&	43.15 	/	0.05 	&	42.64 	/	0.94 	&	1.30 	\\
J0221.1+3556	&	$\rm	F	$	&	0.09 	/	8.68 	&	13.98 	/	0.06 	&	0.14 	/	0.03 	&	1.07 	/	0.04 	&	1.62 	/	0.05 	&	45.75 	/	0.08 	&		&	0.04 	\\
$\rm	S3\;0218+35   	$	&	0.944 	&	0.08 	/	0.81 	&	21.96 	/	0.04 	&	1.03 	/	0.03 	&	1.20 	/	0.04 	&	1.46 	/	0.01 	&	46.69 	/	0.05 	&	46.20 	/	1.71 	&	0.04 	\\
J0222.6+4302	&	$\rm	I	$	&	0.09 	/	1.98 	&	14.65 	/	0.06 	&	0.05 	/	0.01 	&	0.95 	/	0.01 	&	1.49 	/	0.01 	&	45.77 	/	0.09 	&	0.10 	&	0.11 	\\
$\rm	3C\;66A     $	&	0.444 	&	0.16 	/	0.86 	&	23.37 	/	0.02 	&	0.91	/	0.01 	&	1.07 	/	0.01 	&	1.41 	/	0.01 	&	46.24 	/	0.03 	&	46.10 	/	1.84 	&	0.06 	\\
J0232.8+2018	&	$\rm	H	$	&	0.05 	/	1.60 	&	17.67 	/	0.09 	&	0.08 	/	0.01 	&	0.65 	/	0.01 	&	0.98 	/	0.01 	&	44.29 	/	0.13 	&	2.50 	&	0.07 	\\
$\rm	1ES\;0229+200   	$	&	0.139 	&	0.25 	/	1.29 	&	25.20 	/	0.02 	&	0.62 	/	0.01 	&	0.72 	/	0.01 	&	0.75 	/	0.01 	&	44.01 	/	0.03 	&	43.44 	/	0.75 	&	0.16 	\\
J0238.4-3116	&	$\rm	H	$	&	0.06 	/	1.29 	&	16.95 	/	0.06 	&	0.01 	/	0.01 	&	0.67 	/	0.01 	&	1.07 	/	0.01 	&	44.92 	/	0.09 	&		&	0.03 	\\
$\rm	1RXS\;J023832.6-311658 $	&	0.233 	&	0.15 	/	1.93 	&	24.23 	/	0.06 	&	0.64 	/	0.01 	&	0.76 	/	0.01 	&	1.13 	/	0.02 	&	44.28 	/	0.08 	&	44.43 	/	1.21 	&	0.24 	\\
J0303.4-2407	&	$\rm	H	$	&	0.07 	/	3.38 	&	15.40 	/	0.14 	&	0.15 	/	0.01 	&	0.86 	/	0.02 	&	1.28 	/	0.02 	&	45.13 	/	0.19 	&		&	0.14 	\\
$\rm	PKS\;0301-243   $	&	0.266 	&	0.12 	/	0.36 	&	24.15 	/	0.01 	&	0.83 	/	0.01 	&	0.96 	/	0.02 	&	1.13 	/	0.01 	&	45.16 	/	0.02 	&	45.23 	/	1.61 	&	0.02 	\\
J0319.8+1845	&	$\rm	H	$	&	0.08 	/	1.23 	&	16.37 	/	0.05 	&	-0.20 	/	0.01 	&	0.67 	/	0.01 	&	1.18 	/	0.01 	&	44.46 	/	0.06 	&		&	0.05 	\\
$\rm	RBS\;0413  $	&	0.190 	&	0.47 	/	5.75 	&	25.41 	/	0.06 	&	0.63 	/	0.01 	&	0.79 	/	0.01 	&	0.32 	/	0.05 	&	44.29 	/	0.08 	&	44.02 	/	0.94 	&	2.61 	\\
J0349.4-1159	&	$\rm	H	$	&	0.06 	/	1.73 	&	17.94 	/	0.09 	&	-0.06 	/	0.01 	&	0.57 	/	0.01 	&	0.95 	/	0.01 	&	44.81 	/	0.13 	&		&	0.13 	\\
$\rm	1ES\;0347-121  $	&	0.188 	&	0.27 	/	1.57 	&	25.18 	/	0.03 	&	0.54 	/	0.01 	&	0.66 	/	0.01 	&	0.73 	/	0.01 	&	44.21 	/	0.04 	&	43.76 	/	0.84 	&	0.31 	\\
J0416.9+0105	&	$\rm	H	$	&	0.07 	/	1.31 	&	17.12 	/	0.06 	&	-0.05 	/	0.01 	&	0.64 	/	0.01 	&	1.05 	/	0.01 	&	45.15 	/	0.09 	&	2.90 	&	0.04 	\\
$\rm	1ES\;0414+009  $	&	0.287 	&	0.36 	/	1.83 	&	24.59 	/	0.02 	&	0.61 	/	0.01 	&	0.73 	/	0.01 	&	1.07 	/	0.02 	&	44.61 	/	0.03 	&	44.41 	/	1.04 	&	0.36 	\\
J0449.4-4350	&	$\rm	H	$	&	0.09 	/	3.48 	&	15.58 	/	0.10 	&	-0.18 	/	0.01 	&	0.78 	/	0.02 	&	1.35 	/	0.02 	&	45.35 	/	0.15 	&		&	0.15 	\\
$\rm	PKS\;0447-439  $	&	0.205 	&	0.37 	/	1.05 	&	24.20 	/	0.01	&	0.73 	/	0.01 	&	0.91 	/	0.02 	&	1.36 	/	0.01 	&	45.52 	/	0.02 	&	45.29 	/	1.74 	&	0.11 	\\
J0507.9+6737	&	$\rm	H	$	&	0.07 	/	1.46 	&	16.90 	/	0.06 	&	-0.15 	/	0.01 	&	0.63 	/	0.01 	&	1.09 	/	0.01 	&	45.81 	/	0.09 	&	5.60 	&	0.07 	\\
$\rm	1ES\;0502+675   	$	&	0.416 	&	0.18 	/	3.08 	&	25.33 	/	0.08 	&	0.59 	/	0.01 	&	0.73 	/	0.01 	&	0.78 	/	0.03 	&	45.66 	/	0.11 	&	45.36 	/	1.37 	&	0.75 	\\
J0509.4+0542	&	$\rm	I	$	&	0.11 	/	1.93 	&	14.39 	/	0.04 	&	-0.19 	/	0.01 	&	1.00 	/	0.01 	&	1.70 	/	0.01 	&	45.48 	/	0.06 	&		&	0.06 	\\
$\rm	TXS\;0506+056  	$	&	0.336 	&	0.10 	/	0.24 	&	23.28 	/	0.01 	&	0.94 	/	0.01 	&	1.16 	/	0.01 	&	1.27 	/	0.01 	&	45.44 	/	0.02 	&	45.31 	/	1.57 	&	0.01 	\\
J0521.7+2112	&	$\rm	H	$	&	0.08 	/	2.07 	&	15.35 	/	0.07 	&	-0.01 	/	0.01 	&	0.84 	/	0.01 	&	1.34 	/	0.01 	&	44.61 	/	0.10 	&		&	0.08 	\\
$\rm	VER\;J0521+211    $	&	0.108 	&	0.16 	/	1.26 	&	23.81 	/	0.03 	&	0.80 	/	0.01 	&	0.96 	/	0.01 	&	1.28 	/	0.01 	&	44.80 	/	0.05 	&	44.75 	/	1.74 	&	0.11 	\\
J0550.5-3216	&	$\rm	H	$	&	0.05 	/	1.61 	&	17.57 	/	0.10 	&	0.18 	/	0.01 	&	0.69 	/	0.01 	&	0.99 	/	0.01 	&	44.12 	/	0.15 	&		&	0.13 	\\
$\rm	PKS\;0548-322  $	&	0.069 	&	0.25 	/	2.16 	&	25.12 	/	0.04 	&	0.66 	/	0.01 	&	0.76 	/	0.01 	&	0.78 	/	0.02 	&	43.13 	/	0.06 	&	42.76 	/	0.86 	&	0.34 	\\
J0648.7+1516	&	$\rm	H	$	&	0.07 	/	1.81 	&	16.71 	/	0.08 	&	-0.03 	/	0.01 	&	0.68 	/	0.01 	&	1.11 	/	0.01 	&	44.69 	/	0.11 	&		&	0.06 	\\
$\rm	RX\;J0648.7+1516   $	&	0.179 	&	0.50 	/	2.31 	&	24.86 	/	0.02 	&	0.65 	/	0.01 	&	0.78 	/	0.01 	&	0.82 	/	0.02 	&	44.88 	/	0.03 	&	44.41 	/	1.21 	&	0.46 	\\
J0650.7+2503	&	$\rm	H	$	&	0.07 	/	1.28 	&	16.65 	/	0.05 	&	-0.05 	/	0.01 	&	0.68 	/	0.01 	&	1.12 	/	0.01 	&	44.91 	/	0.08 	&	3.20 	&	0.04 	\\
$\rm	1ES\;0647+250  $	&	0.203 	&	0.50 	/	2.07 	&	24.33 	/	0.02 	&	0.65 	/	0.01 	&	0.78 	/	0.01 	&	1.36 	/	0.02 	&	45.11 	/	0.03 	&	44.86 	/	1.45 	&	0.28 	\\
J0710.4+5908	&	$\rm	H	$	&	0.06 	/	2.02 	&	17.86 	/	0.12 	&	0.04 	/	0.01 	&	0.62 	/	0.01 	&	0.96 	/	0.01 	&	44.68 	/	0.16 	&	2.90 	&	0.12 	\\
$\rm	RGB\;J0710+591  $	&	0.125 	&	0.14 	/	0.91 	&	24.86 	/	0.03 	&	0.59 	/	0.01 	&	0.69 	/	0.01 	&	0.95 	/	0.01 	&	43.86 	/	0.04 	&	43.67 	/	1.07 	&	0.10 	\\
J0721.9+7120	&	$\rm	I	$	&	0.11 	/	2.61 	&	14.91 	/	0.06 	&	-0.31 	/	0.01 	&	0.88 	/	0.01 	&	1.59 	/	0.02 	&	45.33 	/	0.09 	&	2.10 	&	0.26 	\\
$\rm	S5\;0716+714  $	&	0.127 	&	0.09 	/	0.85 	&	23.03 	/	0.04 	&	0.82 	/	0.01 	&	1.04 	/	0.01 	&	1.30 	/	0.01 	&	45.07 	/	0.06 	&	45.03 	/	2.00 	&	0.06 	\\
J0739.2+0137	&	$\rm	F	$	&	0.10 	/	1.14 	&	13.97 	/	0.03 	&	0.04 	/	0.01 	&	1.08 	/	0.01 	&	1.70 	/	0.01 	&	44.62 	/	0.04 	&	8.60 	&	0.09 	\\
$\rm	PKS\;0736+017   $	&	0.189 	&	0.16 	/	0.99 	&	21.69 	/	0.03 	&	1.03 	/	0.01 	&	1.22 	/	0.01 	&	1.93 	/	0.01 	&	45.34 	/	0.04 	&	44.40 	/	1.57 	&	0.06 	\\
J0809.8+5218	&	$\rm	H	$	&	0.08 	/	1.09 	&	15.95 	/	0.04 	&	-0.09 	/	0.01 	&	0.75 	/	0.01 	&	1.25 	/	0.01 	&	44.59 	/	0.06 	&	3.30 	&	0.02 	\\
$\rm	1ES\;0806+524  $	&	0.138 	&	0.38 	/	0.98 	&	24.28 	/	0.01 	&	0.71 	/	0.01 	&	0.86 	/	0.01 	&	1.31 	/	0.01 	&	44.60 	/	0.02 	&	44.53 	/	1.58 	&	0.13 	\\
J0847.2+1134	&	$\rm	H	$	&	0.06 	/	2.08 	&	17.46 	/	0.10 	&	-0.03 	/	0.01 	&	0.62 	/	0.01 	&	1.00 	/	0.01 	&	44.56 	/	0.14 	&		&	0.06 	\\
$\rm	RBS\;0723  $	&	0.198 	&	0.56 	/	5.93 	&	25.09 	/	0.05 	&	0.59 	/	0.01 	&	0.71 	/	0.01 	&	0.55 	/	0.05 	&	44.35 	/	0.07 	&	43.94 	/	0.91 	&	1.95 	\\
J0854.8+2006	&	$\rm	L	$	&	0.18 	/	1.38 	&	13.29 	/	0.02 	&	-0.46 	/	0.01 	&	1.39 	/	0.01 	&	2.47 	/	0.01 	&	45.80 	/	0.03 	&	8.80 	&	0.04 	\\
$\rm	OJ\;287  $	&	0.306 	&	0.07 	/	0.82 	&	21.55 	/	0.04 	&	1.30 	/	0.01 	&	1.63 	/	0.01 	&	1.45 	/	0.01 	&	45.54 	/	0.06 	&	45.08 	/	1.64 	&	0.05 	\\
J0958.7+6534	&	$\rm	L	$	&	0.16 	/	2.61 	&	12.98 	/	0.04 	&	-0.21 	/	0.01 	&	1.45 	/	0.01 	&	2.42 	/	0.01 	&	45.18 	/	0.05 	&	5.00 	&	0.08 	\\
$\rm	S4\;0954+65   $	&	0.367 	&	0.07 	/	1.00 	&	21.57 	/	0.05 	&	1.37 	/	0.01 	&	1.66 	/	0.01 	&	1.45 	/	0.01 	&	45.40 	/	0.08 	&	45.02 	/	1.57 	&	0.09 	\\
J1010.2-3119	&	$\rm	H	$	&	0.06 	/	2.57 	&	17.18 	/	0.13 	&	0.07 	/	0.01 	&	0.68 	/	0.01 	&	1.04 	/	0.02 	&	44.37 	/	0.19 	&	2.80 	&	0.10 	\\
$\rm	1RXS\;J101015.9-311909 $	&	0.143 	&	0.25 	/	1.81 	&	24.18 	/	0.03 	&	0.65 	/	0.01 	&	0.76 	/	0.01 	&	1.25 	/	0.02 	&	44.38 	/	0.05 	&	43.85 	/	0.99 	&	0.31 	\\
J1015.0+4926	&	$\rm	H	$	&	0.07 	/	1.47 	&	15.99 	/	0.06 	&	0.01 	/	0.01 	&	0.77 	/	0.01 	&	1.21 	/	0.01 	&	45.07 	/	0.08 	&	3.00 	&	0.04 	\\
$\rm	1ES\;1011+496  $	&	0.212 	&	0.13 	/	0.72 	&	24.55 	/	0.03 	&	0.73 	/	0.01 	&	0.87 	/	0.01 	&	1.04 	/	0.01 	&	45.21 	/	0.04 	&	45.27 	/	1.75 	&	0.05 	\\
J1103.6-2329	&	$\rm	H	$	&	0.07 	/	2.04 	&	16.66 	/	0.09 	&	0.02 	/	0.01 	&	0.70 	/	0.01 	&	1.11 	/	0.01 	&	45.28 	/	0.13 	&		&	0.24 	\\
$\rm	1ES\;1101-232  $	&	0.186 	&	0.32 	/	0.95 	&	24.93 	/	0.01 	&	0.67 	/	0.01 	&	0.79 	/	0.01 	&	0.84 	/	0.01 	&	44.57 	/	0.02 	&	43.91 	/	0.88 	&	0.14 	\\
J1104.4+3812	&	$\rm	H	$	&	0.07 	/	0.98 	&	17.01 	/	0.05 	&	-0.05 	/	0.01 	&	0.65 	/	0.01 	&	1.06 	/	0.01 	&	44.26 	/	0.06 	&	0.50 	&	0.13 	\\
$\rm	Mkr\;421 $	&	0.030 	&	0.12 	/	0.42 	&	25.23 	/	0.02 	&	0.61 	/	0.01 	&	0.74 	/	0.01 	&	0.88 	/	0.01 	&	44.31 	/	0.02 	&	44.22 	/	2.03 	&	0.15 	\\
J1136.4+6736	&	$\rm	H	$	&	0.06 	/	2.08 	&	17.49 	/	0.11 	&	0.04 	/	0.01 	&	0.64 	/	0.01 	&	1.00 	/	0.01 	&	44.39 	/	0.16 	&	2.60 	&	0.12 	\\
$\rm	RX\;J1136.5+6737 $	&	0.136 	&	0.40 	/	3.31 	&	24.21 	/	0.04 	&	0.61 	/	0.01 	&	0.72 	/	0.01 	&	1.38 	/	0.03 	&	43.86 	/	0.05 	&	43.70 	/	1.11 	&	0.65 	\\
J1136.4+7009	&	$\rm	H	$	&	0.06 	/	1.78 	&	16.35 	/	0.08 	&	0.08 	/	0.01 	&	0.75 	/	0.01 	&	1.15 	/	0.01 	&	43.56 	/	0.11 	&	0.80 	&	0.10 	\\
$\rm	Mkr\;180  $	&	0.045 	&	0.11 	/	0.53 	&	24.60 	/	0.02 	&	0.72 	/	0.01 	&	0.84 	/	0.01 	&	1.02 	/	0.01 	&	43.08 	/	0.03 	&	43.02 	/	1.33 	&	0.02 	\\
\hline
\\
\end{tabular}}
\end{adjustwidth}
\end{center}
\end{table*}
\renewcommand\thetable{1}
\begin{table*}[!htp]
\setlength{\abovecaptionskip}{0 cm}
\setlength{\belowcaptionskip}{0.1cm}
\begin{center}
\caption{Sample Data and Fitting Results}{}
\label{tab:1}
\setcounter{table}{1}
\renewcommand{\thetable}{1/arabic{table}}
\centering
\small
\begin{adjustwidth}{-1cm}{-1cm}
\resizebox{\textwidth}{!}{
\begin{tabular}{cccccccccc} \hline
\renewcommand\arraystretch{0.5}
\centering
{4FGL Name}& class & $\beta_{\rm syn}$/($\sigma \times {10^{ - 4}}$)& $\log \nu _{\rm peak}^{\rm syn}$ /$\sigma $ & $\alpha_{\rm r}$/$\sigma$\ & $\alpha_{\rm o}$/$\sigma$& $\alpha_{\rm x}$/$\sigma$ & $\log L_{\rm peak}^{\rm syn}$/$\sigma$ & $D_{\rm r}$ &$\chi _{\rm syn}^2$ \\
Other Name & z & $\beta_{\rm ic}$/($\sigma \times {10^{ - 4}}$)  & $\log \nu _{\rm peak}^{\rm ic}$ /$\sigma $ & $\alpha_{\rm ir}$/$\sigma$ & $\alpha_{\rm uv}$/$\sigma$ & ${\alpha _{\rm \gamma }}$/$\sigma$ & $\log L_{\rm peak}^{\rm ic}$/$\sigma$ & $\log L_{\rm \gamma}$/$\sigma$ &$\chi _{\rm ic}^2$ \\
\normalsize(1) & \normalsize(2) & \normalsize(3) &\normalsize(4) & \normalsize(5) &  \normalsize(6) &  \normalsize(7) &  \normalsize(8) & \normalsize(9) & \normalsize(10) \\
\hline 
J1159.5+2914	&	$\rm	F	$	&	0.18 	/	3.00 	&	13.05 	/	0.04 	&	-0.39 	/	0.01 	&	1.48 	/	0.02 	&	2.58 	/	0.02 	&	46.08 	/	0.06 	&	6.40 	&	0.14 	\\
$\rm	TON\;0599  $	&	0.729 	&	0.09 	/	0.78 	&	21.89 	/	0.04 	&	1.39 	/	0.01 	&	1.72 	/	0.02 	&	1.48 	/	0.01 	&	46.57 	/	0.05 	&	45.96 	/	1.70 	&	0.05 	\\
J1217.9+3007	&	$\rm	I	$	&	0.09 	/	2.03 	&	14.89 	/	0.06 	&	-0.01 	/	0.01 	&	0.91 	/	0.01 	&	1.46 	/	0.01 	&	44.39 	/	0.09 	&	0.40 	&	0.11 	\\
$\rm	1ES\;1215+303  $	&	0.130	&	0.16 	/	0.78 	&	23.72 	/	0.02 	&	0.87 	/	0.01 	&	1.03 	/	0.01 	&	1.30 	/	0.01 	&	44.67 	/	0.03 	&	44.74 	/	1.70 	&	0.05 	\\
J1221.3+3010	&	$\rm	H	$	&	0.06 	/	1.56 	&	17.21 	/	0.07 	&	-0.03 	/	0.01 	&	0.64 	/	0.01 	&	1.03 	/	0.01 	&	45.13 	/	0.11 	&	2.60 	&	0.12 	\\
$\rm	1ES\;1218+304  $	&	0.184 	&	0.32 	/	0.62 	&	24.61 	/	0.01 	&	0.61 	/	0.01 	&	0.73 	/	0.01 	&	1.05 	/	0.01 	&	45.10 	/	0.01 	&	44.78 	/	1.43 	&	0.07 	\\
J1221.5+2814	&	$\rm	H	$	&	0.08 	/	2.03 	&	15.36 	/	0.08 	&	0.07 	/	0.01 	&	0.85 	/	0.01 	&	1.32 	/	0.01 	&	44.51 	/	0.11 	&	0.20 	&	0.18 	\\
$\rm	W\;Comae  $	&	0.102 	&	0.18 	/	1.32 	&	23.77 	/	0.03 	&	0.82 	/	0.01 	&	0.96 	/	0.01 	&	1.33 	/	0.01 	&	44.45 	/	0.05 	&	43.93 	/	1.53 	&	0.14 	\\
J1224.4+2436	&	$\rm	H	$	&	0.09 	/	1.36 	&	15.68 	/	0.04 	&	-0.12 	/	0.01 	&	0.78 	/	0.01 	&	1.31 	/	0.01 	&	44.19 	/	0.06 	&	3.00 	&	0.01 	\\
$\rm	MS\;1221.8+2452  $	&	0.219 	&	0.71 	/	3.25 	&	24.56 	/	0.02 	&	0.73 	/	0.01 	&	0.90 	/	0.01 	&	1.17 	/	0.03 	&	45.06 	/	0.03 	&	44.45 	/	1.29 	&	0.68 	\\
J1224.9+2122	&	$\rm	F	$	&	0.13 	/	5.92 	&	13.87 	/	0.11 	&	-0.23 	/	0.02 	&	1.13 	/	0.03 	&	1.94 	/	0.03 	&	45.89 	/	0.15 	&	6.02 	&	0.27 	\\
$\rm	4C+21.35  	$	&	0.434 	&	0.10 	/	1.65 	&	21.99 	/	0.07 	&	1.07 	/	0.02 	&	1.31 	/	0.03 	&	1.51 	/	0.01 	&	46.54 	/	0.10 	&	45.82 	/	1.92 	&	0.28 	\\
J1230.2+2517	&	$\rm	I	$	&	0.12 	/	2.88 	&	14.61 	/	0.06 	&	-0.33 	/	0.01 	&	0.94 	/	0.01 	&	1.70 	/	0.02 	&	44.73 	/	0.08 	&	3.60 	&	0.05 	\\
$\rm	S3\;1227+25 $	&	0.135 	&	0.14 	/	1.47 	&	23.24 	/	0.05 	&	0.88 	/	0.01 	&	1.11 	/	0.01 	&	1.40 	/	0.01 	&	44.21 	/	0.06 	&	44.29 	/	1.52 	&	0.12 	\\
J1256.1-0547	&	$\rm	F	$	&	0.13 	/	2.25 	&	12.84 	/	0.04 	&	0.01 	/	0.01 	&	1.42 	/	0.01 	&	2.25 	/	0.01 	&	46.60 	/	0.05 	&	18.00 	&	0.28 	\\
$\rm	3C\;279 $	&	0.536 	&	0.12 	/	1.13 	&	21.71 	/	0.04 	&	1.35 	/	0.01 	&	1.60 	/	0.01 	&	1.72 	/	0.01 	&	46.94 	/	0.05 	&	46.11 	/	1.93 	&	0.27 	\\
J1315.0-4236	&	$\rm	H	$	&	0.07 	/	3.13 	&	16.99 	/	0.13 	&	-0.13 	/	0.01 	&	0.63 	/	0.02 	&	1.07 	/	0.02 	&	43.99 	/	0.19 	&	3.00 	&	0.19 	\\
$\rm	1ES\;1312-423  $	&	0.105 	&	0.21 	/	4.14 	&	24.85 	/	0.09 	&	0.59 	/	0.01 	&	0.73 	/	0.02 	&	0.93 	/	0.04 	&	43.36 	/	0.13 	&	43.32 	/	0.79 	&	1.48 	\\
J1427.0+2348	&	$\rm	H	$	&	0.09 	/	2.37 	&	15.47 	/	0.07 	&	-0.10 	/	0.01 	&	0.81 	/	0.01 	&	1.35 	/	0.01 	&	46.20 	/	0.10 	&	2.70 	&	0.11 	\\
$\rm	PKS\;1424+240  $	&	0.604 	&	0.48 	/	0.67 	&	24.10 	/	0.01 	&	0.77 	/	0.01 	&	0.93 	/	0.01 	&	1.56 	/	0.01 	&	46.59 	/	0.01 	&	46.39 	/	1.78 	&	0.08 	\\
J1428.5+4240	&	$\rm	H	$	&	0.08 	/	1.73 	&	17.18 	/	0.07 	&	-0.24 	/	0.01 	&	0.57 	/	0.01 	&	1.05 	/	0.01 	&	44.78 	/	0.10 	&	1.60 	&	0.13 	\\
$\rm	H\;1426+428  $	&	0.129 	&	0.65 	/	2.97	&	24.56 	/	0.02 	&	0.53 	/	0.01 	&	0.68 	/	0.01 	&	1.17 	/	0.03 	&	44.52 	/	0.03 	&	43.79 	/	1.09 	&	0.55 	\\
J1442.7+1200	&	$\rm	H	$	&	0.08 	/	2.27 	&	16.24 	/	0.09 	&	-0.09 	/	0.01 	&	0.71 	/	0.01 	&	1.19 	/	0.01 	&	44.47 	/	0.12 	&	2.10 	&	0.08 	\\
$\rm	1ES\;1440+122 $	&	0.163 	&	0.17 	/	1.24 	&	24.70 	/	0.03 	&	0.68 	/	0.01 	&	0.82 	/	0.01 	&	0.99 	/	0.01 	&	43.91 	/	0.05 	&	43.81 	/	1.00 	&	0.09 	\\
J1443.9+2501	&	$\rm	F	$	&	0.10 	/	3.22 	&	13.39 	/	0.08 	&	0.19 	/	0.01 	&	1.19 	/	0.02 	&	1.79 	/	0.02 	&	44.77 	/	0.12 	&		&	0.08 	\\
$\rm	PKS\;1441+25  $	&	0.939 	&	0.09 	/	1.64 	&	22.51 	/	0.08 	&	1.14 	/	0.01 	&	1.33 	/	0.02 	&	1.37 	/	0.01 	&	46.21 	/	0.11	&	46.27 	/	1.58 	&	0.16 	\\
J1443.9-3908	&	$\rm	H	$	&	0.09 	/	1.64 	&	15.82 	/	0.05 	&	-0.16 	/	0.01 	&	0.75 	/	0.01 	&	1.29 	/	0.01 	&	44.04 	/	0.07 	&	3.00 	&	0.06 	\\
$\rm	PKS\;1440-389  $	&	0.065 	&	0.36 	/	3.43 	&	24.31 	/	0.04 	&	0.71 	/	0.01 	&	0.87 	/	0.01 	&	1.27 	/	0.03 	&	43.94 	/	0.06 	&	43.88 	/	1.50 	&	0.78 	\\
J1512.8-0906	&	$\rm	F	$	&	0.11 	/	1.24 	&	13.67 	/	0.03 	&	-0.03 	/	0.01 	&	1.16 	/	0.01 	&	1.87 	/	0.01 	&	45.50 	/	0.04 	&	14.50 	&	0.02 	\\
$\rm	PKS\;1510-089  $	&	0.360 	&	0.12 	/	0.89 	&	21.72 	/	0.03 	&	1.11 	/	0.01 	&	1.32 	/	0.01 	&	1.72 	/	0.01 	&	46.76 	/	0.04 	&	45.89 	/	2.03 	&	0.06 	\\
J1517.7-2422	&	$\rm	L	$	&	0.12 	/	0.86 	&	13.95 	/	0.02 	&	-0.11 	/	0.01 	&	1.10 	/	0.01 	&	1.82 	/	0.01 	&	43.86 	/	0.03 	&		&	0.01 	\\
$\rm	AP\;Lib   $	&	0.048 	&	0.05 	/	0.58 	&	22.08 	/	0.05 	&	1.04 	/	0.01 	&	1.26 	/	0.01 	&	1.26 	/	0.01 	&	43.75 	/	0.07 	&	43.55 	/	1.62 	&	0.03 	\\
J1555.7+1111	&	$\rm	H	$	&	0.09 	/	0.84 	&	15.78 	/	0.03 	&	-0.20 	/	0.01 	&	0.75 	/	0.01 	&	1.31 	/	0.01 	&	46.04 	/	0.04 	&	5.10 	&	0.04 	\\
$\rm	PG\;1553+113 $	&	0.360 	&	0.61 	/	1.23 	&	24.71 	/	0.01 	&	0.70 	/	0.01 	&	0.87 	/	0.01 	&	0.98 	/	0.01 	&	46.30 	/	0.01 	&	46.07 	/	1.80 	&	0.26 	\\
J1653.8+3945	&	$\rm	H	$	&	0.06 	/	1.43 	&	16.30 	/	0.07 	&	0.09 	/	0.01 	&	0.76 	/	0.01 	&	1.15 	/	0.01 	&	44.04 	/	0.09 	&	1.50 	&	0.16 	\\
$\rm	Mkr\;501  $	&	0.033 	&	0.13 	/	0.85 	&	25.17 	/	0.03 	&	0.72 	/	0.01 	&	0.84 	/	0.01 	&	0.88 	/	0.01 	&	44.03 	/	0.04 	&	43.82 	/	1.80 	&	0.21 	\\
J1725.0+1152	&	$\rm	H	$	&	0.07 	/	2.41 	&	16.46 	/	0.10 	&	-0.01 	/	0.01 	&	0.71 	/	0.01 	&	1.14 	/	0.02 	&	44.92 	/	0.14 	&	1.10 	&	0.10 	\\
$\rm	H\;1722+119  $	&	0.180 	&	0.18 	/	0.78 	&	24.36 	/	0.02 	&	0.68 	/	0.01 	&	0.81 	/	0.01 	&	1.11 	/	0.01 	&	44.77 	/	0.03 	&	44.73 	/	1.46 	&	0.04 	\\
J1728.3+5013	&	$\rm	H	$	&	0.07 	/	1.13 	&	16.37 	/	0.05 	&	0.06 	/	0.01 	&	0.74 	/	0.01 	&	1.15 	/	0.01 	&	43.60 	/	0.07 	&	0.01 	&	0.02 	\\
$\rm	1ES\;1727+502  $	&	0.055 	&	0.06 	/	1.23 	&	25.66 	/	0.09 	&	0.71 	/	0.01 	&	0.83 	/	0.01 	&	0.88 	/	0.01 	&	43.46 	/	0.13 	&	43.41 	/	1.40 	&	0.10 	\\
J1744.0+1935	&	$\rm	H	$	&	0.06 	/	1.32 	&	16.57 	/	0.07 	&	0.18 	/	0.01 	&	0.76 	/	0.01 	&	1.10 	/	0.01 	&	43.77 	/	0.10 	&		&	0.06 	\\
$\rm	1ES\;1741+196  $	&	0.084 	&	0.90 	/	3.11 	&	24.35 	/	0.02 	&	0.73 	/	0.01 	&	0.84 	/	0.01 	&	1.61 	/	0.03 	&	43.92 	/	0.02 	&	43.33 	/	1.12 	&	0.56 	\\
J1751.5+0938	&	$\rm	L	$	&	0.17 	/	2.08 	&	13.23 	/	0.03 	&	-0.37 	/	0.01 	&	1.39 	/	0.01 	&	2.43 	/	0.01 	&	45.44 	/	0.04 	&	11.90 	&	0.08 	\\
$\rm	OT\;081  $	&	0.322 	&	0.07 	/	0.67 	&	21.21 	/	0.04 	&	1.30 	/	0.01 	&	1.62 	/	0.01 	&	1.47 	/	0.01 	&	45.67 	/	0.06 	&	45.01 	/	1.54 	&	0.03 	\\
J2000.0+6508	&	$\rm	H	$	&	0.07 	/	1.44 	&	16.82 	/	0.07 	&	-0.01 	/	0.01 	&	0.68 	/	0.01 	&	1.09 	/	0.01 	&	44.20 	/	0.09 	&	5.20 	&	0.10 	\\
$\rm	1ES\;1959+650  $	&	0.047 	&	0.13 	/	1.29 	&	24.72 	/	0.05 	&	0.65 	/	0.01 	&	0.77 	/	0.01 	&	0.99 	/	0.01 	&	44.05 	/	0.07 	&	43.92 	/	1.74 	&	0.66 	\\
J2009.4-4849	&	$\rm	H	$	&	0.07 	/	1.18 	&	16.48 	/	0.05 	&	0.03 	/	0.01 	&	0.72 	/	0.01 	&	1.13 	/	0.01 	&	44.78 	/	0.07 	&		&	0.09 	\\
$\rm	PKS\;2005-489   $	&	0.071 	&	0.19 	/	0.67 	&	24.42 	/	0.02 	&	0.69 	/	0.01 	&	0.82 	/	0.01 	&	1.10 	/	0.01 	&	43.84 	/	0.02 	&	43.80 	/	1.48 	&	0.07 	\\
J2039.5+5218	&	$\rm	H	$	&	0.07 	/	2.59 	&	16.20 	/	0.11 	&	-0.02 	/	0.01 	&	0.74 	/	0.01 	&	1.18 	/	0.02 	&	43.17 	/	0.15 	&	3.60 	&	0.08 	\\
$\rm	1ES\;2037+521  $	&	0.053 	&	0.42 	/	4.71 	&	24.67 	/	0.05 	&	0.70 	/	0.01 	&	0.84 	/	0.01 	&	1.02 	/	0.04 	&	42.71 	/	0.07 	&	42.74 	/	0.82 	&	1.10 	\\
J2158.8-3013	&	$\rm	H	$	&	0.10 	/	1.31 	&	15.92 	/	0.04 	&	-0.40 	/	0.01 	&	0.68 	/	0.01 	&	1.32 	/	0.01 	&	45.50 	/	0.05 	&		&	0.09 	\\
$\rm	PKS\;2155-304  $	&	0.116 	&	0.23 	/	0.84 	&	24.54 	/	0.02 	&	0.63 	/	0.01 	&	0.83 	/	0.01 	&	1.07 	/	0.01 	&	45.46 	/	0.02 	&	45.08 	/	1.88 	&	0.38 	\\
J2202.7+4216	&	$\rm	L	$	&	0.20 	/	2.42	&	12.95 	/	0.03 	&	-0.55 	/	0.01 	&	1.58 	/	0.01 	&	2.85 	/	0.01 	&	44.51 	/	0.04 	&	4.40 	&	0.10 	\\
$\rm	BL\;Lacertae    $	&	0.069 	&	0.05 	/	1.29 	&	21.28 	/	0.09 	&	1.48 	/	0.01 	&	1.87 	/	0.01 	&	1.36 	/	0.01 	&	44.62 	/	0.13 	&	44.26 	/	1.89 	&	0.21 	\\
J2250.0+3825	&	$\rm	H	$	&	0.07 	/	2.41 	&	16.33 	/	0.11 	&	0.05 	/	0.01 	&	0.74 	/	0.01 	&	1.15 	/	0.01 	&	44.18 	/	0.15 	&	2.00 	&	0.10 	\\
$\rm	B3\;2247+381  $	&	0.119 	&	0.20 	/	1.67 	&	24.77 	/	0.04 	&	0.71 	/	0.01 	&	0.84 	/	0.01 	&	0.97 	/	0.02 	&	44.04 	/	0.05 	&	43.75 	/	1.05 	&	0.24 	\\
J2324.7-4041	&	$\rm	H	$	&	0.10 	/	1.21 	&	15.79 	/	0.04 	&	-0.26 	/	0.01 	&	0.73 	/	0.01 	&	1.32 	/	0.01 	&	44.81 	/	0.05 	&		&	0.02 	\\
$\rm	1ES\;2322-409   $	&	0.174 	&	0.17 	/	1.42 	&	23.87 	/	0.04 	&	0.69 	/	0.01 	&	0.87 	/	0.01 	&	1.27 	/	0.01 	&	44.33 	/	0.05 	&	44.21 	/	1.14 	&	0.13 	\\
J2347.0+5141	&	$\rm	H	$	&	0.06 	/	0.97 	&	16.72 	/	0.05 	&	0.03 	/	0.01 	&	0.70 	/	0.01 	&	1.10 	/	0.01 	&	43.66 	/	0.07 	&	3.60 	&	0.04 	\\
$\rm	1ES\;2344+514   $	&	0.044 	&	0.24 	/	2.00 	&	24.95 	/	0.04 	&	0.67 	/	0.01 	&	0.79 	/	0.01 	&	0.87 	/	0.02 	&	43.88 	/	0.06 	&	43.40 	/	1.44 	&	0.57 	\\
J2359.0-3038	&	$\rm	H	$	&	0.07 	/	1.09 	&	17.18 	/	0.05 	&	-0.06	/	0.01 	&	0.63 	/	0.01 	&	1.04 	/	0.01 	&	44.80 	/	0.07 	&	2.90 	&	0.07 	\\
$\rm	H\;2356-309  $	&	0.165 	&	0.32 	/	0.94 	&	24.38 	/	0.01 	&	0.60 	/	0.01 	&	0.72 	/	0.01 	&	1.20 	/	0.01 	&	44.53 	/	0.02 	&	43.75 	/	1.03 	&	0.18 	\\
J0733.4+5152	&	$\rm	B	$	&	0.07 	/	1.47 	&	16.88 	/	0.06 	&	-0.12 	/	0.01 	&	0.64 	/	0.01 	&	1.09 	/	0.01 	&	43.65 / 0.09 	&		&	0.04 	\\
$\rm	PGC\;2402248  $	&	0.065 	&	0.40 	/	4.44 	&	24.44 	/	0.05 	&	0.60 	/	0.01 	&	0.74 	/	0.01 	&	1.20 	/	0.04 	&   43.14 / 0.07			&		42.68 / 0.81		&	0.99 	\\
\hline%
\\
\end{tabular}}
\end{adjustwidth}
\end{center}
\end{table*}

\renewcommand\thetable{2}
\begin{table*}[!htp]
\centering
\normalsize
\begin{center}
\caption{Jet Radiation Zone Parameters}
\label{tab:2}
\setcounter{table}{2}
\renewcommand{\thetable}{2/arabic{table}}
\renewcommand\arraystretch{0.80}
\begin{adjustwidth}{2.3cm}{1cm}
\scalebox{0.8}{
\begin{tabular}{cccccccc} 
\hline	 			
$\rm TeV\;Source\;Name$&{$\log {D_{\rm Th}}$} & {$\log {D_{\rm KN}}$} & {$\log B$} & {$\log R$} & {$\log {\gamma _{\rm p}}$} \\
\normalsize(1) & \normalsize(2) & \normalsize(3) &\normalsize(4) & \normalsize(5)   &\normalsize(6) \\
\hline
\centering
$\rm	SHBL\;J001355.9-185406  $	&	-0.80 	&		0.98 	&	-2.45 	&	16.89 	&	4.70 	\\
$\rm	RGB\;J0152+017    $	&	-0.18 	&	1.16 	&	-3.34 	&	17.15 	&	4.88 	\\
$\rm	1ES\;0229+200  $	&	-1.02 	& 	1.39 	&	-4.61 	&	17.50 	&	5.62 	\\
$\rm	1RXS\;J023832.6-311658  $	&	-0.65 	&	1.20 	&	-2.47 	&	17.22 	&	4.59 	\\
$\rm	RBS\;0413   	$	&	0.12 &	1.24 	&	-3.91 	&	17.28 	&	5.35 	\\
$\rm	1ES\;0347-121  $	&	-1.09  &	1.14 	&	-3.83 	&	17.12 	&	5.69 	\\
$\rm	1ES\;0414+009  $	&	-0.63  &	1.19 	&	-2.86 	&	17.20 	&	4.93 	\\
$\rm	1ES\;0502+675  $	&	-0.12  &	1.37 	&	-3.75 	&	17.47 	&	5.37 	\\
$\rm	PKS\;0548-322   $	&	-0.86 	&	1.76 	&	-4.91 	&	18.06 	&	5.33 	\\
$\rm	RGB\;J0710+591   $	&	-1.12 	&	1.30 	&	-3.71 	&	17.36 	&	5.35 	\\
$\rm	RBS\;0723     $	&	-0.87  &	1.18 	&	-3.91 	&	17.18 	&	5.52 	\\
$\rm	1RXS\;J101015.9-311909   $	&	-1.08 	&	1.09 	&	-2.87 	&	17.05 	&	4.79 	\\
$\rm	1ES\;1101-232   $	&	-0.04 	&	1.62 	&	-3.62 	&	17.84 	&	4.83 	\\
$\rm	RX\;J1136.5+6737   $	&	-1.20 	&	1.05 	&	-2.67 	&	16.98 	&	4.86 	\\
$\rm	Mkr\;180   $	&	-0.28 	&	1.37 	&	-3.57 	&	17.47 	&	4.72 	\\
$\rm	1ES\;1218+304   $	&	-0.78 	&	1.13 	&	-3.01 	&	17.10 	&	5.02 	\\
$\rm	W\;Comae      $	&	0.27   &	1.36  	&	-2.00 	&	17.46 	&	3.69 	\\
$\rm	MS\;1221.8+2452   $	&	0.06 	&	1.03 	&	-2.97 	&	16.96 	&	4.69 	\\
$\rm	1ES\;1312-423    $	&	-0.59 	&	1.00 	&	-3.12 	&	16.91 	&	5.19 	\\
$\rm	H\;1426+428   $	&	-0.78  &	0.75 	&	-2.24 	&	16.54 	&	5.07 	\\
$\rm	1ES\;1440+122  $	&	0.02   &	1.26 	&	-2.97 	&	17.31 	&	4.71 	\\
$\rm	1ES\;1727+502   $	&	0.07 	&	1.64 	&	-5.32 	&	17.87 	&	5.50 	\\
$\rm	1ES\;1741+196   $	&	-0.67 	&	1.45 	&	-3.79 	&	17.59 	&	4.66 	\\
$\rm	PKS\;2005-489   $	&	-0.13 	&	1.46 	&	-2.81 	&	17.61 	&	4.41 	\\
$\rm	B3\;2247+381    	$	&	-0.16 	&	1.42 	&	-3.71 	&	17.55 	&	4.82 	\\
$\rm	H\;2356-309    $	&	-0.86  &	0.99 	&	-2.53 	&	16.90 	&	4.88 	\\
$\rm	TXS\;0210+515   $	&	-0.62 	&	1.39 	&	-3.52 	&	17.49 	&	4.57 	\\
$\rm	RX\;J0648.7+1516  $	&	-0.40 	&	1.23 	&	-3.55 	&	17.26 	&	5.08 	\\
$\rm	1ES\;2037+521   $	&	-0.22 	&	1.16 	&	-3.36 	&	17.16 	&	4.82 	\\
$\rm	KUV\;00311-1938  $	&	0.39 	&	1.16 	&	-2.21 	&	17.15 	&	4.54 	\\
$\rm	PKS\;0301-243   $	&	0.46 	&	1.75 	&	-2.96 	&	18.04 	&	3.89 	\\
$\rm	PKS\;0447-439   	$	&	0.38 	&	1.17 	&	-1.89 	&	17.16 	&	4.11 	\\
$\rm	1ES\;0647+250   $	&	-0.53 	&	1.06 	&	-2.56 	&	17.01 	&	4.69 	\\
$\rm	1ES\;0806+524   $	&	0.01 	&	1.13 	&	-2.35 	&	17.10 	&	4.37 	\\
$\rm	1ES\;1011+496    $	&	0.13 	&	1.45 	&	-3.11 	&	17.58 	&	4.48 	\\
$\rm	Mkr\;421   $	&	-0.52 	&	1.18 	&	-4.06 	&	17.19 	&	5.46 	\\
$\rm	PKS\;1424+240    $	&	0.48 	&	1.38 	&	-1.97 	&	17.48 	&	3.99 	\\
$\rm	PKS\;1440-389   $	&	0.05 	&	1.00 	&	-2.25 	&	16.91 	&	4.40 	\\
$\rm	PG\;1553+113   $	&	0.51 &	1.35 	&	-2.63 	&	17.45 	&	4.50 	\\
$\rm	Mkr\;501   $	&	-0.02 	&	1.60 	&	-4.55 	&	17.82 	&	5.06 	\\
$\rm	H\;1722+119  $	&	-0.29  &	1.26 	&	-2.70 	&	17.30 	&	4.54 	\\
$\rm	1ES\;1959+650  $	&	-0.55  &	1.16 	&	-3.35 	&	17.16 	&	5.00 	\\
$\rm	PKS\;2155-304   $	&	0.34 	&	1.01 	&	-1.90 	&	16.93 	&	4.46 	\\
$\rm	1ES\;2322-409    $	&	0.11 	&	0.92 	&	-1.12 	&	16.80 	&	3.98 	\\
$\rm	VER\;J0521+211   $	&	0.27 	&	1.23 	&	-1.85 	&	17.26 	&	3.77 	\\
$\rm	1ES\;2344+514   	$	&	-0.56 	&	1.20 	&	-4.06 	&	17.22 	&	5.24 	\\
$\rm	S2\;0109+22    $	&	1.06   &	1.18 	&	-0.76 	&	17.18 	&	3.22 	\\
$\rm	3C\;66A   $	&	0.76 	&	1.48 	&	-1.24 	&	17.63 	&	3.13 	\\
$\rm	TXS\;0506+056   $	&	1.01 	&	1.23 	&	-0.42 	&	17.26 	&	2.97 	\\
$\rm	S5\;0716+714  $	&	0.56   &	 0.89 	&	0.36 	&	16.75 	&	2.98 	\\
$\rm	1ES\;1215+303   	$	&	0.54 	&	1.24 	&	-1.72 	&	17.28 	&	3.58 	\\
$\rm	S3\;1227+25    $	&	0.81 	&	0.98 	&	-0.05 	&	16.88 	&	3.05 	\\
$\rm	OJ\;287   $	&	1.33 	&	0.86 	&	2.84 	&	16.70 	&	1.29 	\\
$\rm	S4\;0954+65  $	&	1.36 	&	0.93 	&	2.28 	&	16.81 	&	1.32 	\\
$\rm	AP\;Lib    $	&	0.64 	&	0.77 	&	1.26 	&	16.56 	&	2.11 	\\
$\rm	OT\;081    $	&	1.07 	&	0.67 	&	3.17 	&	16.42 	&	1.16 	\\
$\rm	BL\;Lacertae   $	&	1.23 	&	0.55 	&	3.14 	&	16.25 	&	1.16 	\\
$\rm	S3\;0218+35   $	&	2.12 	&	1.21	&	0.66 	&	17.23 	&	2.07 	\\
$\rm	PKS\;0736+017   $	&	0.41 	&	0.80 	&	1.51 	&	16.62 	&	1.87 	\\
$\rm	TON\;0599   $	&	1.50 	&	1.01 	&	1.98 	&	16.92 	&	1.59 	\\
$\rm	4C+21.35     $	&	0.84 	&	0.86 	&	1.71 	&	16.70 	&	1.97 	\\
$\rm	3C\;279   $	&	1.75 	&	1.42 	&	1.85 	&	17.54 	&	1.13 	\\
$\rm	PKS\;1441+25   $	&	0.97 	&	1.33 	&	-0.57 	&	17.41 	&	2.42 	\\
$\rm	PKS\;1510-089   	$	&	0.70 	&	0.85 	&	1.63 	&	16.69 	&	1.79 	\\
$\rm    PGC\;2402248    $    & -0.74    & 0.82     & -2.52    & 16.65    & 4.93 \\
\hline
\end{tabular}}\\
\end{adjustwidth}
\end{center}
{\footnotesize{Note. Column (2) gives the physical parameters of the Doppler factor ($\log {D_{\rm Th}}$) obtained from the Thomson scattering region; Columns (3)\raisebox{0.5mm}{-}(6) give the physical parameters obtained from the KN scattering region \raisebox{0.5mm}{------} namely, the Doppler factor ($\log {D_{\rm KN}}$), magnetic field strength ($\log B$), radiative zone radius ($\log R$), and peak Lorentz factor ($\log {\gamma _{\rm p}}$).}}
\end{table*}

\begin{figure}[!htp]
   \setlength{\abovecaptionskip}{0.1cm}
   \setlength{\belowcaptionskip}{5cm}
   \flushleft
   \includegraphics[bb=55 80 900 680,width=30cm,height=8.8in,clip]{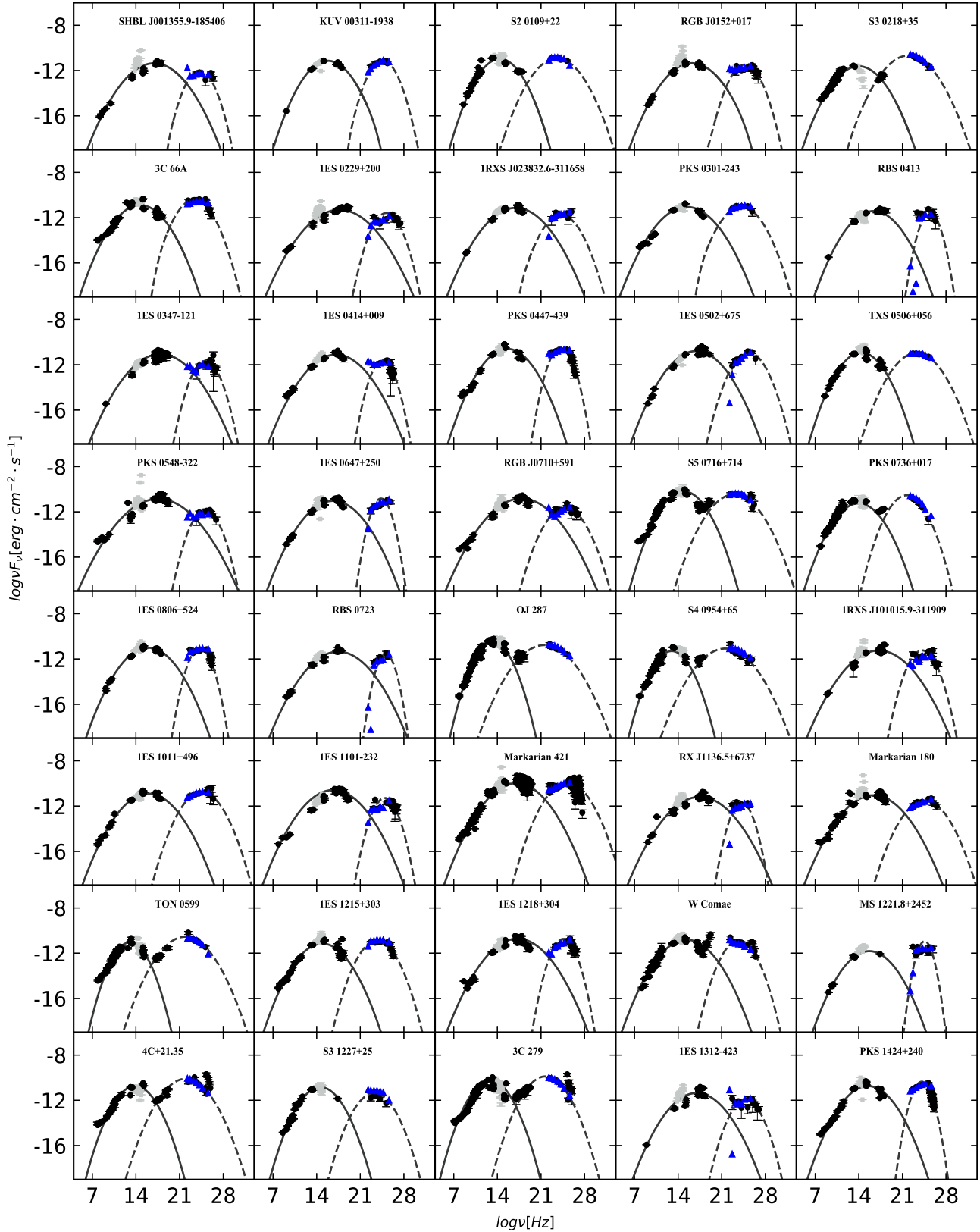}
   \caption{The SEDs of Fermi TeV blazars. The dots represent the data from the SSDC, and the square points represent the data of the thermal radiation bump. The triangles represent the data from the 4FGL (\citealt{2020ApJS..247...33A}). The solid line is the fitting line in the synchrotron region, and the dashed line is the fitting line in the ICs region. }
   \label{Fig1}
\end{figure}

\begin{figure}[!htp]
   \setlength{\abovecaptionskip}{0.1cm}
   \setlength{\belowcaptionskip}{10cm}
   \flushleft
   \includegraphics[bb=55 80 900 680,width=30cm,height=5.6in,scale=10,clip]{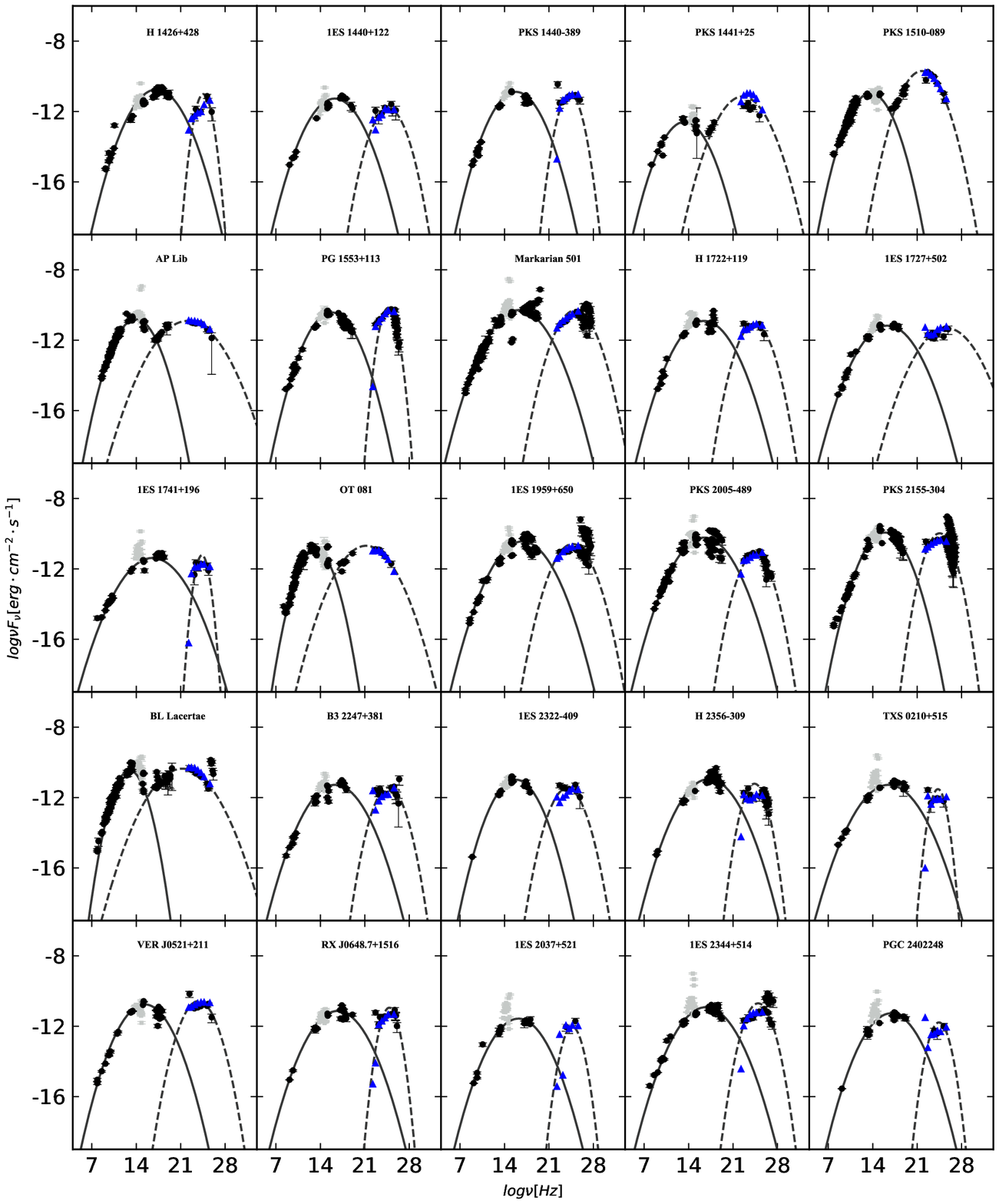}
   \caption{The SEDs of the remaining Fermi TeV blazars. }
   \label{Fig2}
\end{figure}

\section{Results} \label{sect:results}
\begin{figure}[!htp]
  \begin{minipage}[t]{0.495\linewidth}
  \centering
   \includegraphics[width=70mm]{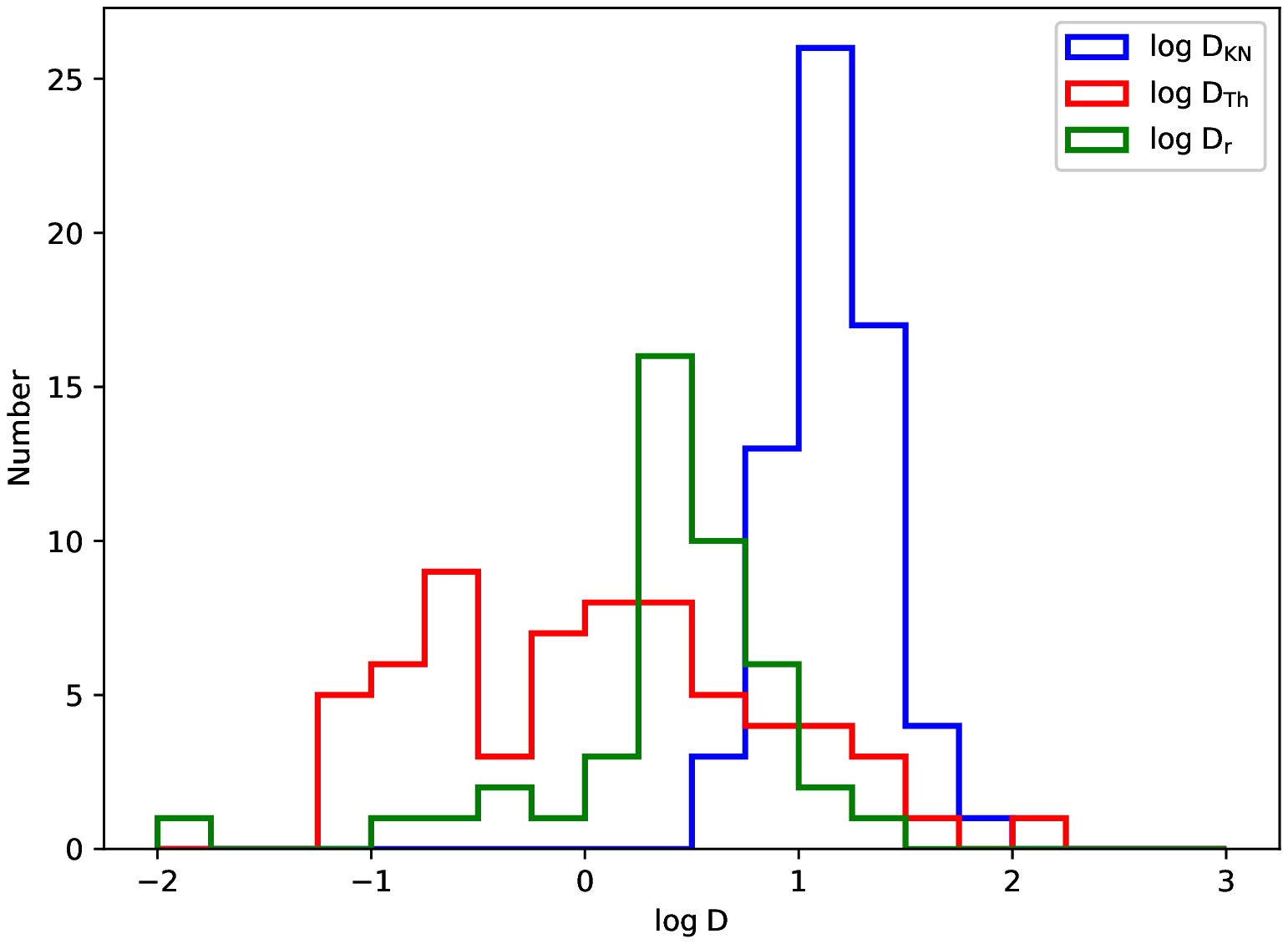}
   \caption{{\small The Doppler factor distributions for the TeV blazars. } }
   \label{Fig3}
  \end{minipage}%
  \begin{minipage}[t]{0.495\textwidth}
  \centering
   \includegraphics[width=70mm]{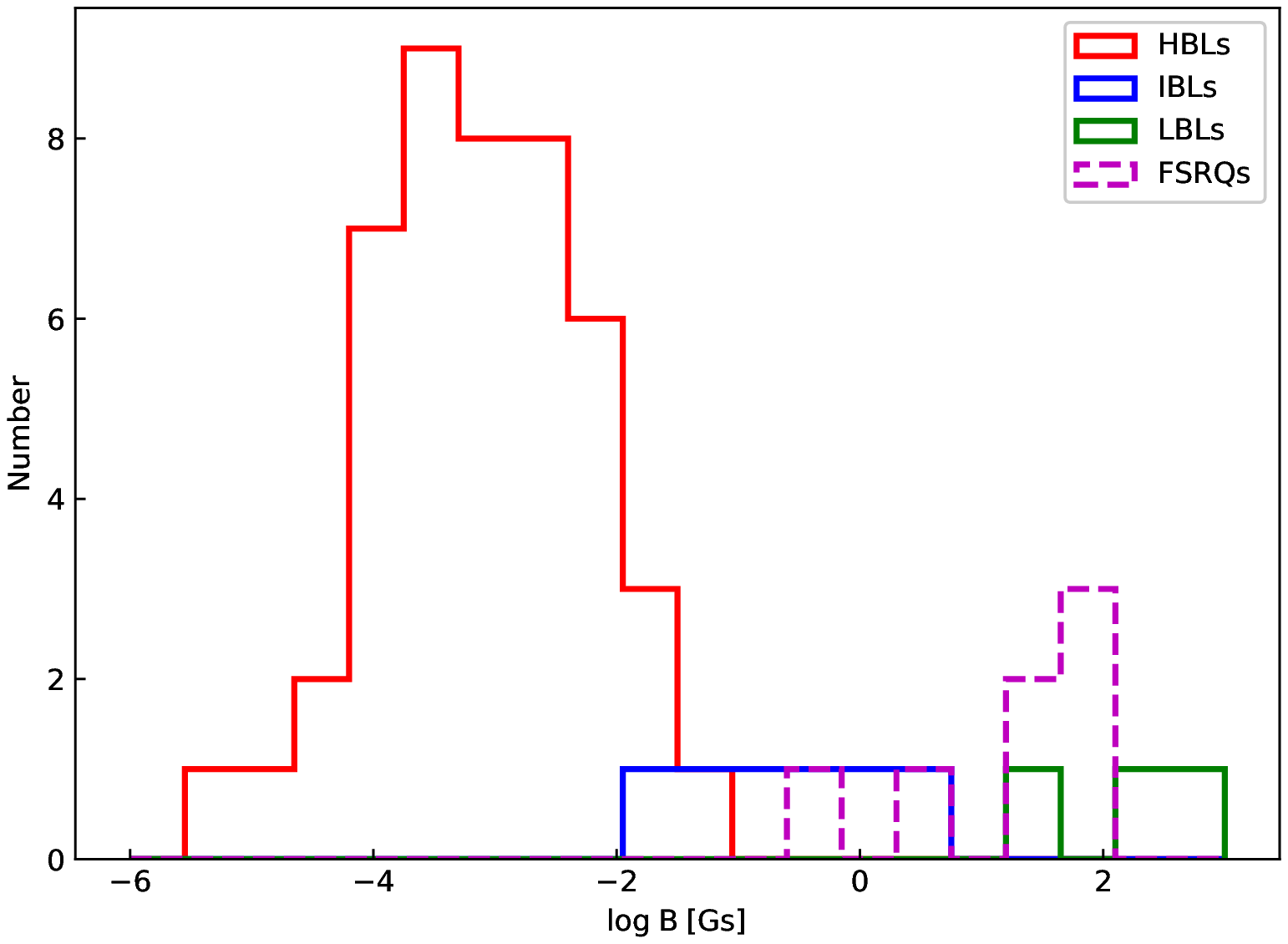}
  \caption{{\small Distribution of magnetic field strength for different types of TeV blzars. }}
  \label{Fig4}
  \end{minipage}%
  \\
  \begin{minipage}[t]{0.495\linewidth}
  \centering
   \includegraphics[width=70mm]{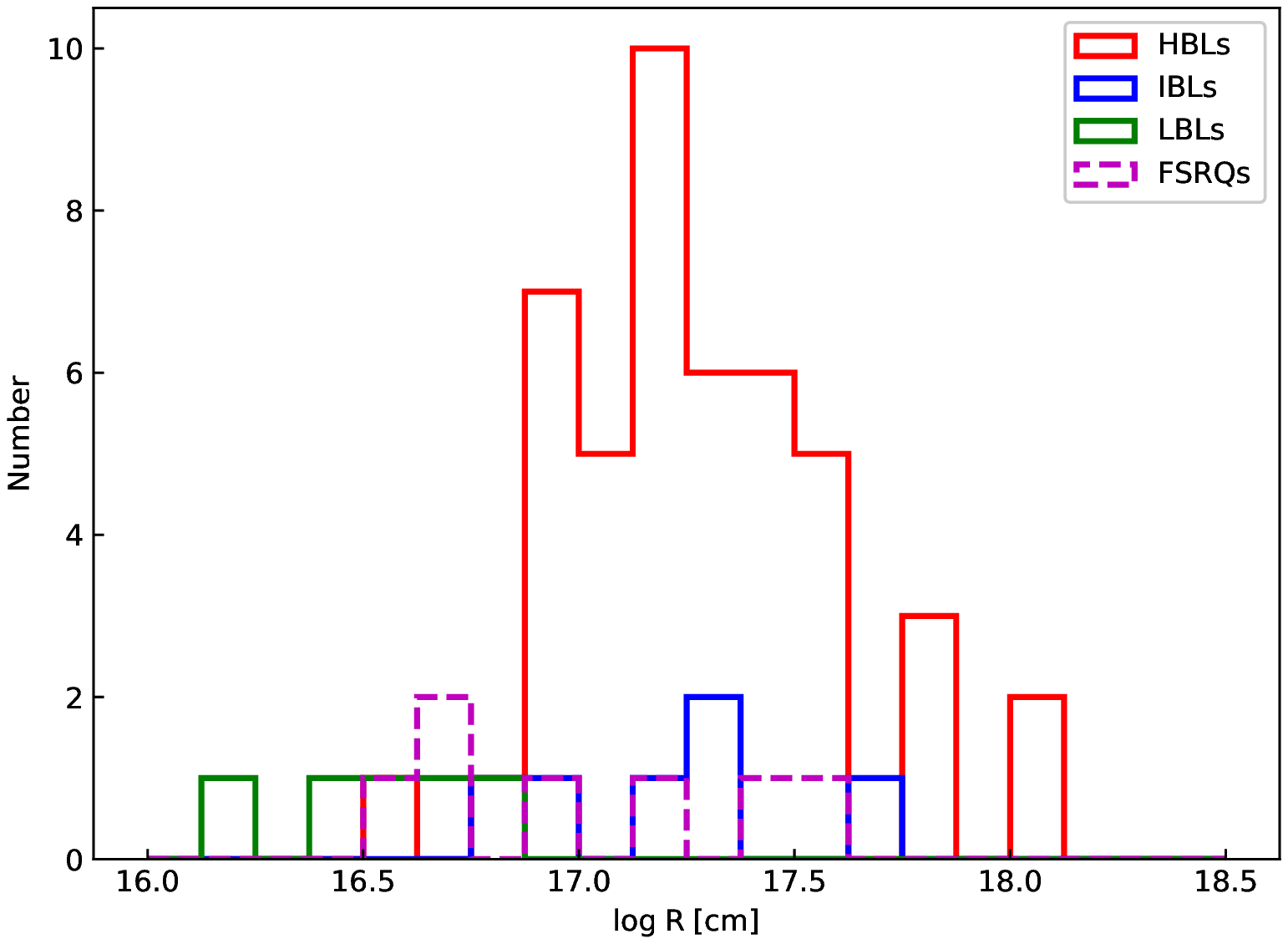}
  \caption{{\small Distribution of radiative zone radius for different types of TeV blzars. } }
  \label{Fig5}
  \end{minipage}%
    \begin{minipage}[t]{0.495\linewidth}
  \centering
   \includegraphics[width=70mm]{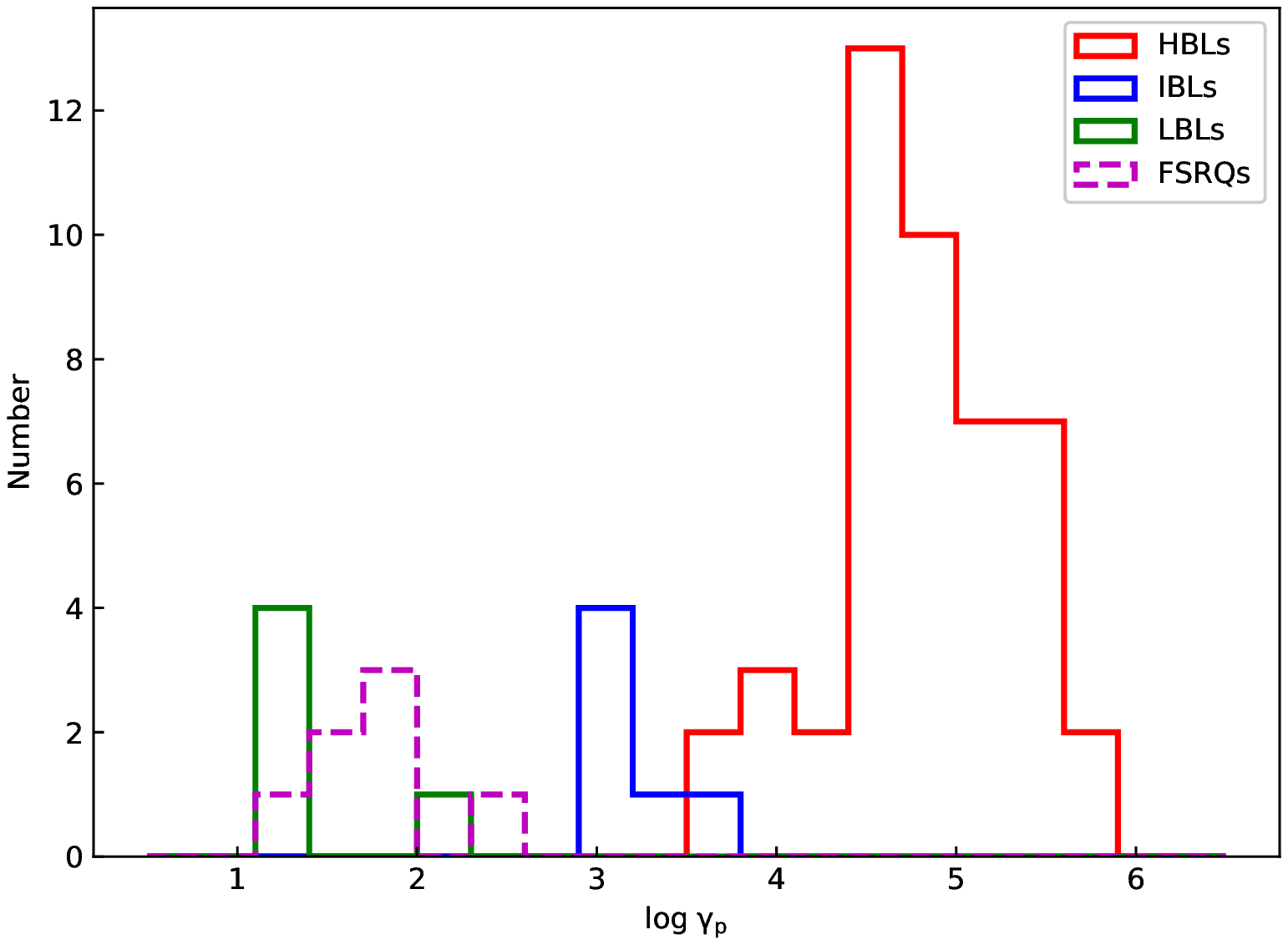}
   \caption{{\small Distribution of peak Lorentz factor for different types of TeV blzars. } }
   \label{Fig6}
  \end{minipage}%
  \setlength{\abovecaptionskip}{-0cm}
  \setlength{\belowcaptionskip}{-0cm}
\end{figure}

In our samples of 64 identified classified TeV blazars with redshift, there are 57 BL Lac objects and 7 FSRQs. According to the peak frequency of synchrotron radiation, the sample can be further divided into 46 HBLs, 6 IBLs, 5 LBLs, and 7 FSRQs. Among these, the FSRQs are all low-synchrotron-peaked. Based on our sample, the statistical results of the fitting parameters are listed in Table~\ref{tab:3}. The histogram distributions of $\log {D}$ are shown in Figure~\ref{Fig3}, where three different definitions of the Doppler factor are represented. Figures~\ref{Fig4}$-$~\ref{Fig6} show the statistical histograms of the radiation parameters obtained under the KN scattering region conditions between the different types of TeV blazars. The correlation between the parameters and the linear regression analysis results are shown in Table~\ref{tab:4}.

\renewcommand\thetable{3}
\begin{table*}[!htp]
\flushleft{}
\caption{Statistical Values for the Fitting Results}
\label{tab:3}
\setcounter{table}{3}
\renewcommand{\thetable}{3/arabic{table}}
\renewcommand\arraystretch{0.8}
\small
\scalebox{0.8}{
\resizebox{\textwidth}{!}{
\begin{tabular}{cp{1cm}<{\centering}p{1cm}<{\centering}p{1cm}<{\centering}p{1cm}<{\centering}p{1cm}<{\centering}
p{1cm}<{\centering}p{1cm}<{\centering}} 
\hline		
{$\rm Parameters$} &{$\rm Items$} & {$\rm HBLs$}& {$\rm IBLs$} &$\rm LBLs$& {$\rm FSRQs$}  &$\rm All$\\
\tiny(1) & \tiny(2) & \tiny(3) &\tiny(4) & \tiny(5) &  \tiny(6) &  \tiny(7) \\
\hline
\multirow{3}{*}{$\beta_{\rm syn}$} &Mean &0.07 & 0.11 &0.16 &$0.12$ & 0.09 \\
  &Max &0.10 & 0.12 &0.20 &$0.18$ & 0.20 \\
  &Min &0.05 & 0.09 &0.12 &$0.09$ & 0.05 \\
\multirow{3}{*}{$\beta_{\rm ic}$} &Mean &0.31 & 0.12 &0.06 &$0.11$ & 0.25\\
  &Max &0.90 & 0.16 &0.07 &$0.16$ & 0.90\\
  &Min &0.06 & 0.09 &0.05 &$0.08$ & 0.05\\
\multirow{3}{*}{$\log \nu _{\rm peak}^{\rm syn}$} &Mean &16.53 & 14.64 &13.28 &$13.54$ & 15.77 \\
  &Max &17.94 & 14.91 &13.95 &$13.98$ & 17.94 \\
  &Min &15.35 & 14.39 &12.95 &$12.84$ & 12.84 \\
\multirow{3}{*}{$\log \nu _{\rm peak}^{\rm ic}$} &Mean &24.61 & 23.37 &21.54 &$21.92$ & 23.96 \\
  &Max &25.66 & 23.72 &22.08 &$22.51$ & 25.66 \\
  &Min &23.77 & 23.03 &21.21 &$21.69$ & 21.21 \\
\multirow{3}{*}{$\alpha_{\rm r}$} &Mean &-0.04 & -0.18 &-0.34 &$-0.04$ & -0.08 \\
  &Max &0.18 & 0.05 &-0.11 &$0.19$ & 0.19 \\
  &Min &-0.40 & -0.33 &-0.55 &$-0.39$ & -0.55 \\
\multirow{3}{*}{$\alpha_{\rm ir}$} &Mean &0.67 & 0.89 &1.30 &$1.16$ & 0.79 \\
  &Max &0.83 & 0.94 &1.48 &$1.39$ & 1.48 \\
  &Min &0.53 & 0.82 &1.04 &$1.03$ & 0.53 \\
\multirow{3}{*}{$\alpha_{\rm o}$} &Mean &0.71 & 0.95 &1.38 &$1.22$ & 0.84 \\
  &Max &0.86 & 1.00 &1.58 &$1.48$ & 1.58 \\
  &Min &0.57 & 0.88 &1.10 &$1.07$ & 0.57 \\
\multirow{3}{*}{$\alpha_{\rm uv}$} &Mean &0.81 & 1.10 &1.61 &$1.39$ & 0.96 \\
  &Max &0.96 & 1.16 &1.87 &$1.72$ & 1.87 \\
  &Min &0.66 & 1.03 &1.26 &$1.20$ & 0.66 \\
\multirow{3}{*}{$\alpha_{\rm x}$} &Mean &1.15 & 1.61 &2.40 &$1.96$ & 1.38 \\
  &Max &1.35 & 1.74 &2.85 &$2.58$ & 2.85 \\
  &Min &0.95 & 1.46 &1.82 &$1.62$ & 0.95 \\
\multirow{3}{*}{${\alpha _{\rm \gamma }}$} &Mean &1.06 & 1.32 &1.40 &$1.60$ & 1.17 \\
  &Max &1.61 & 1.41 &1.47 &$1.93$ & 1.93 \\
  &Min &0.32 & 1.22 &1.26 &$1.37$ & 0.32 \\
\multirow{3}{*}{$\log L_{\rm peak}^{\rm syn}$} &Mean &44.58 & 45.15 &44.96 &$45.60$ & 44.77 \\
  &Max &46.20 & 45.77 &45.80 &$46.60$ & 46.60 \\
  &Min &43.17 & 44.39 &43.86 &$44.62$ & 43.17 \\
\multirow{3}{*}{$\log L_{\rm peak}^{\rm ic}$} &Mean &44.39 & 45.15 &45.00 &$46.44$ & 44.73 \\
  &Max &46.59 & 46.24 &45.67 &$46.94$ & 46.94 \\
  &Min &42.71 & 44.21 &43.75 &$45.34$ & 42.71 \\
\hline
\end{tabular}}}\\
{\footnotesize{Note. Column (1) gives the parameter, where $\log \nu _{\rm peak}^{\rm syn}$ and $\log \nu _{\rm peak}^{\rm ic}$ in units of $\rm hertz$, $\log L_{\rm peak}^{\rm syn}$ and $\log L_{\rm peak}^{\rm ic}$ are in units of ergs per second; Column (2) indicates the mean value (Mean), maximum value (Max), and minimum value (Min) of the parameter items.} }
\end{table*}

\renewcommand\thetable{4}
\begin{table*}[!htp]
\centering
\caption{Results of Correlation Text and Linear Regression Analysis}
\label{tab:4}
\setcounter{table}{4}
\renewcommand{\thetable}{4/arabic{table}}
\renewcommand\arraystretch{1.2}
\begin{adjustwidth}{-1.5cm}{-1cm}
\resizebox{\textwidth}{!}{
\begin{tabular}{p{3cm}<{\centering}p{3cm}<{\centering}p{2cm}<{\centering}p{1cm}<{\centering}p{1cm}<{\centering}p{1cm}<{\centering}p{1cm}<{\centering}
p{2cm}<{\centering}p{2cm}<{\centering}} 
\hline		
{$\rm y \sim x$} & {$a \sim \sigma_{a}$} & {$b \sim \sigma_{b}$} &${r^\prime }$ & {$r$}& $N$ & {$\rho $}&$ P$ \\
\tiny(1) & \tiny(2) & \tiny(3) &\tiny(4) & \tiny(5) & \tiny(6) & \tiny(7) &  \tiny(8) \\
\hline
\footnotesize{$\log {\rm{ }}{\nu ^\prime }_{\rm peak}^{\rm ic}{\rm{ }} \sim {\rm{ }}\log {\rm{ }}{\nu ^\prime }_{\rm peak}^{\rm syn}$} &$12.80 \pm 0.70$ &$0.70 \pm 0.04$ & 0.79 &0.82 &64 &$\rm 0.79$ &$ < {10^{ - 4}}$\\
\footnotesize{${\rm{ }}\log {\rm{ }}{L^\prime }_{\rm peak}^{\rm ic} \sim \log {\rm{ }}L_\gamma ^\prime {\rm{ }}$}  & $3.60 \pm 1.30$ &$0.95 \pm 0.03$ & 0.93 &0.93 &64 &$0.95$ &$ < {10^{ - 4}}$\\
\footnotesize{${\alpha _{\rm r}}{\rm{ }} \sim {\rm{ }}\log {\rm{ }}\nu _{\rm peak}^{\prime \rm syn}$}  &$- 0.60 \pm 0.20$ &$0.04 \pm 0.02$ & 0.07 &0.17 &64 &$0.22$ &$ > 0.05$\\
\footnotesize{${\alpha _{\rm ir}}{\rm{ }} \sim {\rm{ }}\log {\rm{ }}\nu _{\rm peak}^{\prime \rm syn}$}  &$3.10 \pm 0.13$ &$- 0.16 \pm 0.01$ & 0.83 &0.89 &64 &$-0.96$ &$ < {10^{ - 4}}$\\
\footnotesize{${\alpha _{\rm o}}{\rm{ }} \sim {\rm{ }}\log {\rm{ }}\nu _{\rm peak}^{\prime \rm syn}$}  &$3.30 \pm 0.10$ &$ - 0.17 \pm 0.01$ & 0.83 &0.89 &64  &$-0.97$ &$ < {10^{ - 4}}$\\
\footnotesize{${\alpha _{\rm uv}}{\rm{ }} \sim {\rm{ }}\log {\rm{ }}\nu _{\rm peak}^{\prime \rm syn}$}  &$3.80 \pm 0.20$ &$ - 0.20 \pm 0.01$ & 0.81 &0.89 &64 &$-0.98$ &$ < {10^{ - 4}}$\\
\footnotesize{${\alpha _{\rm x}}{\rm{ }} \sim {\rm{ }}\log {\rm{ }}\nu _{\rm peak}^{\prime \rm syn}$}  &$5.60 \pm 0.30$ &$ - 0.30 \pm 0.02$ & 0.76 &0.87 &64 &$-0.96$ &$ < {10^{ - 4}}$\\
\footnotesize{${\alpha _{\rm \gamma}}{\rm{ }} \sim {\rm{ }}\log {\rm{ }}\nu _{\rm peak}^{\prime \rm ic}$}  &$6.20 \pm 0.50$ &$ - 0.22 \pm 0.02$ & 0.59 &0.44 &64 &$-0.86$ &$ < {10^{ - 4}}$\\
\footnotesize{${\rm{ }}1/{\beta _{\rm syn}} \sim {\log {\rm{ }}\nu _{\rm peak}^{\prime \rm syn}{\rm{ }} }$} &$ \tiny{-19.20 \pm 2.90}$ &$2.20 \pm 0.20$ & 0.66 &0.80 &64 &$0.81$ &$ < {10^{ - 4}}$\\
\footnotesize{${\rm{ }}1/{\beta _{\rm ic}} \sim {\log {\rm{ }}\nu _{\rm peak}^{\prime \rm ic}{\rm{ }} }$}  &$70.80 \pm 9.00$ &$ - 2.80 \pm 0.40$ & 0.44 &0.42 &64 &$-0.54$ &$ < {10^{ - 4}}$\\
\footnotesize{$\log {\rm{ }}{\nu ^\prime }_{\rm peak}^{\rm ic}{\rm{ }} \sim \log {\rm{ }}{\gamma _{\rm p}}$}  &$19.80 \pm 0.10$ &$0.76 \pm 0.02$ & 0.96 &0.97 &64 &$0.94$ &$ < {10^{ - 4}}$\\
\footnotesize{$\log B{\rm{ }} \sim \log {\rm{ }}{\gamma _{\rm p}}$}  &$3.60 \pm 0.80$&$ - 1.40 \pm 0.20$ & $0.60$ &- & 46 &$-0.79$ &$ < {10^{ - 4}}$\\
\footnotesize{$\log R{\rm{ }} \sim \log {\rm{ }}{\gamma _{\rm p}}$}  &$17.20 \pm 0.50$&$ 0.01 \pm 0.10$ & $0.02$ &- & 46 &$0.03$ &$> 0.05$\\
\hline
\end{tabular}}\\
{\footnotesize{Note. Column (1) gives the relation $\rm y = \textit{a} + bx$; Column (2) gives the intercept and the corresponding uncertainty; Column (3) gives the slope and the corresponding uncertainty; Column (4) gives the linear correlation coefficient in the co-moving frame; Column (5) gives the linear correlation coefficient in the observer's coordinate system; Column (6) gives the number of the sources; Column (7) gives the Spearman correlation coefficient; and Column (8) gives the chance probability. } }
\end{adjustwidth}
\end{table*}

According to our results, we can infer the following: (1) The spectral index increases gradually from the infrared to the high-energy band in the synchrotron regions and the spectral index is different for different types of TeV blazars. The mean spectral indices obtained are as follows: ${\left\langle {{\alpha _{\rm r}}} \right\rangle _{\rm BL{\rm{ }}Lacs}} =  - 0.19$, ${\left\langle {{\alpha _{\rm ir}}} \right\rangle _{\rm BL{\rm{ }}Lacs}} = 0.95$, ${\left\langle {{\alpha _{\rm o}}} \right\rangle _{\rm BL{\rm{ }}Lacs}} = 1.01$, ${\left\langle {{\alpha _{\rm uv}}} \right\rangle _{\rm BL{\rm{ }}Lacs}} = 1.17$, and ${\left\langle {{\alpha _{\rm x}}} \right\rangle _{\rm BL{\rm{ }}Lacs}} = 1.72$. These values are close to the results of \cite{2017ApJ...835L..38F} and \cite{2000ApJ...537...80C}. The mean curvatures of the synchrotron bump and ICs bump in the peak satisfy the following relations: $\left\langle {{\beta _{\rm ic}}} \right\rangle  = 2.87\left\langle {{\beta _{\rm syn}}} \right\rangle $, and ${\left\langle {{\beta _{\rm ic}}} \right\rangle _{\rm HBLs}} = 4.35{\left\langle {{\beta _{\rm syn}}} \right\rangle _{\rm HBLs}}$. The latter of the above results is more similar to that of \cite{2004A&A...413..489M,2004A&A...422..103M}, indicating that the curvature relation of TeV HBLs conforms to the condition of the KN scattering region. (2) There are many ways to estimate the Doppler factor at present. In our model, when we derive the Doppler factor, we assume that $D  \approx \Gamma$, which requires that the Doppler factor be greater than 1. Therefore, it can be seen from the Doppler factor histogram that KN scattering region calculation is more suitable. Therefore, Equation (\ref{Eq:17}) is used to calculate the parameters of the jet radiation zone. Within the sample, the Doppler factor is in the range of $3.58 \le {D} \le 57.65$. The median values of ${D}$ of the HBLs, IBLs, LBLs, FSRQs, and total TeV blazars are 15.98, 16.09, 5.83, 10.12, and 15.19, respectively. This result is consistent with that of \cite{2018ApJS..235...39C}, which is within a reasonable range of $1 \le {D} \le 100$. The magnetic field strength is in the range of $4.82 \times {10^{ - 6}}\;{\rm{ G}} \le B \le 1.47 \times {10^3}\;{\rm{ G}}$. The median values of $B$ of the HBLs, IBLs, LBLs, FSRQs, and total TeV blazars are $1.03 \times {10^{ - 3}}$, 0.26, $6.96 \times {10^2}$, 42.88 and 0.003 $\rm G$, respectively. \cite{2012ApJ...752..157Z} estimated the magnetic field strength of GeV-TeV BL Lac objects, and gave a typical value of $\rm 0.1{\rm{ }}\;G \le B \le 0.6{\rm{ }}\;G$. In our results, the magnetic field strengths of LBLs and FSRQs are overestimated. The radiative zone radius is in the range of $1.76 \times {10^{16}}\;{\rm{ }}\rm cm \le R \le 1.14 \times {10^{18}}\;{\rm{ }}\rm cm$. The median values of $R$ of the HBLs, IBLs, LBLs, FSRQs, and total TeV blazars are $1.66 \times {10^{17}}$, $1.67 \times {10^{17}}$, $3.65 \times {10^{16}}$, $8.34 \times {10^{16}}$ and $1.53 \times {10^{17}}$ $\rm cm$, respectively. Our results are close to the result of \cite{2012ApJ...752..157Z}, which is within the reasonable estimation range of ${10^{15}}\;\rm cm \le R \le {10^{17}}\;\rm cm$. The peak Lorentz factor is in the range of $13.57 \le {\gamma _{\rm p}} \le 4.86 \times {10^5}$. The median values of ${\gamma _{\rm p}}$ of the HBLs, IBLs, LBLs, FSRQs, and total TeV blazars are $6.34 \times {10^4}$, $1.23 \times {10^3}$, $19.36$, $73.91$, and $3.58 \times {10^4}$, respectively. The range of the initial Lorentz factor ${\gamma _{\rm 0}}$ is $100 \le {\gamma _{\rm 0}} \le 1000$ and the relationship between ${\gamma _{\rm p}}$ and ${\gamma _{\rm 0}}$ is ${\gamma _{\rm p}} = {\gamma _0}{10^{ - \frac{s}{{2r}}}}$.  We can roughly give the reasonable range of the peak Lorentz factor as ${10^4} \le {\gamma _{\rm p}} \le {10^5}$ (\citealt{2015ApJ...807...51Z,2015MNRAS.448.1060G,2018ApJS..235...39C}). (3) Most of the TeV blazars are HBLs, so we can reasonably estimate their physical parameters based on the above results, as follows: for the HBL sources, the reasonable range of jet radiation zone parameters is $5.6 \le {D} \le 31.6$, $6.3 \times {10^{ - 5}}\;\rm G \le B \le 1.1 \times {10^{ - 2}}\;\rm G$, $7.5 \times {10^{16}}\;\rm cm \le R \le 4.2 \times {10^{17}}\;\rm cm$, and $2.5 \times {10^4} \le {\gamma _{\rm p}} \le 4.0 \times {10^5}$, as shown in Figures~\ref{Fig3}$-$~\ref{Fig6}.

\subsection{Intrinsic Synchrotron Peak Frequency versus Intrinsic ICs Peak Frequency}
\begin{figure}[!htp]
  \begin{minipage}[t]{0.495\linewidth}
  \centering
   \includegraphics[width=80mm]{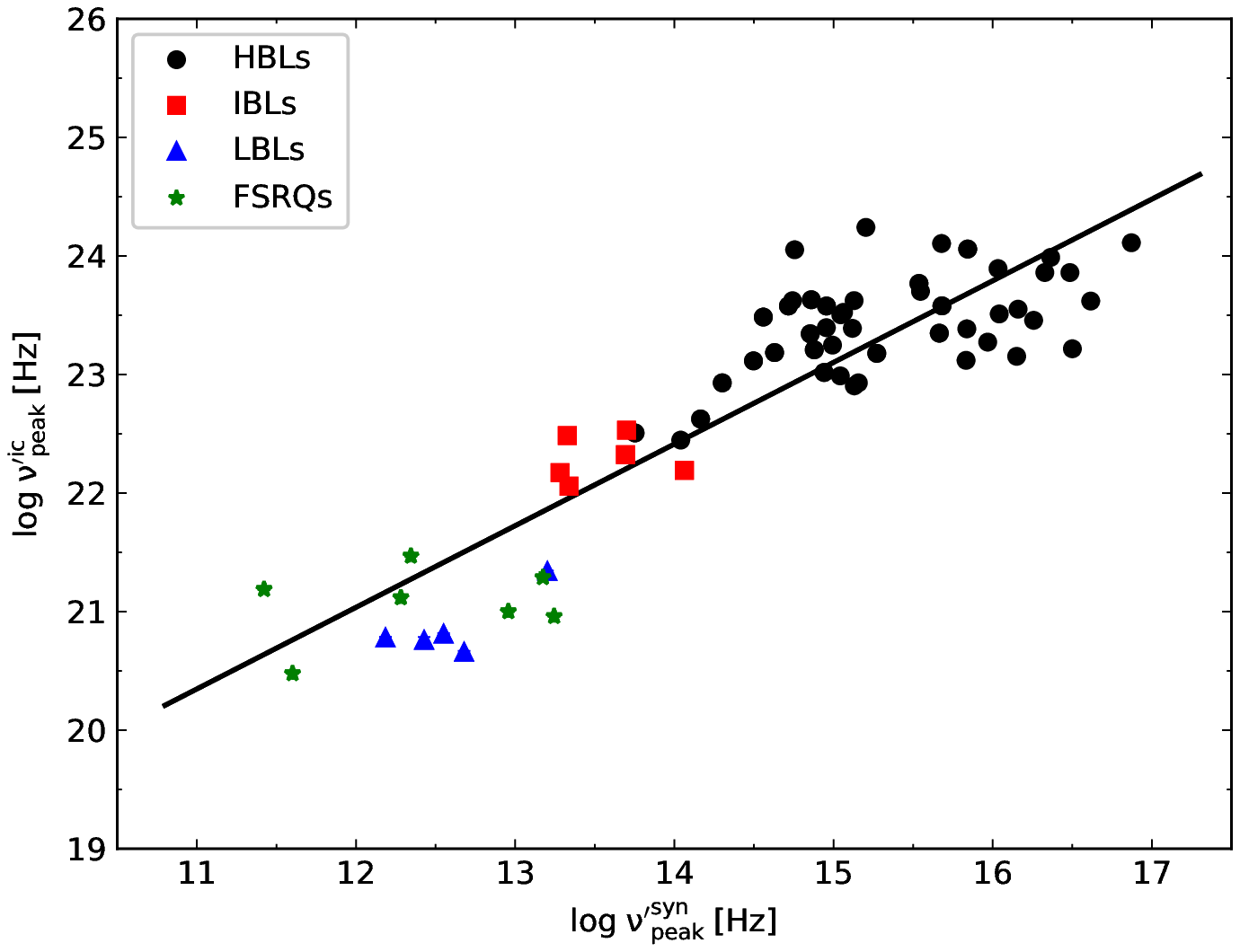}
   \caption{{\small The relation between intrinsic synchrotron peak frequency and intrinsic ICs peak frequency. } }
   \label{Fig7}
  \end{minipage}%
  \begin{minipage}[t]{0.495\textwidth}
  \centering
   \includegraphics[width=80mm]{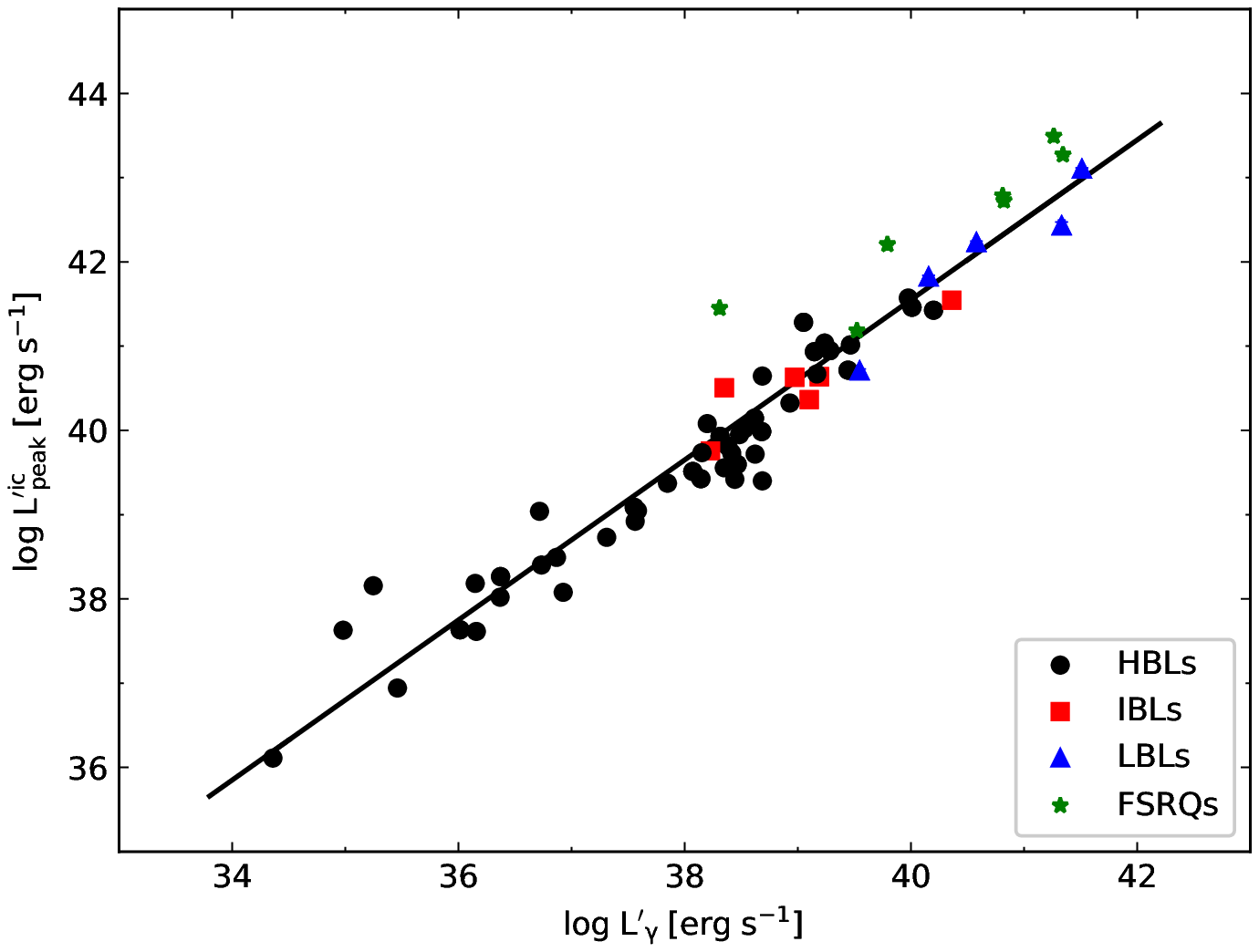}
  \caption{{\small The relation between intrinsic $\gamma$-ray luminosity and intrinsic ICs peak luminosity. }}
  \label{Fig8}
  \end{minipage}%
\end{figure}

We now consider the relation between intrinsic synchrotron peak frequency and intrinsic ICs peak frequency. The results for our sample are shown in Figure~\ref{Fig7} and described as follows: (1) Sources with high intrinsic synchrotron peak frequency have high intrinsic ICs peak frequency for TeV blazars. (2) According to the Spearman correlation test and linear regression analysis, there is a strong positive correlation between them and the empirical relation can be obtained by linear regression analysis. The results are listed in Table~\ref{tab:4}. (3) In the observer's coordinate system, we compare the relation $\frac{{\nu _{\rm peak}^{\rm ic}}}{{\nu _{\rm peak}^{\rm syn}}} \simeq \frac{4}{3}{\left( {\gamma _{\rm peak}^{\rm SSC}} \right)^2}\;\;,\gamma _{\rm peak}^{\rm SSC} \le 2 \times {10^4}$ obtained here with that of \cite{2010ApJ...716...30A} and find that the two are similar, the relative error being $6.2\%$. Considering that most of the sources in the sample are HBLs, this indicates that the radiation of TeV HBLs can well satisfy the SSC model.

\subsection{Intrinsic $\gamma$-Ray Luminosity versus Intrinsic ICs Peak Luminosity}
At present, the range of peak luminosities of the ICs energy spectrum can usually be determined by performing an energy spectrum fitting. In the observations, the $\gamma$-ray photons of blazar radiation have a higher energy, and we consider whether there is a certain intrinsic connection between the two luminosities. The relation between intrinsic $\gamma$-ray luminosity at an energy of 20GeV and ICs peak luminosity is shown in Figure~\ref{Fig8}. According to the results of the Spearman correlation test and linear regression analysis (Table~\ref{tab:4}), we draw the following conclusions: (1) There is a strong positive correlation between the luminosities, and their empirical relation can be obtained according to linear regression analysis, as shown in Table~\ref{tab:4}. The Spearman correlation coefficient is 0.95, and the slope is close to 1, indicating that the two luminosity values are very close to each other. (2) In the observer's coordinate system, we can get the following empirical formula:
\begin{equation}
\log L_{\rm peak}^{\rm ic} = \left( {1.01 \pm 0.04} \right)\log {L_\gamma } + \left( {0.08 \pm 1.54} \right)
\label{Eq:20}
\end{equation}
where the Spearman test yields a correlation coefficient  $\rho  = 0.96$ and the chance probability is $P < {10^{ - 4}}$. (3) The observed quantity ${L_\gamma }$, which can be obtained using Equation (\ref{Eq:18}), has a range of $4.37 \times {10^{42}}\;\rm erg\;{s^{ - 1}} \le {L_\gamma } \le 2.46 \times {10^{46}}\;erg\;{s^{ - 1}}$ in our sample; therefore, the ICs peak luminosity can be estimated within the range of $8.66 \times {10^{42}}\;\rm erg\;{s^{ - 1}} \le {L_{peak}^{ic} } \le 5.09 \times {10^{46}}\;erg\;{s^{ - 1}}$  according to Equation (\ref{Eq:20}).

\subsection{Intrinsic Peak Frequency versus the Spectral Index}
\begin{figure}[!htp]
  \begin{minipage}[t]{0.315\textwidth}
  \centering
   \includegraphics[width=60mm]{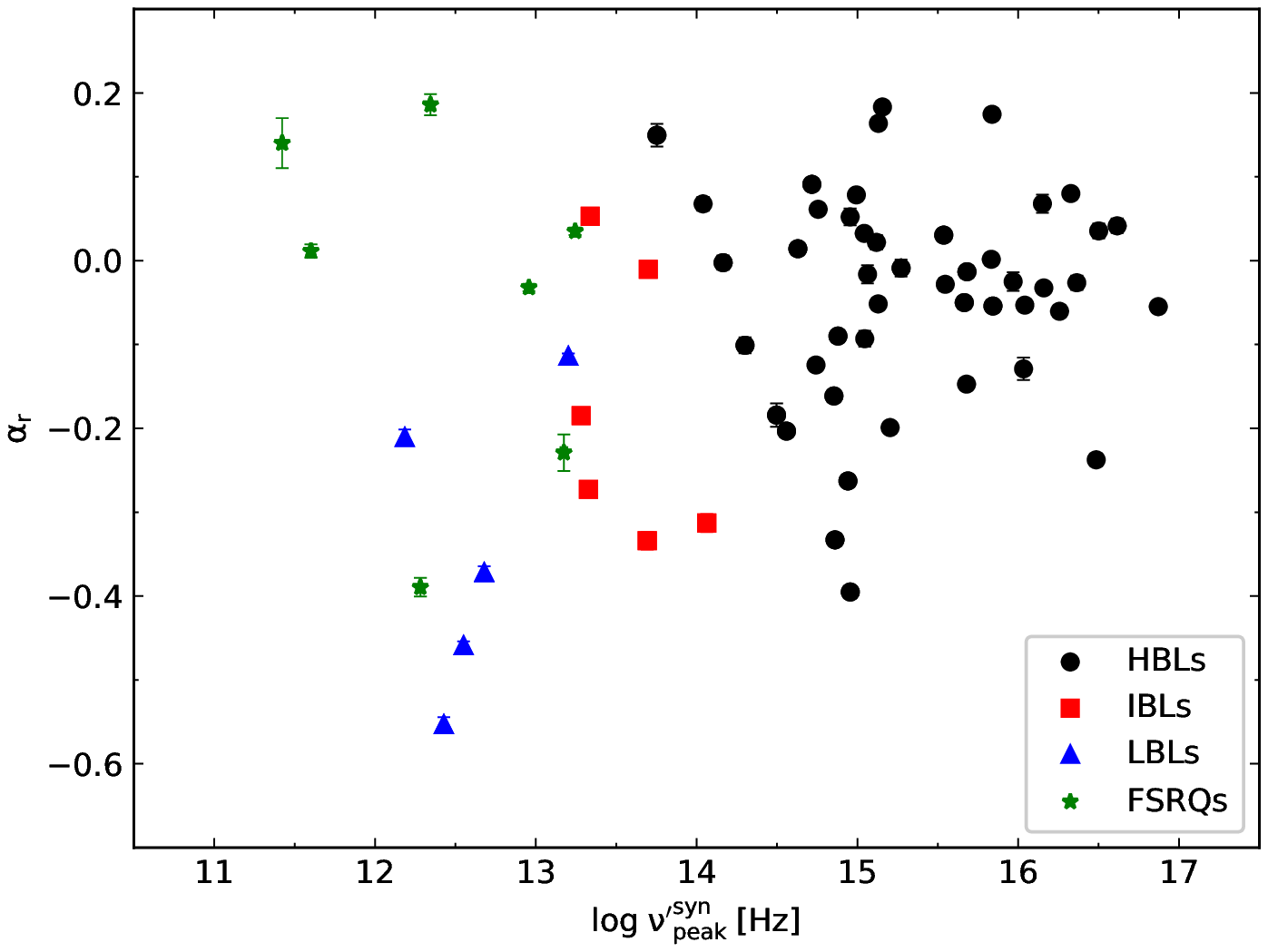}
  \end{minipage}%
  \begin{minipage}[t]{0.315\textwidth}
  \centering
   \includegraphics[width=60mm]{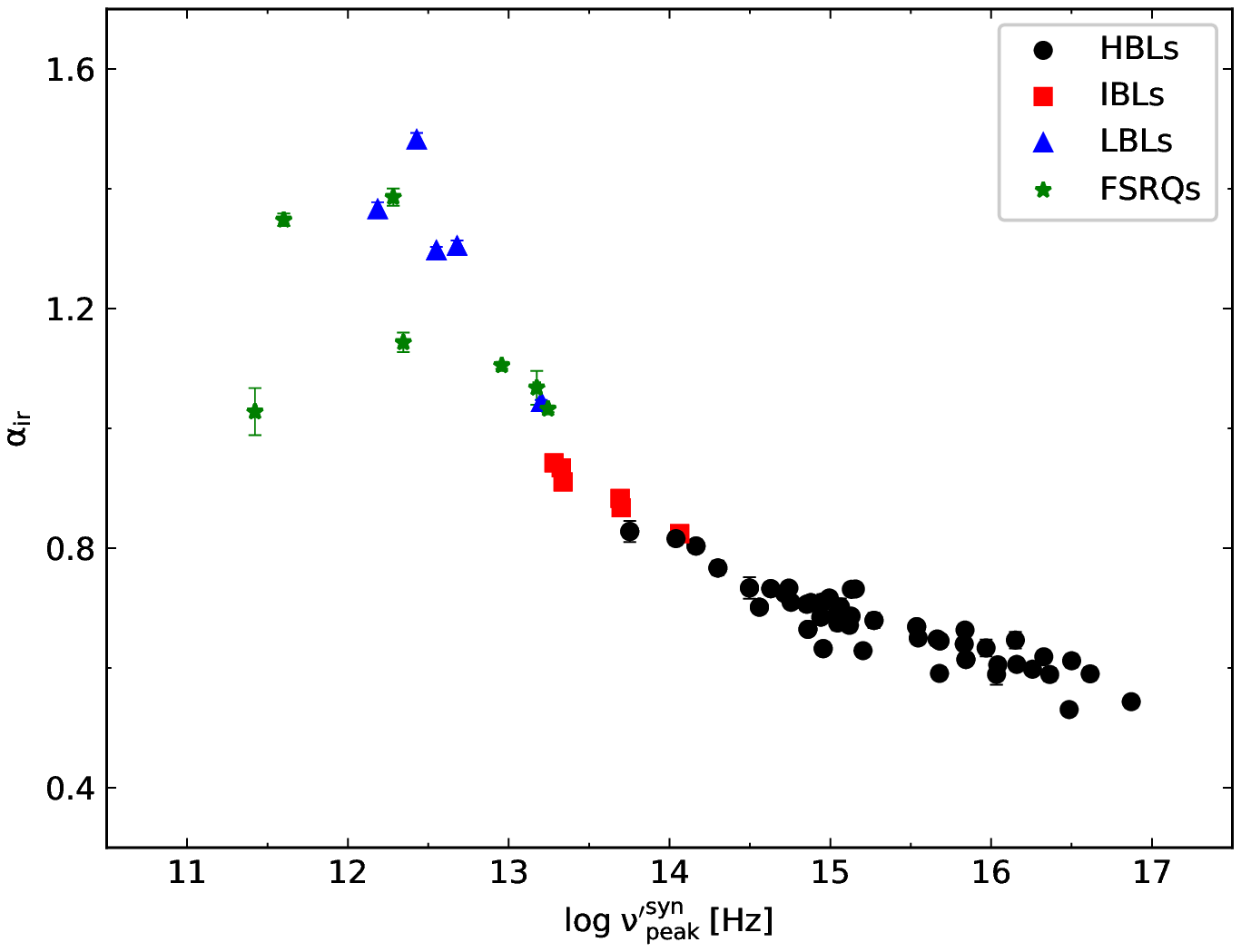}
  \end{minipage}%
  \begin{minipage}[t]{0.315\textwidth}
  \centering
   \includegraphics[width=60mm]{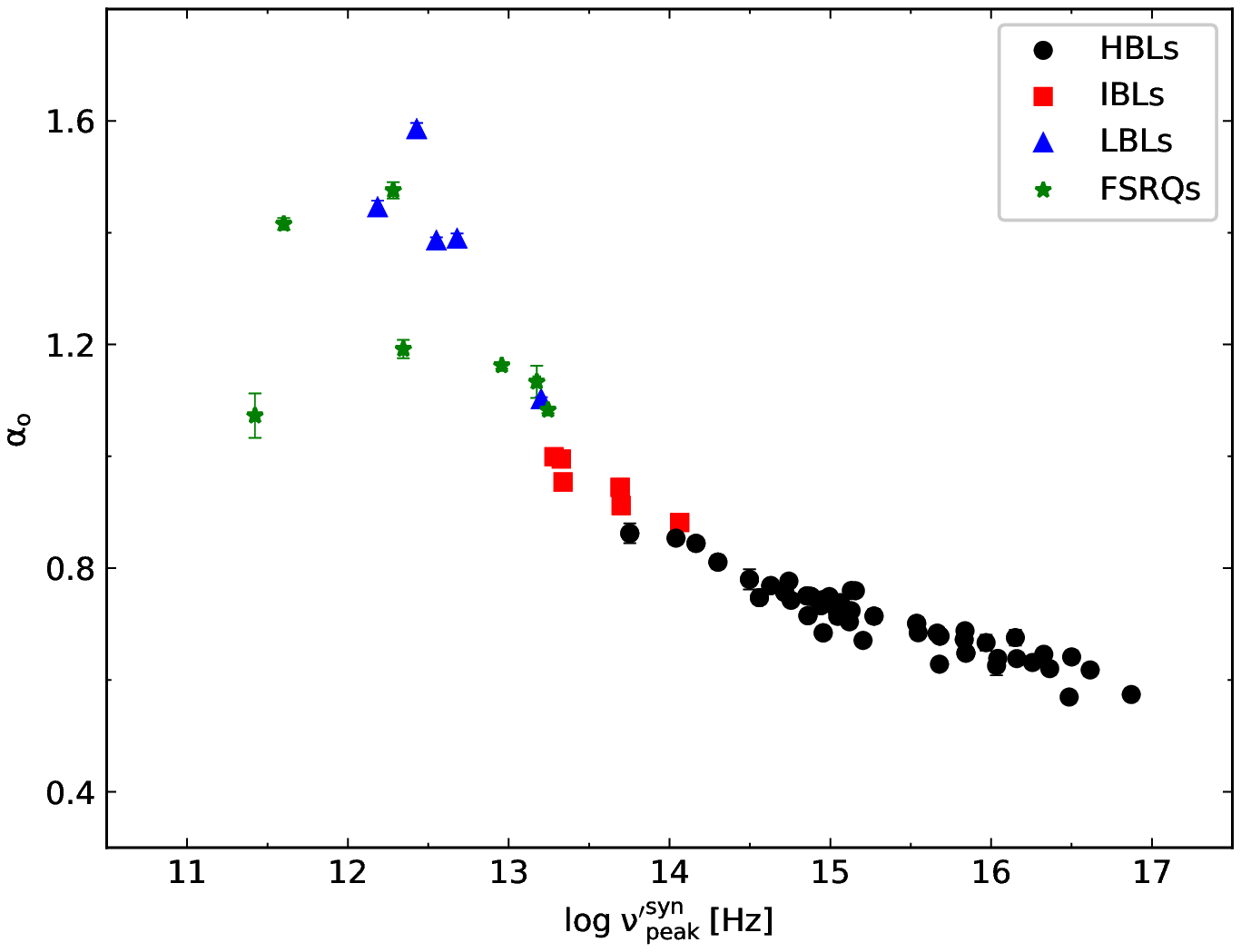}
  \end{minipage}%
  \\
  \begin{minipage}[t]{0.315\textwidth}
  \centering
   \includegraphics[width=60mm]{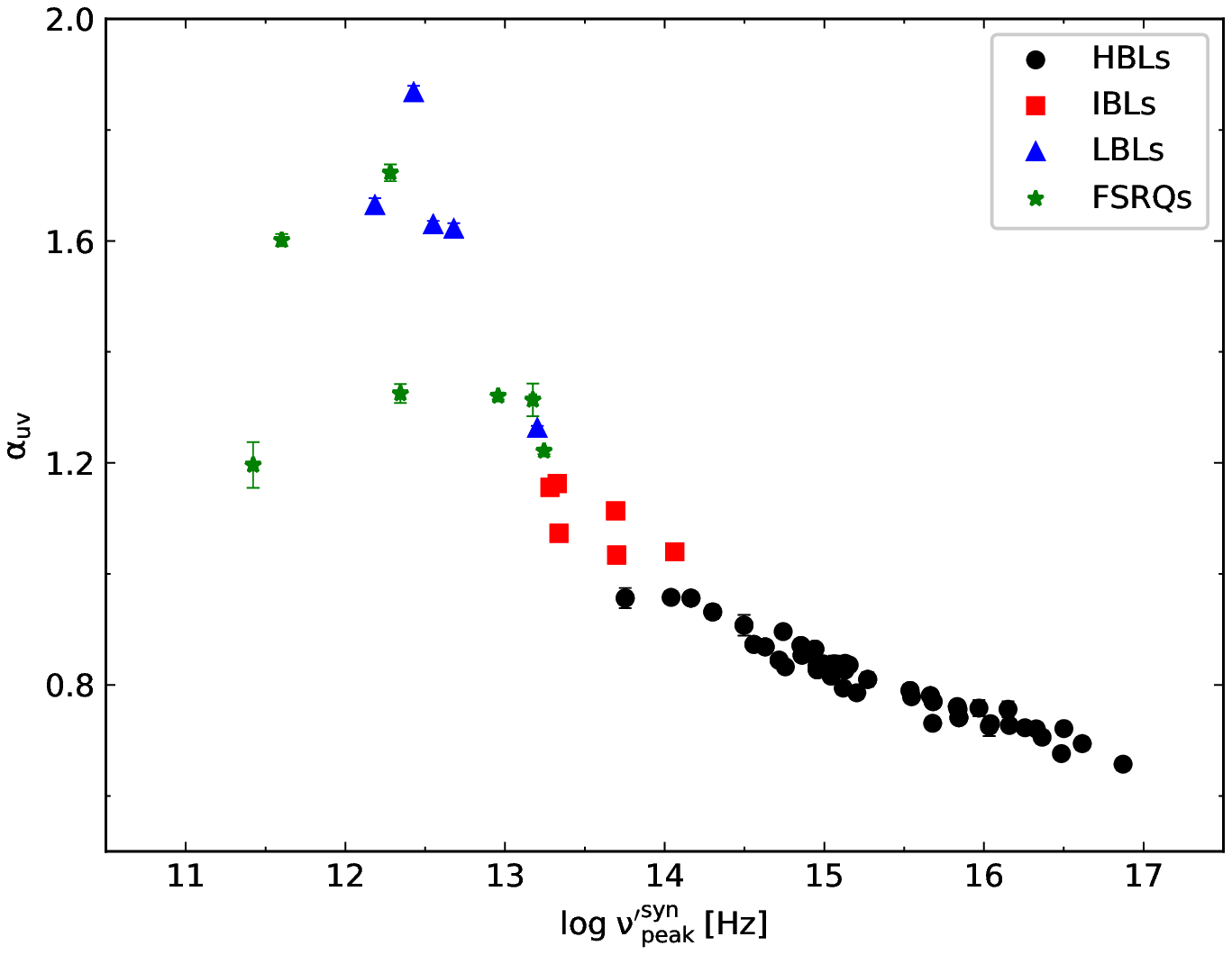}
  \end{minipage}%
  \begin{minipage}[t]{0.315\textwidth}
  \centering
   \includegraphics[width=60mm]{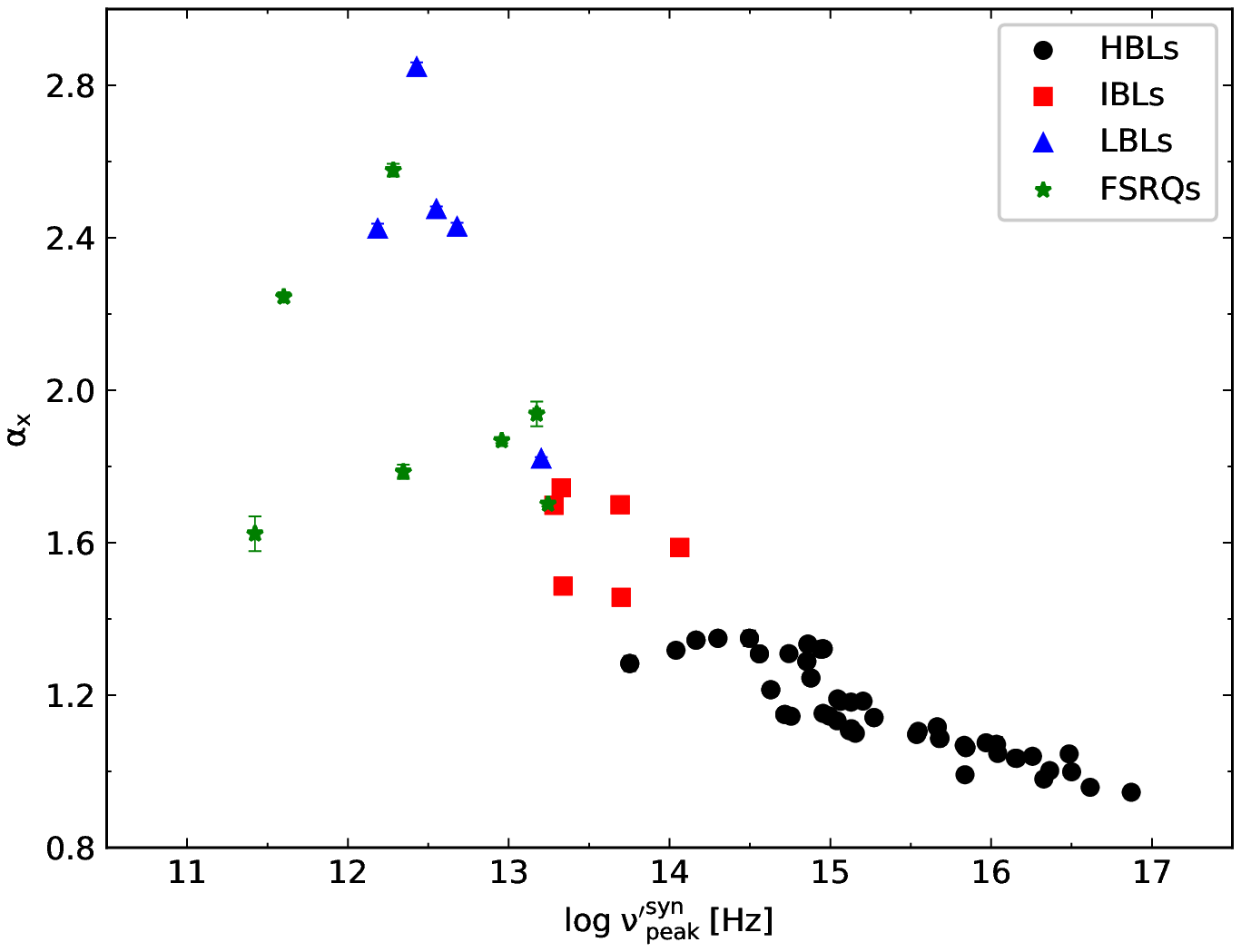}
  \end{minipage}%
  \begin{minipage}[t]{0.315\textwidth}
  \centering
   \includegraphics[width=60mm]{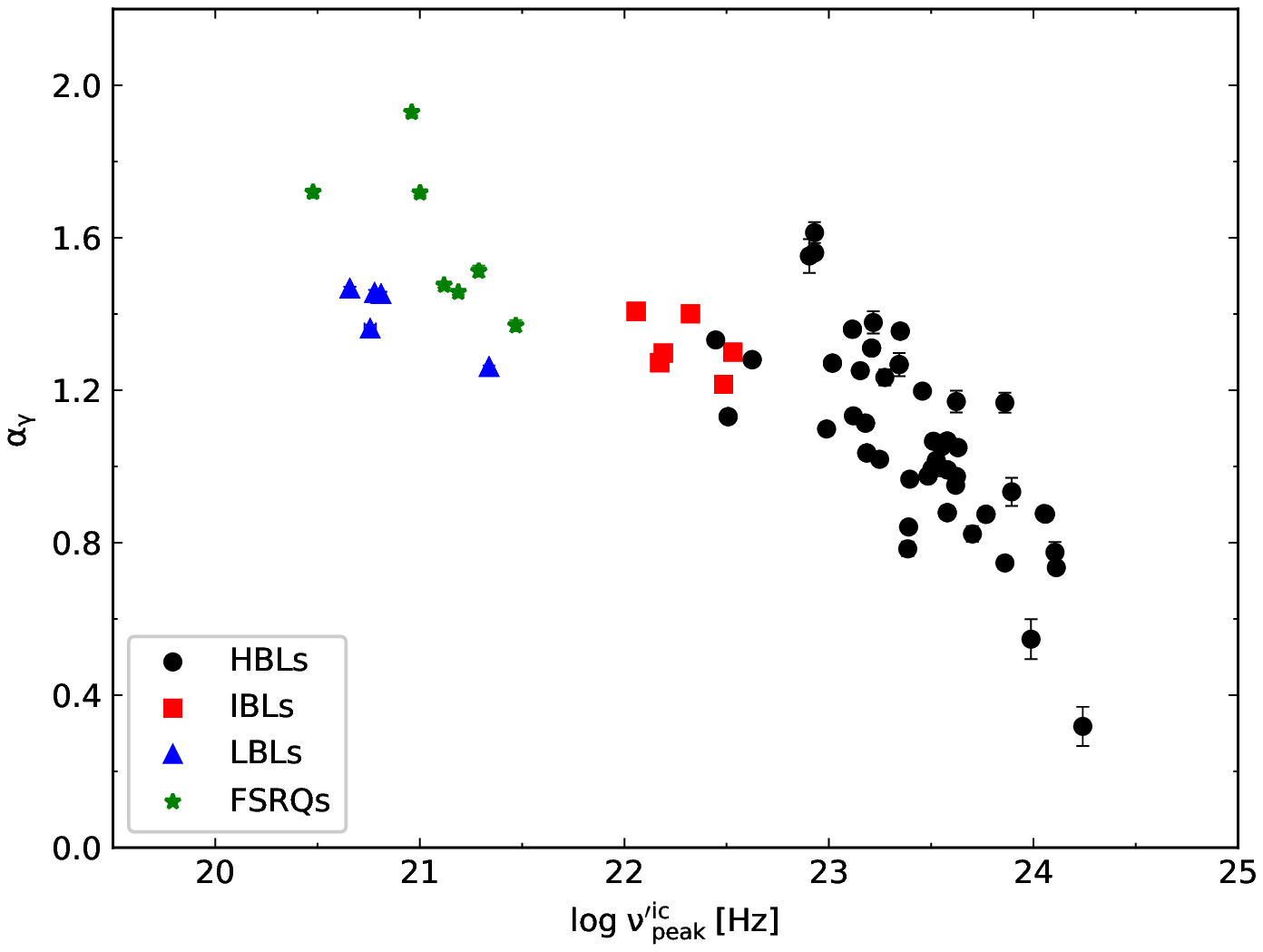}
  \end{minipage}%
  \setlength{\abovecaptionskip}{-0cm}
  \setlength{\belowcaptionskip}{-0cm}
  \caption{{\small The relation between intrinsic peak frequency and the spectral index.} }
  \label{Fig9}
\end{figure}

Figure~\ref{Fig9} shows the relation between intrinsic peak frequency and the spectral index of each band in different types of TeV blazars. Since the peak frequency depends strongly on the Doppler factor (\citealt{2008A&A...488..867N}), we use Doppler corrections and redshift corrections to study the intrinsic relation between peak frequency and the spectral index of each band. In order to clarify the correlation between ${\nu _{\rm peak}}$ and ${\alpha _\nu }$, the Spearman correlation test and a linear regression analysis have been conducted and the results are shown in Table~\ref{tab:4}. We get the following results: (1) There is not any correlation between the spectral indices and intrinsic synchrotron peak frequencies at radio bands. (2) The spectral indices depend strongly on the intrinsic synchrotron peak frequencies, i.e., the spectra of the source become hard when the intrinsic synchrotron peak frequencies increase in the infrared, optical, ultraviolet, and X-ray bands. (3) There is a strong negative correlation between intrinsic ICs peak frequency and the spectral index of $\gamma$-ray bands. It should be noted that their negative correlation is quite different from that of the synchrotron region, showing an inverse L-shaped reduction. (4) If radio bands are excluded, there is a negative correlation between the intrinsic peak frequency and the spectral index of each band. This indicates that the radio radiation generated by synchrotron radiation in the jet is optically thick, and the radio photons are emitted only when the depth of light decreases downstream of the jet. The infrared, optical, ultraviolet, and X-ray bands are all optically thin. This shows that the radiating region radius of the radio band is different from that of the other bands. There is radiation loss in each band, but the loss mechanism in the synchrotron radiation region is obviously different from that in the ICs region.

\subsection{Intrinsic Peak Frequency versus Curvature}
\begin{figure}[!htp]
  \begin{minipage}[t]{0.495\linewidth}
  \centering
   \includegraphics[width=80mm]{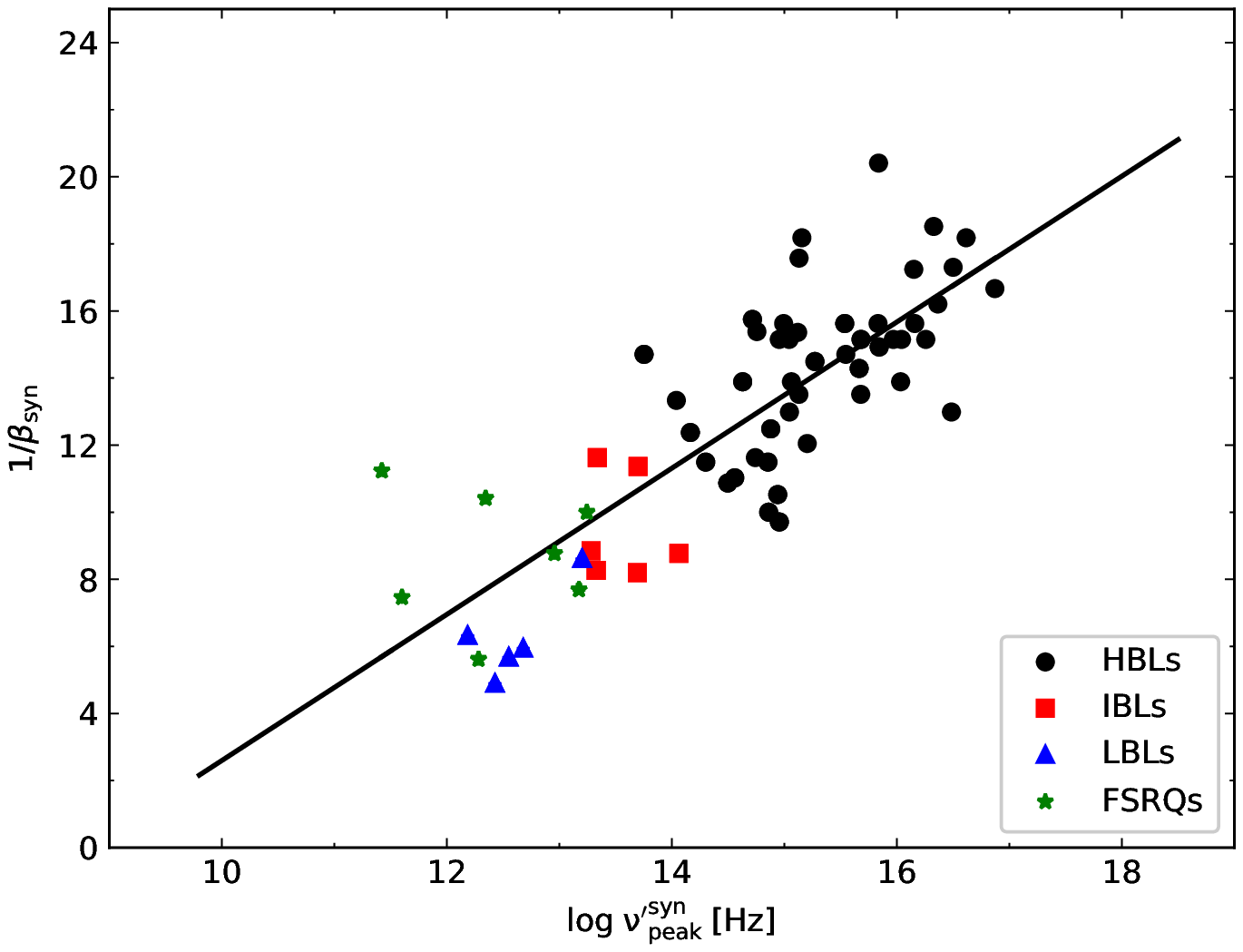}
   \caption{{\small The relation between the reciprocal of the curvature of the synchrotron bump at the peak and the intrinsic synchrotron peak frequency.} }
   \label{Fig10}
  \end{minipage}%
  \begin{minipage}[t]{0.495\textwidth}
  \centering
   \includegraphics[width=80mm]{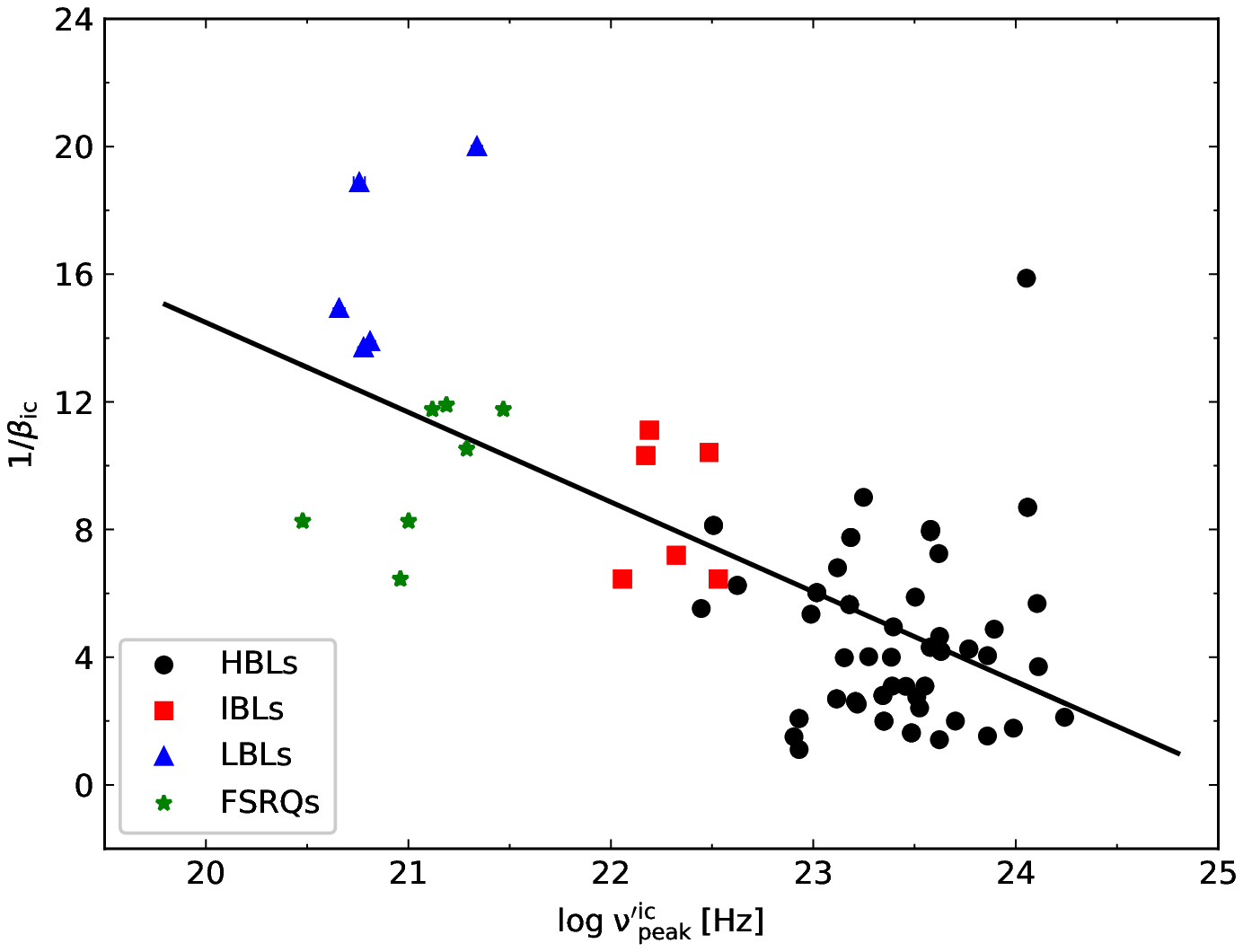}
  \caption{{\small The relation between the reciprocal of the curvature of the ICs bump at the peak and the intrinsic ICs peak frequency.}}
  \label{Fig11}
  \end{minipage}%
\end{figure}

As an important parameter of the energy spectrum, the curvatures of the synchrotron bump and ICs bump are essential for studying the particle acceleration mechanism. Equation (\ref{Eq:10}) is used to define the energy spectrum curvature parameter $\beta$ at the peak. In the process of stochastic acceleration, the decrease of curvature leads to the shift of peak energy toward higher energies. \cite{2014ApJ...788..179C} studied the curvature of the blazars in depth and gave three possible explanations for the acceleration mechanism according to the relation between the synchrotron peak frequency and the curvature of the synchrotron bump. From the result of our fit, we are able to investigate possible relations between the intrinsic peak frequency and the curvature of the energy spectrum at the peak. Based on the Spearman correlation test and linear regression analysis in Table~\ref{tab:4}, the following results can be obtained: (1) As shown in Figure~\ref{Fig10}, there is a strong positive correlation between the reciprocal of the curvature of the synchrotron bump at the peak and the intrinsic synchrotron peak frequency. We obtain an empirical formula between them through linear regression analysis. This is consistent with the result of \cite{2011ApJ...739...66T}. (2) In the observer's coordinate system, the relation between the synchrotron peak frequency and the reciprocal of the curvature of the synchrotron bump is obtained as shown in the following empirical formula:
\begin{equation}
\frac{1}{{{\beta _{\rm syn}}}} = \left( {2.31 \pm 0.14} \right)\log \nu _{\rm peak}^{\rm syn} + \left( { - 23.65 \pm 2.27} \right)
\label{Eq:21}
\end{equation}
where the Spearman test yields a correlation coefficient $\rho  = 0.88$ and the chance probability is $P < {10^{ - 4}}$. Comparing this with the result of \cite{2014ApJ...788..179C}, it can be seen that the slope value of Equation (\ref{Eq:21}) there corresponds to three different values, $5/2$, $10/3$, and $2$, which relate, respectively, to three different mechanisms, i.e., the energy-dependent acceleration probability mechanism, the fluctuation of fractional acceleration gain mechanism, and the stochastic acceleration mechanism(\citealt{2004A&A...413..489M,2014ApJ...788..179C}).
The particle acceleration mechanism of our sample is closer to the energy-dependent acceleration probability and to the stochastic acceleration mechanism. This shows that there is obvious energy gain diffusion in random acceleration, which leads to momentum diffusion.
(3) Unlike the former, the reciprocal of the curvature of the ICs bump at the peak and the intrinsic ICs peak frequency have a weak negative correlation, as shown in Figure~\ref{Fig11}, and the empirical formula between them is obtained through linear regression analysis, as shown in Table~\ref{tab:4}. This indicates that the variation of ICs curvature is more complex, which is consistent with our results when the ICs process is in the strong KN scattering region. However, when the ICs process is in the Thomson scattering region, ${\beta _{\rm ic,Th}} \approx 0.5{\beta _{\rm syn}}$ (\citealt{2011ApJ...736..128P,2009A&A...504..821P,2011ApJ...739...66T}).
In the KN scattering region, $1/{\beta _{{\rm{ic}}}}$ and $\log \nu _{{\rm{peak}}}^{{\rm{'ic}}}$ have a large dispersion, i.e., a weak correlation, due to the KN suppression. We believe that FSRQs and LBLs are more likely to radiate in the Thomson scattering region, while HBLs radiate in the KN scattering region. Moreover, the seed photon sources of FSRQs and most LBLs are complex, and the contribution of the external Compton (EC) process will make the results more complicated.

\subsection{Intrinsic ICs Peak Frequency versus Peak Lorentz Factor}
\begin{figure}[!htp]
  \vspace{-0.4 cm}  
  \setlength{\belowcaptionskip}{-0.5 cm}   
  \begin{minipage}[t]{0.495\linewidth}
  \centering
   \includegraphics[width=80mm]{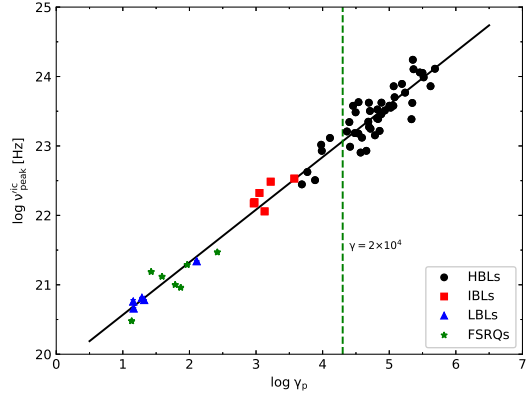}
   \caption{{\small The relation between intrinsic ICs peak frequency and peak Lorentz factor. }}
   \label{Fig12}
  \end{minipage}%
  \centering
\end{figure}

Figure~\ref{Fig12} shows the relation between the peak frequency of the intrinsic ICs spectrum ($\log \nu _{\rm peak}^{\rm ic}$) and the peak Lorentz factor ($\log {\gamma _{\rm p}}$) of the nonthermal electrons. The Spearman correlation test and linear regression analysis show the following conclusions: (1) There is a strong positive linear correlation between $\log \nu _{\rm peak}^{\rm ic}$ and $\log {\gamma _{\rm p}}$. According to \cite{2010ApJ...716...30A}, the Thomson scattering region and the KN scattering region transform in the region $\gamma {\rm{ = }}2 \times {10^4}$. It is found that most of the sources in the sample are in the region of $\gamma  > 2 \times {10^4}$, which indicates that the TeV HBLs are in the KN scattering region during the ICs process.
In the observer's coordinate system, $\log \nu _{\rm peak}^{\rm ic}$ and $\log {\gamma _{\rm p}}$ have satisfied the following relation:
\begin{equation}
\log \nu _{\rm peak}^{\rm ic} = \left( {0.87 \pm 0.02} \right)\log {\gamma _{\rm p}} + \left( {20.45 \pm 0.09} \right)\\
\label{Eq:22}
\end{equation}
where the Spearman test yields a correlation coefficient $\rho  = 0.95$ and the chance probability is $P < {10^{ - 4}}$. This result is consistent with the result of \cite{2006A&A...448..861M}, indicating that when the peak frequency of the ICs spectrum increases with the increase of the peak Lorentz factor, the KN suppression becomes more efficient.

\subsection{Magnetic Field Strength versus Peak Lorentz Factor}
\begin{figure}[!htp]
  \vspace{-0.4 cm}  
  \setlength{\belowcaptionskip}{-0.5 cm}   
  \begin{minipage}[t]{0.495\linewidth}
  \centering
   \includegraphics[width=80mm]{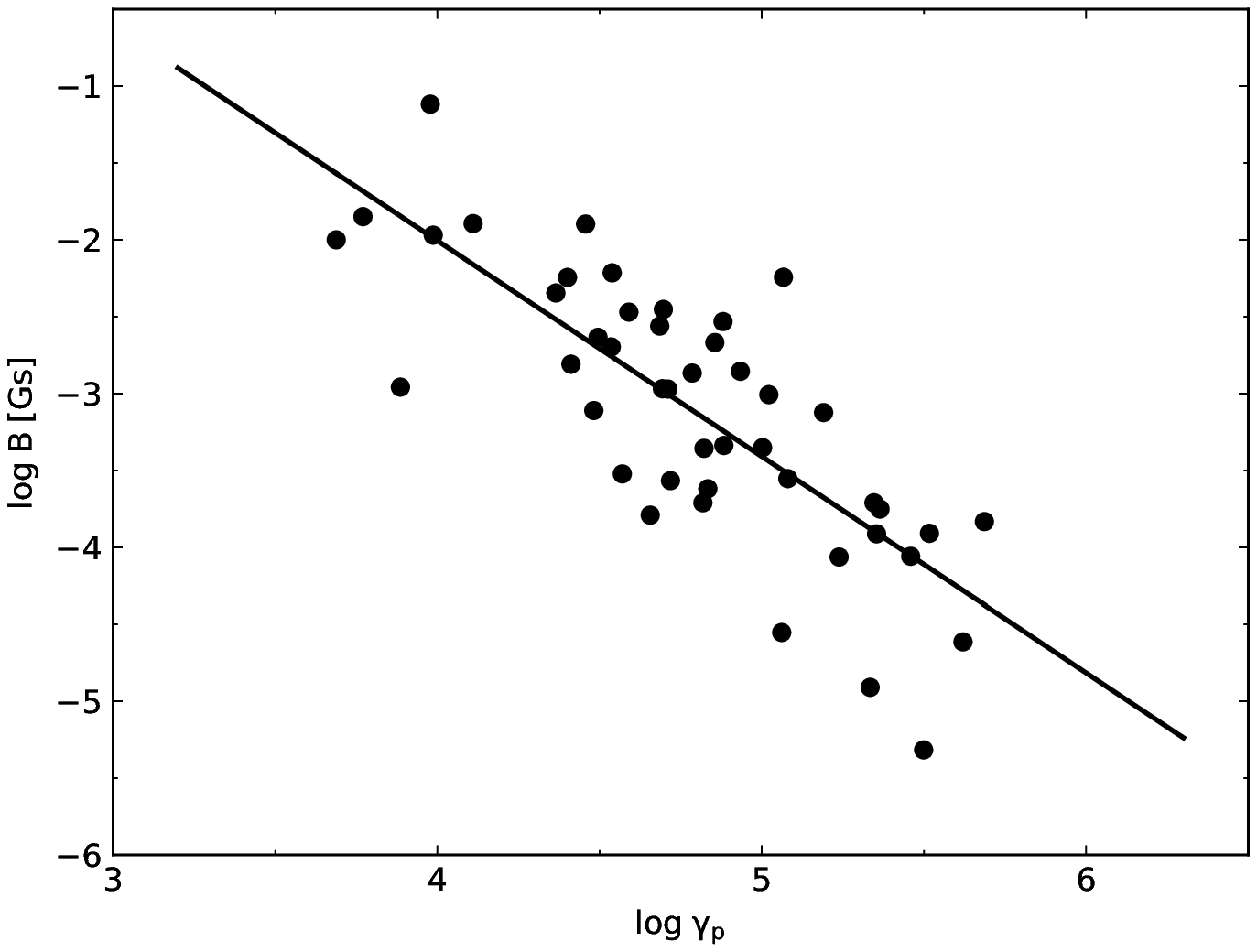}
   \caption{{\small The relation between magnetic field strength and peak Lorentz factor for TeV HBLs. }}
   \label{Fig13}
  \end{minipage}%
  \begin{minipage}[t]{0.495\textwidth}
  \centering
   \includegraphics[width=80mm]{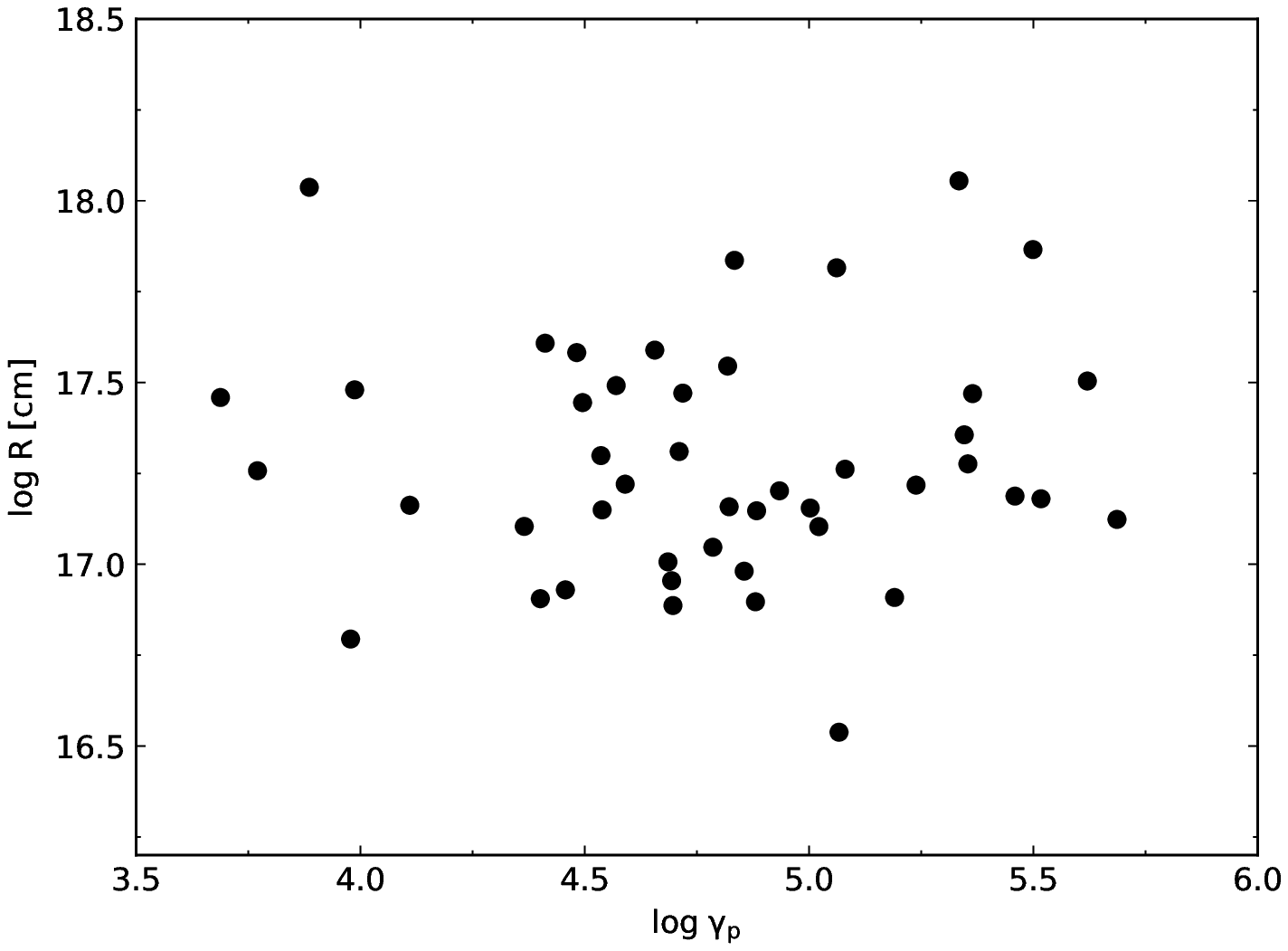}
  \caption{{\small The relation between radiative zone radius and peak Lorentz factor for TeV HBLs.}}
  \label{Fig14}
  \end{minipage}%
\end{figure}

As shown in Figure~\ref{Fig13}, we selected the TeV HBLs in the sample to conduct a correlation test and linear regression analysis between the ${\rm log} B$ and $\log {\gamma _{\rm p}}$ of nonthermal electrons (see Table~\ref{tab:4}). We can draw the following conclusions: There is a strong linear negative correlation between ${\rm log} B$ and $\log {\gamma _{\rm p}}$, and the Spearman correlation coefficient is $-0.79$. We can obtain the following relation:
\begin{equation}
B = {10^{3.61 \pm 0.81}}\gamma _{\rm{P}}^{ - 1.40 \pm 0.81}
\label{Eq:23}
\end{equation}
This shows that the greater the peak Lorentz factor of nonthermal electrons of TeV HBLs, the lower the intensity of the jet magnetic field.
According to Equation (\ref{Eq:A16}) and Equation (\ref{Eq:A17}), there is a relation between the jet magnetic field strength ($\rm B$) and the peak Lorentz factor of nonthermal electrons ($ {\gamma _{\rm p}}$) in the KN scattering region:
$B = K\left( {\frac{{\nu _{{\rm{peak}}}^{{\rm{syn}}}}}{{\nu _{{\rm{peak}}}^{{\rm{ic}}}}}} \right)\gamma _{\rm{P}}^{ - 1}$
where $K = 14.09 \times {10^{ - 3/2r}}$ ($r \approx 0.36$ is constant). Compared with Equation (\ref{Eq:23}), the relative error of the exponential part of $\gamma _{\rm{p}}$ is $4\% $. This suggests that the value of $\nu _{\rm peak}^{\rm syn}/\nu _{\rm peak}^{\rm ic}$ will somehow affect the relation between ${\rm B}$ and ${\gamma _{\rm p}}$.

\subsection{Radiative Zone Radius versus Peak Lorentz Factor}
As shown in Figure~\ref{Fig14}, we selected the TeV HBLs in the sample to conduct a correlation test and linear regression analysis between the radiating region radius ($\rm{ log} R$) and the peak Lorentz factor ($\log {\gamma _{\rm p}}$) of nonthermal electrons (see Table~\ref{tab:4}). The following conclusions were drawn: (1) There is no correlation between $\rm{ log} R$ and $\log {\gamma _{\rm p}}$, the Spearman correlation coefficient is $0.03$, and the chance probability $P>0.05$.
(2) According to Equation (\ref{Eq:A16}) and Equation (\ref{Eq:A17}), there exists the following relation between the radius of the radiation region ($R$) and the peak Lorentz factor (${\gamma _{\rm p}}$) of the nonthermal electrons:
\begin{equation}
\left\{ \begin{array}{l}
R = {K_1}\left( {\frac{{L_{{\rm{peak}}}^{{\rm{ic}}}}}{{L_{{\rm{peak}}}^{{\rm{syn}}}}}} \right) \cdot {n^{ - 1}} \cdot \gamma _{\rm{p}}^{ - 2}{\rm{ }}\;\;\;\;\;(\rm{Thomson\;regime})\\
R = {K_2}\left( {\frac{{L_{{\rm{peak}}}^{{\rm{ic}}}}}{{L_{{\rm{peak}}}^{{\rm{syn}}}}}} \right) \cdot {n^{ - 1}} \cdot \gamma _{\rm{P}}^2 \cdot B{\rm{ }}\;\;(\rm{KN\;regime})
\end{array} \right.
\label{Eq:24}
\end{equation}
where ${K_1} = 9.10 \times {10^{24}}$, and ${K_2} = 5.98 \times {10^{10}}$.
In the Thomson scattering region, there is a linear relation between $\rm{ log} R$ and $\log {\gamma _{\rm p}}$. However, in the KN scattering region, $\rm R$ is not only related to ${\gamma _{\rm p}}$, but also related to the magnetic field strength $\rm B$; therefore $\rm R$ and $\rm B$ have an intrinsic degeneracy, and it is impossible to estimate the change of $\log {\gamma _{\rm p}}$ according only to $\rm{ log} R$.\\

\vspace{2 ex}
\section{Conclusions} \label{sect:discusion}
In this paper, the average-state multiwavelength energy spectrum data of 65 Fermi TeV blazars were collected and the SED was fitted with a log-parabolic model to obtain the relevant physical parameters. We used the log-parabolic model to adjust the parameters repeatedly so that the reduced chi-square values $\chi _{\rm syn}^{\rm 2}$ and $\chi _{\rm ic}^{\rm 2}$ were minimized in the coordinate space of $\log \nu  - \log \nu {f_\nu }$, and we obtained the average reduced chi-square values of $\left\langle {\chi _{{\rm{syn}}}^{\rm{2}}} \right\rangle  = 0.09$ and $\left\langle {\chi _{{\rm{ic}}}^{\rm{2}}} \right\rangle = 0.40$. In order to further study the intrinsic parametric characteristics and to estimate the physical parameters of the jet radiation zone, we selected a samples of 64 classified TeV blazars with redshift. Near the SED peak frequency, the synchrotron region and the ICs region show a log-parabolic shape, and we can calculate the parameters at the peak frequency. We have a total of 17 parameters, including 12 model fitting parameters and 5 indirect calculation parameters, which are listed in Table~\ref{tab:1} and ~\ref{tab:2}.  We calculated the errors for each model fitting parameter, and the fitting errors were all within the range of $1\sigma$. We conducted statistical analysis of these parameters and arrived at the following conclusions:

(1) TeV HBLs can better meet the conditions of the KN scattering region and their parameters can be reasonably estimated by Equation (\ref{Eq:17}), as follows: the Doppler factor, $5.6 \le {D} \le 31.6$; the magnetic field strength, $6.3 \times {10^{ - 5}}\;\rm G \le B \le 1.1 \times {10^{ - 2}}\;\rm G$; the radius of the radiative zone, $7.5 \times {10^{16}}\;\rm cm \le R \le 4.2 \times {10^{17}}\;\rm cm$; and the peak Lorentz factor, $2.5 \times {10^4} \le {\gamma _{\rm p}} \le 4.0 \times {10^5}$.

(2) The radiation of TeV HBLs satisfies the SSC model well and there is a strong linear positive correlation between the intrinsic synchrotron peak frequency and intrinsic ICs peak frequency. According to the definition of the Compton dominance parameter ${A_{\rm C}} = L_{\rm peak}^{\rm IC}/L_{\rm peak}^{\rm syn}$ by \cite{2013ApJ...763..134F}, we can estimate the mean value of ${A_{\rm C}}$ of different types of sources in the sample as ${\left\langle {{A_{\rm C}}} \right\rangle _{\rm HBLs}} = 0.94$, ${\left\langle {{A_{\rm C}}} \right\rangle _{\rm IBLs}} = 1.30$, ${\left\langle {{A_{\rm C}}} \right\rangle _{\rm LBLs}} = 1.19$, and ${\left\langle {{A_{\rm C}}} \right\rangle _{\rm FSRQs}} = 9.41$. According to \cite{2010MNRAS.402..497G}, ${A_{\rm C}} = L_{\rm peak}^{\rm IC}/L_{\rm peak}^{\rm syn} = {U_{\rm rad}}/{U_{\rm B}}$, where ${U_{\rm B}}$ and ${U_{\rm rad}}$ are the energy density of the magnetic field and the radiation field, respectively. Since ${\left\langle {{A_{\rm C}}} \right\rangle _{\rm HBLs}}$ is very small and ${\left\langle {{A_{\rm C}}} \right\rangle _{\rm FSRQs}} \approx 10$, the radiation energy of HBLs is in good agreement with the SSC model, while that of FSRQs is in good agreement with the EC radiation model. When the peak frequency of the ICs spectrum increases with the increase of the peak Lorentz factor, the suppression of the KN effect becomes more efficient.

(3) There is a strong linear positive correlation between the intrinsic ICs peak luminosity and intrinsic $\gamma$-ray luminosity and between the intrinsic ICs peak frequency and peak Lorentz factor. We can estimate the parameter values of ${\rm{ }}\log {\rm{ }}{L }_{\rm peak}^{\rm ic}$ and $\log {\rm{ }}{\nu  }_{\rm peak}^{\rm ic}{\rm{ }}$ in terms of $\log {\rm{ }}L_\gamma  {\rm{ }}$ and $\log {\rm{ }}{\gamma _{\rm p}}$ by using Equation (\ref{Eq:20}) and (\ref{Eq:22}).

(4) If radio bands are excluded, the negative correlation between the intrinsic peak frequency and the spectral index of each band indicates that the radiation in the radio band is optically thick, while the radiation in the infrared, optical, ultraviolet, and X-ray bands is optically thin. This indicates that the radiating region radius of the radio band is different from that of the other bands. There is radiation loss in each band, but the loss mechanism in the synchrotron region is obviously different from that in the ICs region. The spectra of the source become hard when the intrinsic synchrotron peak frequencies increase in the synchrotron region. The $\gamma$-ray spectrum index shows an inverse L-shaped decrease with the peak frequency in the ICs region.

(5) There is a strong positive correlation between the reciprocal of the curvature of the synchrotron bump at the peak ($1/{\beta _{\rm syn}}$) and the peak frequency of intrinsic synchrotron radiation ($\log \nu _{\rm peak}^{'\rm syn}$). According to Equation (\ref{Eq:21}), we can use the energy-dependent acceleration probability and stochastic acceleration mechanism to explain this correlation, i.e., there is obvious energy gain diffusion in stochastic acceleration, which leads to momentum diffusion. There is a weak negative correlation between the reciprocal of the curvature of the ICs bump at the peak (1/$\beta_{\rm ic}$) at the peak and the ICs peak frequency ($\log \nu _{\rm peak}^{'\rm ic}$). We consider that ICs of FSRQs and LBLs is more likely to occur in the Thomson scattering region, while that of HBLs is more likely to occur in the KN scattering region. At the same time, the contribution of EC scattering may influence the results to some extent. In addition, if the combination of acceleration and cooling effect is considered, the interpretation of ICs curvature will be more complicated.

(6) The larger the peak Lorentz factor of nonthermal electrons in TeV HBLs, the lower the intensity of the jet magnetic field. The value of $\nu _{\rm peak}^{\rm syn}/\nu _{\rm peak}^{\rm ic}$ has some effect on the relationship between the two parameters.

(7) In the KN scattering region, $\rm R$ is not only related to ${\gamma _{\rm p}}$, but also related to the magnetic field strength $\rm B$; therefore $\rm R$ and $\rm B$ have an intrinsic degeneracy, and it is impossible to estimate the change of $\log {\gamma _{\rm p}}$ according only to $\rm{ log} R$.

Due to the limited number of Fermi TeV blazars, and to the fact that the energy spectrum data are long-interval average SEDs and most of them are from HBLs, our analysis conclusions may be more suitable to objects of the HBL type. Therefore, the influence of EC scattering has not been fully considered. According to the formulas in \cite{2018ApJS..235...39C} and \cite{2011ApJ...735..108C}, we can get the peak frequency and peak luminosity of the EC process as follows: $\nu _{\rm peak}^{'\rm ec} \approx (4/3){\nu _{\rm ext}}\gamma _{\rm p,Th}^2\Gamma $ and $L_{\rm peak,Th}^{'\rm ec} \approx (17{e^2}/(6\pi m_e^2{c^2}))({u_{\rm ext}}/\nu _{\rm ext}^2){(\nu _{\rm peak,Th}^{\rm ec}/\nu _{\rm peak}^{\rm syn})^2}L_{\rm peak}^{'\rm syn}$, where $\Gamma $ is the bulk Lorentz factor, we take $\Gamma  \approx D_{\rm Th}$, and ${\gamma _{\rm p,Th}}$ is the peak Lorentz factor in the Thomson scattering regime. We substitute in Equation (\ref{Eq:13}) and Equation (\ref{Eq:16}) to derive the following formula:
\begin{equation}
\begin{array}{l}
B = 212.62{\left( {\frac{{\nu _{{\rm{peak}}}^{{\rm{ec}}}}}{{{{10}^{22}}{\rm{Hz}}}}} \right)^{ - \frac{{1 + 2\xi }}{{2(1 + \xi )}}}}{\left( {\frac{{{{10}^{15}}{\rm{Hz}}}}{{\nu _{{\rm{peak}}}^{{\rm{syn}}}}}} \right)^{ - \frac{{1 + 2\xi }}{{2(1 + \xi )}}}}{\left( {\frac{{L_{{\rm{peak}}}^{{\rm{syn}}}}}{{{{10}^{45}}{\rm{erg}}\;{{\rm{s}}^{{\rm{ - 1}}}}}}} \right)^{\frac{1}{{2(1 + \xi )}}}}{\left( {\frac{{{{10}^{45}}{\rm{erg}}\;{{\rm{s}}^{ - 1}}}}{{L_{{\rm{peak}}}^{{\rm{ssc}}}}}} \right)^{\frac{1}{{4(1 + \xi )}}}}\\
{\rm{       }}{\left( {\frac{{\Delta t}}{{1{\rm{ day}}}}} \right)^{ - \frac{1}{{2(1 + \xi )}}}}{\left( {\frac{{{\nu _{{\rm{ext}}}}}}{{2 \times {{10}^{15}}{\rm{Hz}}}}} \right)^{\frac{{1 + 2\xi }}{{2(1 + \xi )}}}}{10^{\frac{{1 - 2\xi }}{{10{\beta _{{\rm{syn}}}}(1 + \xi )}}}}\beta _{{\rm{syn}}}^{ - \frac{1}{{8(1 + \xi )}}}{\left( {1 + z} \right)^{\frac{{\xi  - 1}}{{\xi  + 1}}}}\;{\rm{  Gs}}
\end{array}
\label{Eq:25}
\end{equation}
where ${\nu _{\rm ext}}$ is the modified dominant frequency of external radiation in the jet. If we consider that the broad-line region (BLR) provides the external photon field, we can take ${\nu _{\rm ext - BLR}} \approx 2 \times {10^{15}}{\rm Hz}$. The high-energy observation data points of average-state multiwavelength SED collected in our sample are few, so the position of the EC peak frequency $\nu _{\rm peak}^{\rm ec}$ cannot be given. According to Equation (\ref{Eq:25}), we estimate the physical parameters of LBLs and FSRQs in the case of EC scattering, as shown in Table~\ref{tab:5}.{
\renewcommand\thetable{5}
\begin{table*}[!htp]
\centering
\normalsize
\begin{center}
\caption{Physical parameters of LBLs and FSRQs in EC case}
\label{tab:5}
\setcounter{table}{5}
\renewcommand{\thetable}{5/arabic{table}}
\renewcommand\arraystretch{0.80}
\begin{adjustwidth}{4.2cm}{1cm}
\scalebox{0.8}{
\begin{tabular}{ccccccc} 
\hline				
$\rm TeV\;Source\;Name$&{$\rm{Class}$} & {$\log {\rm{ }}\nu _{{\rm{peak}}}^{{\rm{ec}}}$} & {$\log B$} & {$\log {\gamma _{\rm p}}$} \\
\normalsize(1) & \normalsize(2) & \normalsize(3) &\normalsize(4) & \normalsize(5) \\
\hline
\centering
$	\rm	OJ\;287 $	&	LBLs	&	23.02	&	0.30	&	3.13	\\
$	\rm	S4\;0954+65 $ &	LBLs	&	23.19	&	-0.03	&	3.21	\\
$	\rm	AP\;Lib  $	&	LBLs	&	21.61	&	0.92	&	2.77	\\
$	\rm	OT\;081 $	&	LBLs	&	22.61	&	0.16	&	3.06	\\
$	\rm	BL\;Lacertae $	&	LBLs	&	23.35	&	-0.21   &	3.35	\\
$	\rm	S3\;0218+35 $	&	FSRQs	&	22.39	&	1.49    &	2.42	\\
$	\rm	PKS\;0736+017 $	&	FSRQs	&	20.70   &	1.41    &	2.43    \\
$	\rm	TON\;0599  $	&	FSRQs	&	23.74       &	-0.25	&	3.41 	\\
$	\rm	4C+21.35 $	&	FSRQs	&	22.02	&	1.07	&	2.88	\\
$	\rm	3C\;279 $	&	FSRQs	&	23.20	&	-0.20	&	3.01    \\
$	\rm	PKS\;1441+25 $	&	FSRQs	&	22.27	&	0.07  &	2.94	\\
$	\rm	PKS\;1510-089 $	&	FSRQs	&	21.58	&	0.67	&	2.72 \\
\hline
\end{tabular}}\\
\end{adjustwidth}
\end{center}
{\footnotesize{Note. Column (1) gives the TeV source name. Column (2) gives the class designation for the source. Column (3)-(5) give the peak frequency of EC scattering obtained from the Thomson scattering region $\log {\rm{ }}\nu _{{\rm{peak}}}^{{\rm{ec}}}{\rm{ (Hz)}}$, the magnetic field strength $\log B\;(\rm {G})$, and the peak Lorentz factor $\log {\gamma _{\rm p}}$.}}
\end{table*}}\\
According to the calculation results, for LBLs and FSRQs, if the EC effect is taken into account, the average magnetic field strength of LBLs and FSRQs is ${\left\langle {\log {\rm{ }}B} \right\rangle _{{\rm{LBLs}}}} = 0.23{\rm{ G}}$ and ${\left\langle {\log {\rm{ }}B} \right\rangle _{{\rm{FSRQs}}}} = 0.61{\rm{ G}}$; the average values of the peak Lorentz factors are ${\left\langle {\log {\rm{ }}{\gamma _{\rm{p}}}} \right\rangle _{{\rm{LBLs}}}} = 3.10$ and ${\left\langle {\log {\rm{ }}{\gamma _{\rm{p}}}} \right\rangle _{{\rm{FSRQs}}}} = 2.83$. Compared with the results in \cite{2018ApJS..235...39C}, in the Thomson case, when the influence of the EC process on LBLs and FSRQs is considered, the values of the magnetic field strength and peak Lorentz factor can reach the reasonable parameter range of $0 \le \log B \le 1$ and $2 \le \log {\gamma _p} \le 3$ (\citealt{2015MNRAS.448.1060G}). In addition, the log-parabola model does not consider the combination of acceleration and cooling effects on the curvature of the ICs energy spectrum, and we leave this issue for future work.

\section*{Acknowledgements}

We thank the anonymous referee for valuable comments and
suggestions. This work was partially supported by the National Natural Science Foundation of China (grant Nos.11763005 and 11873043), and the Science and Technology Foundation of Guizhou Province (QKHJC[2019]1290). This work is also supported by the Graduate Research Foundation of Yunnan Normal University. The authors would like to express their gratitude to EditSprings (https://www.editsprings.com/) for the expert linguistic services provided.
\appendix                  

\section{Derivation of jet radiation zone parameters} \label{sect:appendix}
In order to further study the physical properties of the jet radiation zone, four parameters are introduced in Section \ref{sect:parameters}, which are as follows: the Doppler factor $D$; the peak Lorentz factor $\gamma_{\rm p}$ of nonthermal electrons; the radiation zone radius $R$; and the magnetic field of the jet $B$.

In the case of stochastic acceleration, and monoenergetic and instantaneous injection, the EED in the jet can be written as (\citealt{1962SvA.....6..317K,2006A&A...448..861M,2009A&A...504..821P,2011ApJ...736..128P,2018PASP..130h3001Z})
\begin{equation}
N(\gamma ) = {N_0}{\left( {\frac{\gamma }{{{\gamma _0}}}} \right)^{ - s - r\log \left( {\frac{\gamma }{{{\gamma _0}}}} \right)}}
\label{Eq:A1}
\end{equation}
where $\gamma _{\rm 0}$ is the initial electron Lorentz factor, $s$ is the electron spectrum index, $r$ is the electron spectrum curvature, ${N_0} = \frac{n}{{{\gamma _0}}}{10^{ - \frac{{{{(s - 1)}^2}}}{{4r}}}}\sqrt {\frac{r}{\pi }} \frac{1}{{\sqrt {\ln 10} }}$ is the normalization constant in units of $\rm c{m^{ - 3}}$, $n = \int {N(\gamma )} d\gamma$ is the particle number density, and $\gamma$ is the electronic Lorentz factor.

The traditional single-zone uniform SSC model assumes that there is a homogeneous spherical radiation zone (or blob) with radius $R$ moving along the jet orientation at a relativistic velocity $v = {\beta ^\prime } c$ ($c$ is the speed of light, and $ \Gamma  = 1/\sqrt {1 - {{\beta ^\prime } ^2}} $ is the bulk Lorentz factor), surrounded by a magnetic field $B$ (\citealt{2017ApJ...837...38K,2010MNRAS.402..497G,2018A&A...616A..63A}). The Doppler factor is denoted by
\begin{equation}
D = \frac{1}{{\Gamma (1 - {\beta ^\prime } \cos \theta )}}
\label{Eq:A2}
\end{equation}

We assume that when the relativistic jet has a small viewing angle of $\theta  \le 1/\Gamma $ , the above equation is written as  $D \approx \Gamma$.

In the co-moving coordinate system, the luminosity can be expressed as (\citealt{2011ApJ...736..128P,2018ApJS..235...39C})
\begin{equation}
L^\prime(\nu ^\prime ) = \frac{4}{3}\pi {R^3}\frac{{\nu^\prime {j_{\nu^\prime} }}}{{{\tau _{\nu^\prime} }}}\left[ {1 - \frac{{1 - {e^{ - 2{\tau _{\nu^\prime} }}}(1 + 2{\tau _{\nu^\prime} })}}{{2{\tau _{\nu^\prime}} ^2}}} \right]
\label{Eq:A3}
\end{equation}
where $j_{\nu^\prime}$ is the emission coefficient, ${\tau _{\nu^\prime} } = {k_{\nu^\prime} }R$ is the optical depth of the source, and ${k_{\nu^\prime} }$ is the absorption coefficient. For optically thin sources (${\tau _{\nu^\prime} } \ll 1$), the above equation can be written as $L^\prime(\nu^\prime ) = \frac{4}{3}\pi {R^3} \cdot 4\pi \nu^\prime j(\nu^\prime )$.

The radiation is usually optically thin at the peak frequency (\citealt{2018ApJS..235...39C}). According to ${L^\prime}(\nu^\prime ) = 4\pi d_L^2\nu {f^\prime}(\nu^\prime )$, we can get the following formula:
\begin{equation}
f^\prime(\nu^\prime ) = \frac{1}{3}\frac{{4\pi {R^3}}}{{d_L^2}}j(\nu^\prime )
\label{Eq:A4}
\end{equation}

For the emission coefficient of synchrotron radiation, we have
\begin{equation}
{j_{\rm syn}}(\nu^\prime ) = \frac{1}{{4\pi }}\int {N(\gamma )\frac{{d{P_{\rm syn}}}}{{d\nu^\prime }}} d\gamma
\label{Eq:A5}
\end{equation}
where ${P_{{\rm{syn}}}} = \frac{4}{3}{\sigma _T}c{\beta ^2}{\gamma ^2}{u_{\rm{B}}}$ is the emitted power by the synchrotron radiation of a single electron, ${u_{\rm B}} = \frac{{{B^2}}}{{8\pi }}$ is the energy density of the magnetic field, and ${\sigma _{\rm T}} = \frac{{8\pi {e^4}}}{{3{m^2}{c^4}}}$ is the Thomson cross section. For relativistic electrons, we take $\beta  \approx 1$ (\citealt{1979rpa..book.....R}).

In Equation (\ref{Eq:A5}), we introduce a delta approximated by
\begin{equation}
\frac{{d{P_{\rm syn}}}}{{d\nu^\prime }} = {P_{\rm syn}}\delta (\nu^\prime  - {\gamma ^2}{\nu _{c}})
\label{Eq:A6}
\end{equation}
where ${\nu _c} = 0.29(3/2){\nu _{\rm L}}\sin \theta$ is the characteristic frequency; ${\nu _{\rm L}} = \frac{{eB}}{{2\pi {m_e}c}}$ is the Larmor frequency of relativistic electrons, and $\theta $ is the angle between $\vec v$ and $\vec B$ (\citealt{1979rpa..book.....R,2011ApJ...736..128P}).

Substitute Equation (\ref{Eq:A1}), Equation (\ref{Eq:A6}), and Equation(\ref{Eq:A5}) into Equation (\ref{Eq:A4}), and we get the synchrotron radiation flux:
\begin{equation}
{f^\prime_{\rm syn}}(\nu^\prime ) = {G_1} \cdot B\frac{{{R^3}}}{{d_{L}^2}}\sqrt {\frac{\nu^\prime }{{{\nu _c}}}} N\left( {\sqrt {\frac{\nu^\prime }{{{\nu _c}}}} } \right)
\label{Eq:A7}
\end{equation}
where ${G_1} \approx 0.156\frac{{{\sigma _T}m{c^2}}}{e}$ is a constant.

According to Equation (\ref{Eq:A4}), the synchrotron luminosity is calculated as
\begin{equation}
{L^\prime_{\rm syn}}(\nu^\prime ) = {G_2} \cdot n\sqrt r {10^{ - \frac{{{{(s - 1)}^2}}}{{4r}}}}\gamma _0^2{R^3}{B^2}{\left( {\frac{\nu^\prime }{{{\nu^\prime _0}}}} \right)^{ - \frac{{(s - 3)}}{2} - \frac{r}{4}\log \left( {\frac{\nu^\prime }{{{\nu^\prime _0}}}} \right)}}
\label{Eq:A8}
\end{equation}
where ${G_2} \approx 0.051{\sigma _T}c$ is a constant and ${\nu^\prime _0} = \gamma _0^2{\nu _c}$.

For ICs, we consider that the ambient soft photons originate from synchrotron in the jet and have the same electron number distribution as those in SSC radiation. The SSC emission coefficient can be expressed as
\begin{equation}
{j_{\rm ic}}({\nu ^\prime }) = \frac{1}{{4\pi }}\int {N(\gamma )} \frac{{d{P_{\rm ic}}}}{{d{\nu ^\prime }}}d\gamma
\label{Eq:A9}
\end{equation}
where ${P_{\rm ic}} = \frac{4}{3}c{\sigma _{\rm T}}{\gamma ^2}{\beta ^2}{u_{\rm ph}}$ is the ICs emitted power by an electron, ${u_{\rm ph}} = {m_e}{c^2}\int {{\varepsilon _{\rm syn}}n({\varepsilon _{\rm syn}})} d{\varepsilon _{\rm syn}}$ is the photon energy density, ${\varepsilon _{\rm syn}} = \frac{{h{\nu _{\rm syn}}}}{{{m_e}{c^2}}}$ is the synchrotron photon energy, and ${n({\varepsilon _{syn}})}$ is the seed photon number density. In the Equation (\ref{Eq:A9}), we introduce a delta approximated by
\begin{equation}
\frac{{d{P_{\rm ic}}}}{{d{\nu ^\prime }}} = P{}_{\rm ic}\delta \left( {{\nu ^\prime } - \frac{4}{3}{\gamma ^2}\nu _{\rm peak}^{\prime \rm syn}} \right)
\label{Eq:A10}
\end{equation}
where $\nu _{\rm peak}^{\prime \rm syn} = {\nu _{\rm c}}\gamma _0^2{10^{\frac{{3 - s}}{r}}}$ is the synchrotron radiation peak frequency.

Based on the scattered photon energy, ICs is divided into the Thomson scattering region (${E_{\rm ph}} \ll {m_e}{c^2}$) and the KN scattering region (${E_{\rm ph}} \gg {m_e}{c^2}$).

For the optically thin case, we substitute Equation (\ref{Eq:A1}), Equation (\ref{Eq:A10}), and Equation (\ref{Eq:A9}) into Equation (\ref{Eq:A4}), and we get the SSC radiation flux in the Thomson scattering region as
\begin{equation}
f_{\rm ic,Th}^\prime ({\nu ^\prime }) = {{\sigma _T}c} \cdot B\frac{{{R^3}}}{{d_L^2}}\frac{{{u_{\rm ph}}}}{{\nu _{\rm peak}^{\prime \rm syn}}}\sqrt {\frac{3}{4}\frac{{{\nu ^\prime }}}{{{\nu _{\rm c}}}}} N\left( {\sqrt {\frac{3}{4}\frac{{{\nu ^\prime }}}{{{\nu _{\rm c}}}}} } \right)
\label{Eq:A11}
\end{equation}
where $u_{\rm ph}^{\rm Th}(\nu _0^\prime ) = h\int {{\nu ^\prime }{{\bar N}_{\rm syn}}(\nu _0^\prime )} \Omega ({\nu ^\prime },\nu _0^\prime ,\gamma )d\nu _0^\prime$ is the synchrotron photon energy density, $\Omega ({\nu ^\prime },\nu _0^\prime ,\gamma )$ is the Compton kernel function given by \cite{2019ApJ...873....7Z}, and ${{\bar N}_{\rm syn}}(\nu _0^\prime )= \frac{3}{4}\frac{R}{c}\frac{{{j_{\rm syn}}(\nu _0^\prime )}}{{h\nu _0^\prime }}$ is the average synchrotron photon density (\citealt{1979rpa..book.....R,2011ApJ...736..128P}).
In the Thomson scattering region, the SSC radiation luminosity can be expressed as
\begin{equation}
L_{\rm ic,Th}^\prime ({\nu ^\prime }) = {G_3} \cdot {n^2}\sqrt r {10^{ - \frac{{{{(s - 1)}^2}}}{{4r}}}}{10^{\frac{{2 - s}}{r}}}\gamma _0^4{R^4}{B^2}{\left( {\frac{{{\nu ^\prime }}}{{\hat \nu _0^\prime }}} \right)^{ - \frac{{(s - 3)}}{2} - \frac{r}{4}\log \left( {\frac{{{\nu ^\prime }}}{{\hat \nu _0^\prime }}} \right)}}
\label{Eq:A12}
\end{equation}
where ${G_3} = 0.479\frac{{{e^4}r_0^2}}{{m_e^2{c^3}}}$ is a constant and $\hat \nu _0^\prime  = \frac{4}{3}{\nu _c}\gamma _0^4{10^{\frac{{3 - s}}{r}}}$.

For extreme KN scattering, we rewrite Equation (\ref{Eq:A9}) as
\begin{equation}
{j_{\rm ic,KN}}({\nu ^\prime }) = h\int {d\gamma N(\gamma )\int {{\nu ^\prime }} } {\bar N_{\rm syn}}(\nu _0^\prime )\Omega ({\nu ^\prime },\nu _0^\prime ,\gamma )d\nu _0^\prime
\label{Eq:A13}
\end{equation}
where we introduce a delta approximated by $\Omega ({\nu ^\prime },\nu _0^\prime ,\gamma ) = \frac{{2\pi r_0^2c{\nu ^\prime }}}{{{\gamma ^2}}}\delta \left( {\gamma  - \frac{{h{\nu ^\prime }}}{{m{c^2}}}} \right)$, ${{\bar N}_{\rm syn}}(\nu _0^\prime )= \frac{3}{4}\frac{R}{c}\frac{{{j_{\rm syn}}(\nu _0^\prime )}}{{h\nu _0^\prime }}$ is the average synchrotron photon density, and ${r_0} = \frac{{{e^2}}}{{{m_e}{c^2}}}$ is the classical radius of the electron (\citealt{1979rpa..book.....R,2011ApJ...736..128P}).

In the KN scattering region, the SSC radiation luminosity can be expressed as
\begin{equation}
L_{\rm ic,KN}^\prime ({\nu ^\prime }) = {G_4} \cdot {n^2}\sqrt r {10^{ - \frac{{{{(s - 1)}^2}}}{{4r}}}}{R^4}B{\left( {\frac{{{\nu ^\prime }}}{{\hat \nu _a^\prime }}} \right)^{ - (s - 1) - r\log \left( {\frac{{{\nu ^\prime }}}{{\hat \nu _a^\prime }}} \right)}}
\label{Eq:A14}
\end{equation}
where ${G_4} = 130.428\frac{{{e^3}}}{h}r_0^2$ is a constant and $\hat \nu _a^\prime  = \frac{{{\gamma _0}m{c^2}}}{h}$ (\citealt{2018ApJS..235...39C}).

From the equation
\begin{equation}
\log \left( {\frac{{N(\gamma )}}{{{N_0}}}} \right) =  - r{\log ^2}\left( {\frac{\gamma }{{{\gamma _0}}}} \right) - s\log \left( {\frac{\gamma }{{{\gamma _0}}}} \right)
\label{Eq:A15}
\end{equation}
where setting ${\gamma _0} = {\gamma _{\rm p}}$, we can get ${\gamma _{\rm p}} = {\gamma _0}{10^{ - \frac{s}{{2r}}}}$ and ${N_{\rm p}} = {N_0}{10^{\frac{{{s^2}}}{{4r}}}}$.

We find that Equation (\ref{Eq:A8}), Equation (\ref{Eq:A12}) and Equation (\ref{Eq:A14}) all have the form of Equation (\ref{Eq:A1}).
According to the result of Equation (\ref{Eq:A15}), we can get $\nu _{\rm peak}^{\prime \rm syn}$, $\nu _{\rm peak}^{\prime \rm ic}$, $L_{\rm peak}^{\prime \rm syn}$, $L_{\rm peak,Th}^{\prime \rm ic}$ and $L_{\rm peak,KN}^{\prime \rm ic}$.

In the Thomson scattering region, we get
\begin{equation}
\left\{ \begin{array}{l}
\nu _{\rm peak}^{\prime \rm syn} = {c_1} \cdot B\gamma _{\rm p}^2{10^{\frac{1}{r}}}\\
L_{\rm peak}^{\prime \rm syn} = {c_2} \cdot n\sqrt r \gamma _{\rm p}^2{R^3}{B^2}\\
\nu _{\rm peak,Th}^{\prime \rm ic} = {c_3} \cdot B\gamma _{\rm p}^4{10^{\frac{2}{r}}}\\
L_{\rm peak,Th}^{\prime \rm ic} = {c_4} \cdot {n^2}\sqrt r \gamma _{\rm p}^4{R^4}{B^2}
\end{array} \right.
\label{Eq:A16}
\end{equation}
where ${c_1} = 9.942 \times {10^5}$, ${c_2} = 8.239 \times {10^{ - 16}}$, ${c_3} = 1.326 \times {10^6}$, and ${c_4} = 9.056 \times {10^{ - 41}}$ are constant.

In the KN scattering region, we get
\begin{equation}
\left\{ \begin{array}{l}
\nu _{\rm peak}^{\prime \rm syn} = {c_1} \cdot B\gamma _{\rm p}^2{10^{\frac{1}{r}}}\\
L_{\rm peak}^{\prime \rm syn} = {c_2} \cdot n\sqrt r \gamma _{\rm p}^2{R^3}{B^2}\\
\nu _{\rm peak,KN}^{\prime \rm ic} = {c_5} \cdot {\gamma _{\rm p}}{10^{ - \frac{1}{{2r}}}}\\
L_{\rm peak,KN}^{\prime \rm ic} = {c_6} \cdot {n^2}\sqrt r {R^4}B
\end{array} \right.
\label{Eq:A17}
\end{equation}
where ${c_5} = 1.236 \times {10^{20}}$, and ${c_6} = 1.378 \times {10^{ - 26}}$ are constant.

According to the research of \cite{2004A&A...422..103M,2008A&A...478..395M,2008A&A...489.1047M}, the relation between the electron spectrum curvature ($r$) and the curvature of the synchrotron bump ($\beta_{\rm syn}$) is $r \approx 5{\beta _{\rm syn}}$. Thus, if we solve Equation (\ref{Eq:A16}), in the Thomson scattering region, we can get
\begin{equation}
{\gamma _{\rm p}} = 8660.3{\left( {\frac{{\nu _{\rm peak,Th}^{\prime {\rm ic}}}}{{{{10}^{23}}\rm Hz}} \cdot \frac{{{{10}^{15}}\rm Hz}}{{\nu _{\rm peak}^{\prime {\rm syn}}}}} \right)^{\frac{1}{2}}}{10^{ - \frac{1}{{10{\beta _{\rm syn}}}}}}
\label{Eq:A18}
\end{equation}
\begin{equation}
B = 13.4109{\left( {\frac{{\nu _{\rm peak}^{\prime {\rm syn}}}}{{{{10}^{15}}\rm Hz}}} \right)^2}\left( {\frac{{{{10}^{23}}\rm Hz}}{{\nu _{\rm peak,Th}^{\prime {\rm ic}}}}} \right) 
\label{Eq:A19}
\end{equation}
\begin{equation}
R = 2.7235 \times {10^{16}}\left( {\frac{{\nu _{\rm peak,Th}^{\prime {\rm ic}}}}{{{{10}^{23}}\rm Hz}}} \right){\left( {\frac{{{{10}^{15}}\rm Hz}}{{\nu _{\rm peak}^{\prime {\rm syn}}}}} \right)^2}\left( {\frac{{L_{\rm peak}^{\prime {\rm syn}}}}{{{{10}^{45}}\rm erg\ {s^{ - 1}}}}} \right){\left( {\frac{{{{10}^{45}}\rm erg\ {s^{ - 1}}}}{{L_{\rm peak,Th}^{\prime {\rm ic}}}}} \right)^{\frac{1}{2}}} 
\label{Eq:A20}
\end{equation}

By substituting Equation (\ref{Eq:13}), Equation (\ref{Eq:15}) and Equation (\ref{Eq:A20}) into $R = c\Delta t\frac{D_{\rm var} }{{(1 + z)}}$ (e.g., \cite{1979rpa..book.....R}), we obtain
\begin{equation}
\begin{array}{l}
{D_{\rm Th}} = {\left( {247.2179} \right)^{\frac{1}{{4\xi }}}}{(1 + z)^{\frac{{1 - \xi }}{\xi }}}{\left( {\frac{{{{10}^{45}}{\rm{erg}}\;{{\rm{s}}^{{\rm{ - 1}}}}}}{{L_{{\rm{peak}},{\rm{Th}}}^{\rm{ic}}}}} \right)^{\frac{1}{{4\xi }}}}{\left( {\frac{{L_{\rm peak}^{\rm syn}}}{{{{10}^{45}}\rm erg\ {s^{ - 1}}}}} \right)^{\frac{1}{{2\xi }}}} {\left( {\frac{{\nu _{\rm peak,Th}^{\rm ic}}}{{{{10}^{23}}\rm Hz}}} \right)^{\frac{1}{{2\xi }}}}{\left( {\frac{{{{10}^{15}}\rm Hz}}{{\nu _{\rm peak}^{\rm syn}}}} \right)^{\frac{1}{\xi }}}{\left( {\frac{{1\ \rm day}}{{\Delta t}}} \right)^{\frac{1}{{2\xi }}}}{\left( {{\beta _{\rm syn}}} \right)^{\frac{1}{{8\xi }}}}
\end{array}
\label{Eq:A21}
\end{equation}
where $\xi  = 1 + \frac{{2{\alpha _{\nu _{\rm peak}^{\rm syn}}} - {\alpha _{\nu _{\rm peak}^{\rm ic}}}}}{4}$, and $\Delta {t}$ is the minimum timescale of variation.

Then, if we solve Equation (\ref{Eq:A17}), in the KN scattering region we can get
\begin{equation}
{\gamma _{\rm p}} = 809.332\left( {\frac{{\nu _{\rm peak,KN}^{\prime {\rm ic}}}}{{{{10}^{23}}\rm Hz}}} \right){10^{\frac{1}{{10{\beta _{\rm syn}}}}}}
\label{Eq:A22}
\end{equation}
\begin{equation}
{B} = 1.5356 \times {10^3}\left( {\frac{{\nu _{\rm peak}^{\prime {\rm syn}}}}{{{{10}^{15}}\rm Hz}}} \right){\left( {\frac{{{{10}^{23}}\rm Hz}}{{\nu _{\rm peak,KN}^{\prime {\rm ic}}}}} \right)^2}{10^{ - \frac{2}{{5{\beta _{\rm syn}}}}}}
\label{Eq:A23}
\end{equation}
\begin{equation}
{R} = 7.6456 \times {10^{13}}{\left( {{\beta _{\rm syn}}} \right)^{ - \frac{1}{4}}}{10^{\frac{2}{{5{\beta _{\rm syn}}}}}}\left( {\frac{{\nu _{\rm peak,KN}^{\prime \rm ic}}}{{{{10}^{23}}\rm Hz}}} \right){\left( {\frac{{{{10}^{15}}\rm Hz}}{{\nu _{\rm peak}^{\prime {\rm syn}}}}} \right)^{\frac{3}{2}}}\left( {\frac{{L_{\rm peak}^{\prime {\rm syn}}}}{{{{10}^{45}}\rm erg\ {s^{ - 1}}}}} \right){\left( {\frac{{{{10}^{45}}\rm erg\ {s^{ - 1}}}}{{L_{\rm peak,KN}^{\prime {\rm ic}}}}} \right)^{\frac{1}{2}}}
\label{Eq:A24}
\end{equation}

By substituting Equation (\ref{Eq:13}), Equation (\ref{Eq:15}) and Equation (\ref{Eq:A24}) into $R = c\Delta t\frac{D_{\rm var} }{{(1 + z)}}$ (e.g., \cite{1979rpa..book.....R}), we obtain
\begin{equation}
\begin{array}{l}
{D_{\rm KN}} = {(0.0295)^{\frac{1}{\zeta }}}\left( {\left( {1 + z} \right)^{\frac{{\zeta  - 2}}{\zeta }}} \right){\left( {\frac{{L_{\rm peak}^{\rm syn}}}{{{{10}^{45}}\rm erg\ {s^{ - 1}}}}} \right)^{\frac{1}{\zeta }}}{\left( {\frac{{{{10}^{45}}\rm erg\ {s^{ - 1}}}}{{L_{\rm peak,KN}^{\rm ic}}}} \right)^{\frac{1}{{2\zeta }}}}{\left( {\frac{{{{10}^{15}}\rm Hz}}{{\nu _{\rm peak}^{\rm syn}}}} \right)^{\frac{3}{{2\zeta }}}}{\left( {\frac{{\nu _{\rm peak,KN}^{\rm ic}}}{{{{10}^{23}}\rm Hz}}} \right)^{\frac{1}{\zeta }}}{\left( {\frac{{1\rm day}}{{\Delta t}}} \right)^{\frac{1}{\zeta }}}{10^{\frac{2}{{5{\beta _{\rm syn}}\zeta }}}}{\left( {{\beta _{\rm syn}}} \right)^{ - \frac{1}{{4\zeta }}}}
\end{array}
\label{Eq:A25}
\end{equation}
where $\zeta  = \frac{{5 + 2{\alpha _{\nu _{\rm peak}^{\rm syn}}} - {\alpha _{\nu _{\rm peak}^{\rm ic}}}}}{2}$, and $\Delta {t}$ is the minimum timescale.
%




\end{document}